\documentclass[fleqn,usenatbib]{mnras}
\usepackage{newtxtext,newtxmath}

\usepackage[T1]{fontenc}

\DeclareRobustCommand{\VAN}[3]{#2}
\let\VANthebibliography\thebibliography
\def\thebibliography{\DeclareRobustCommand{\VAN}[3]{##3}\VANthebibliography}

\usepackage{graphicx}	
\usepackage{amsmath}	

\usepackage{amssymb}	
\usepackage{tablefootnote}
\usepackage{multirow}
\usepackage{float}
\usepackage{gensymb}
\usepackage{enumerate}
\usepackage{algorithm}
\usepackage{algpseudocode} 
\usepackage{amsmath}
\usepackage{pgfplots}
\usepackage{threeparttable}
\usepackage{bbding}
\usepackage{tabularx}
\usepackage{color}
\usepackage{caption}
\usepackage[]{hyperref}

\title[Classification methods on astronomical spectra]{Data mining techniques on astronomical spectra data. II : Classification Analysis}

\author[H.Yang, L.Zhou, J.Cai, et al.]{Haifeng Yang$^{1}$,
Lichan Zhou$^{1}$,
Jianghui Cai$^{1,2}$\thanks{E-mail: jianghui@tyust.edu.cn},
Chenhui Shi$^{1}$,
Yuqing Yang$^{1}$,
Xujun Zhao$^{1}$,
Juncheng Duan$^{1}$ \and and
Xiaona Yin$^{1}$
\\
$^{1}$School of Computer Science and Technology, Taiyuan University of Science and Technology, Taiyuan 030024, China\\
$^{2}$School of Computer Science and Technology, North University of China, Taiyuan 030051, China
}

\date{Accepted XXX. Received YYY; in original form ZZZ}

\pubyear{2022}

\begin{document}

\label{firstpage}
\pagerange{\pageref{firstpage}--\pageref{lastpage}}
\maketitle

\begin{abstract}
Classification is valuable and necessary in spectral analysis, especially for data-driven mining. Along with the rapid development of spectral surveys, a variety of classification techniques have been successfully applied to astronomical data processing. However, it is difficult to select an appropriate classification method in practical scenarios due to the different algorithmic ideas and data characteristics. Here, we present the second work in the data mining series - a review of spectral classification techniques. This work also consists of three parts: a systematic overview of current literature, experimental analyses of commonly used classification algorithms and source codes used in this paper.  
Firstly, we carefully investigate the current classification methods in astronomical literature and organize these methods into ten types based on their algorithmic ideas. For each type of algorithm, the analysis is organized from the following three perspectives. (1) their current applications and usage frequencies in spectral classification are summarized; (2) their basic ideas are introduced and preliminarily analysed; (3) the advantages and caveats of each type of algorithm are discussed. 
Secondly, the classification performance of different algorithms on the unified data sets is analysed. Experimental data are selected from the LAMOST survey and SDSS survey. 
Six groups of spectral data sets are designed from data characteristics, data qualities and data volumes to examine the performance of these algorithms. Then the scores of nine basic algorithms are shown and discussed in the experimental analysis.  
Finally, nine basic algorithms source codes written in python and manuals for usage and improvement are provided.

\end{abstract}




\section{Introduction}
Classification of astronomical spectra is an essential part of astronomical research. It can provide valuable information about the formation and evolution of the Universe. With the implementation of sky survey projects \citep{2015RAA....15.1089L, Zhao_2012}, a large number of methods have been applied to automatically handle various astronomical classification tasks \citep{yang2021spectral, baron2019machine, luo2004design, luo2013data, yang2021spectral,10.1016/j.ins.2022.03.027,10.1016/j.eswa.2022.117018,10.1016/j.eswa.2019.112846,10.1145/3522592}. However, classification methods achieve different results on different data, so it is difficult to evaluate the classification performance and determine the application scenarios.

In this paper, we investigate lots of classification methods on astronomical spectra data and organize them into ten types. Each type of them is displayed based on its usage frequencies in astronomical tasks. And we mainly discuss its application scenarios, main ideas, merits and caveats. Then, we construct six collections of data sets to provide a unified measurement platform.
For the astronomical classification tasks ( A/F/G/K stars classification, star/galaxy/quasar classification and rare object identification), we construct data sets from three criteria including data characteristics, signal-to-noise ratio (S/N), data volumes. Then we compare the performance of nine basic classification methods on the aforementioned data sets and give an objective appraisal of the classification results.
Besides, the source codes of each testing algorithm help researchers to study further and a brief manual about usage and revision tips of our program is provided in this work.

The rest of this paper is organized as follows. In Section 2, classification methods on astronomical spectra data are briefly introduced from application scenarios, main ideas, merits, and caveats. In Section 3, experiments on three tasks of A/F/G/K stars classification, star/galaxy/quasar classification and rare object identification are carried out. Section 4 represents python source codes of the above experiments and a manual about how to use and revise our codes. Finally, a discussion is drawn and our future work is discussed in Section 5.

\section{Investigation of Classification Methods on Astronomical Spectra Data}

\begin{figure*}
\centering
\includegraphics[width=12.879cm,height=7.29cm]{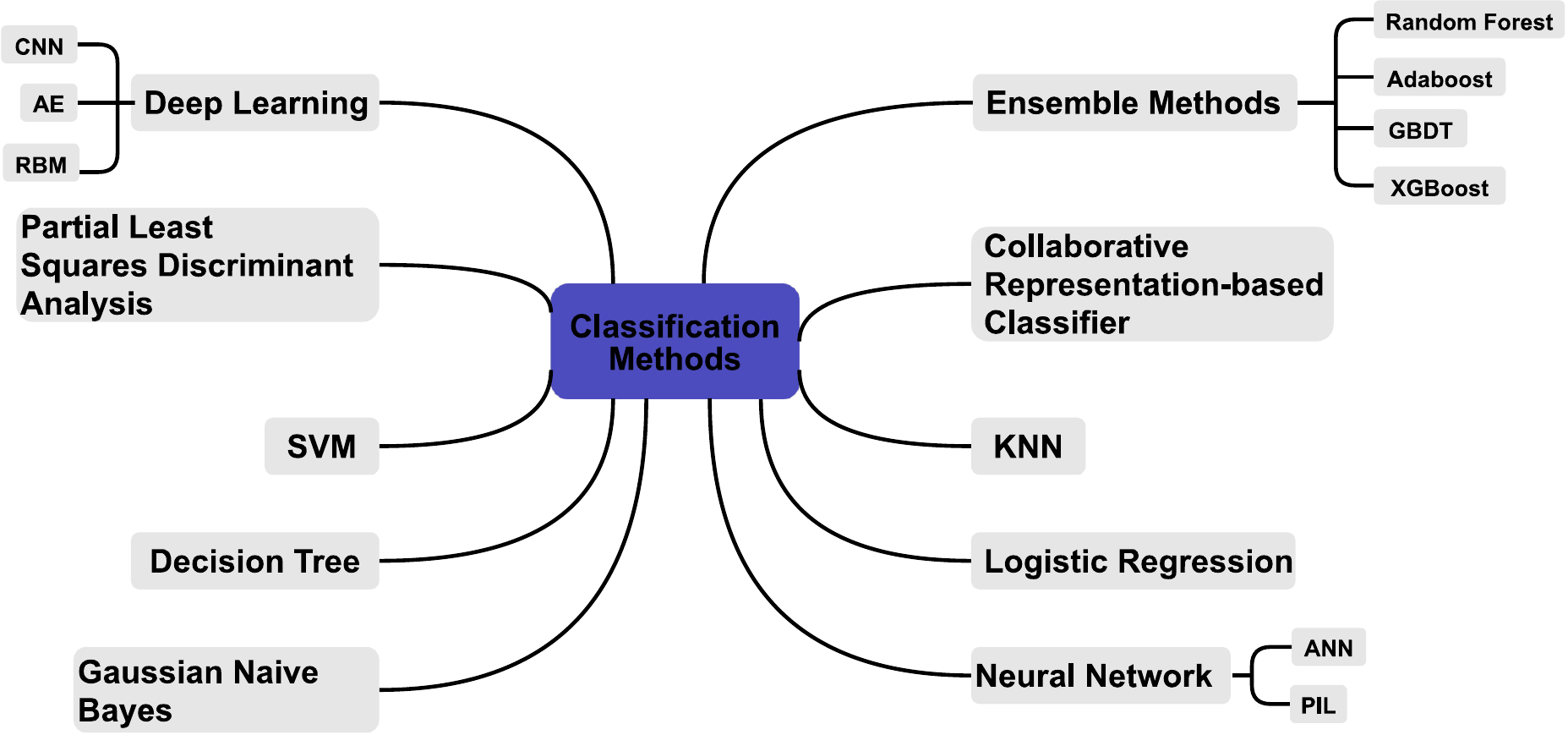}
\caption{Classification methods on astronomical spectra data. We pay more attention on main ideas, advantages, caveats and application scenarios of these methods.}
\label{fig:methods_mind}   
\end{figure*}

\begin{figure*}
\centering
\includegraphics[width=12.62cm,height=7.27cm]{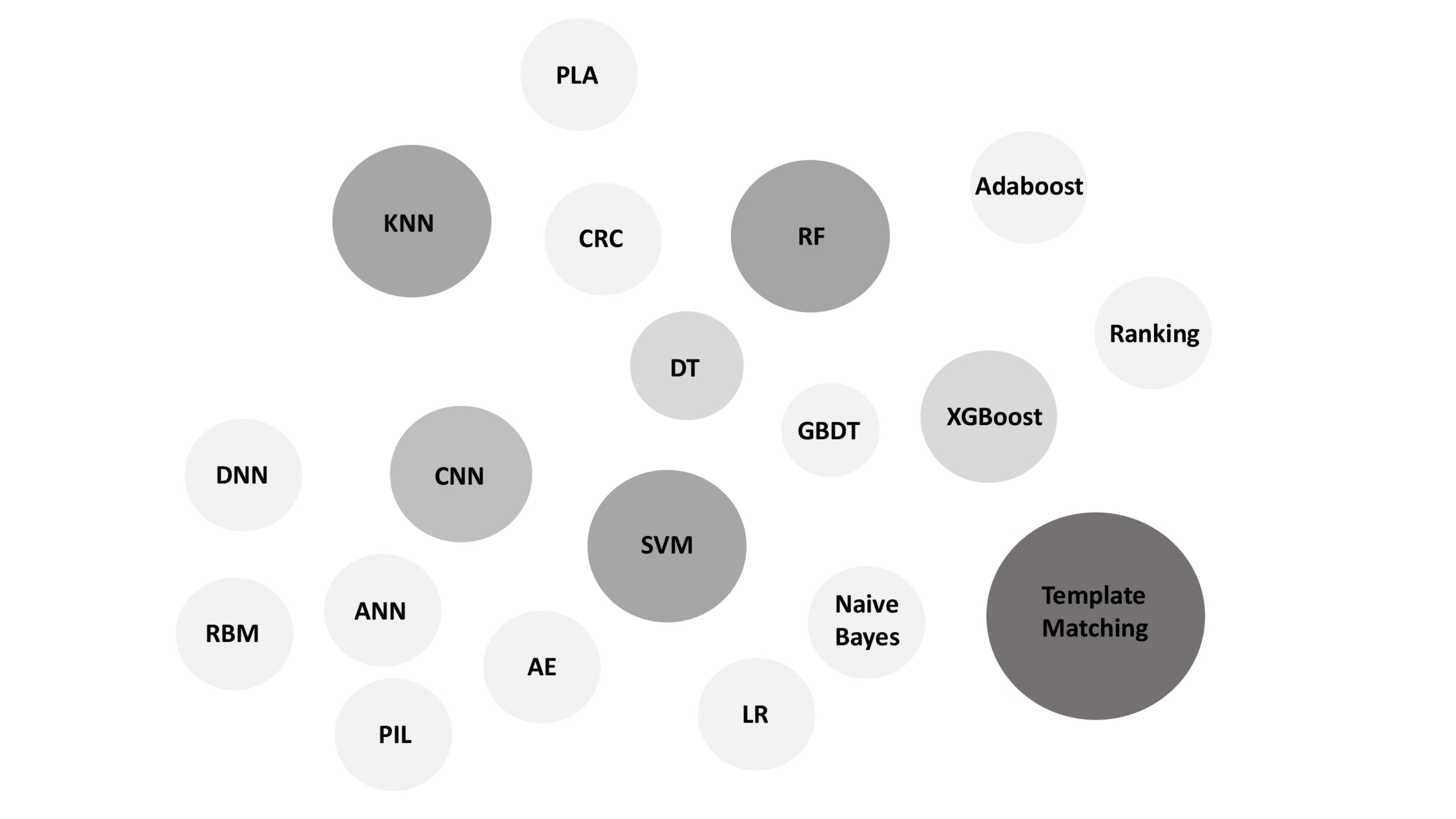}
\caption{Classification methods on astronomical spectra data. The size of circles means the usage frequencies of each type of algorithm in our paper. And the color of circles is consistent with their sizes. That is, bigger and deeper circles mean that this type of algorithm is more frequently applied in astronomical tasks.}
\label{fig:classification_wordcloud}   
\end{figure*}

The commonly used classification methods on astronomical spectra are shown in Fig. \ref{fig:methods_mind}. Each type of methods has its own characteristics and applicable data sets. And some of them have been widely used for spectral classification, like template matching, K-nearest neighbor (KNN) based classification algorithms and support vector machine (SVM) based classification algorithms, but some of them are rarely used, like logistic regression (LR) based classification algorithms and collaborative representation based classifier (CRC) (Fig. \ref{fig:classification_wordcloud}). Here, for each type of investigated methods, we analyse its application scenarios on astronomical spectra and give some objective appraisals. Then we introduce the main ideas, advantages and caveats of these methods.



\subsection{Template matching}
Template matching is a flexible and relative straightforward technique. The classification process of template matching is to build a template database for each class, then divide the unknown data into the most similar template data \citep{template_matching}. In astronomy, template matching matches spectral lines with templates and there is no training stage. So it has been widely applied in celestial object classification, redshift estimation, stellar parameters estimation and other projects \citep{subbarao2002sloan, sdssimagepipeline, Westfall_2019, 2015RAA....15.1089L, Zhao_2012}.
Table \ref{Investigate_template} shows the main astronomical spectral investigations of template matching.

\begin{table*}
\centering
\caption{Investigations of template matching on astronomical spectra data}
\label{Investigate_template}
\begin{threeparttable}
\resizebox{\linewidth}{!}{
\begin{tabular}{cccc} 
\hline
Merits                                                                                                                                                                                                                                                    & Caveats                                                                                                                                                                                             & References                                                                                                                                                                                                                                                                                                                                                                                                                                                                                                                                                                           \\ 
\hline
\multicolumn{1}{l}{\begin{tabular}[c]{@{}l@{}}Straightforward and simple\\ \\ \\ Applied to stellar spectra, rare objects, etc$^{1}$\\ \\ \\Fast because of without training stage\end{tabular}} & \multicolumn{1}{l}{\begin{tabular}[c]{@{}l@{}}Poor performance on low-quality spectra \\ \\ \\Spectra without templates can not be classified well\\ \\ \\Poor performance on unbalanced data\end{tabular}} & \multicolumn{1}{l}{\begin{tabular}[c]{@{}l@{}}{\citet{template_matching}, \citet{Duan_2009}}, \\{\citet{2012SPIE.8451E..37D}, \citet{ramirez2020spectral}},\\{\citet{2019PASP..131g7001W}, \citet{Almeida_2010}, \citet{2018A&A...616A.135M}}, \\{\citet{wang2018spectral}, \citet{2016RAA....16..110L}, \citet{2016A&A...593A..58J}}, \\{\citet{Sako_2018}, \citet{Zhong_2015-M}},\\{\citet{Zhong_2015}, \citet{2012AJ....144..144B}}, \\{\citet{2014AJ....147..101W}, \citet{Gray_2014}}, \\{\citet{2011PASP..123..638M}, \citet{2021A&A...649L...8K}}, \\{\citet{Kesseli_2017}, \citet{2019MNRAS.483.3196C}},\\ {\citet{2017MNRAS.471.2013A}, \citet{2016NewA...44...66Z}, \citet{2015MNRAS.450.2615S}},\\ {\citet{10.1093/mnras/stab1238}, \citet{2019ApJS..245...33G}}\end{tabular}}  \\
\hline
\end{tabular}
}
\begin{tablenotes}
    \footnotesize
    \item[1] : Subtypes of O star, Subtypes of B star, galaxy/others, etc.
    \end{tablenotes}
    \end{threeparttable}
\end{table*}

Template matching is often used to classify stars, galaxies and quasars and further analyze other properties of spectra. \citet{Duan_2009} used spectral line matching to identify the observed spectra class and achieved a high accuracy about 92.9\%, 97.9\% and 98.8\% for stars, galaxies and quasars, respectively. They also obtained a byproduct: high precision of redshift. \citet{Gray_2014} used template matching for Morgan-Keenan (MK) classification and built an expert computer program imitating human classifiers. It was automatic and had comprehensible results. \citet{wang2018spectral} used the line intensity to classify spectra \citep{2018A&A...616A.135M,2019PASP..131g7001W}.

Template matching is also used to find peculiar objects like supernovas, M dwarfs, B stars and M giants, Double-peak emission line galaxies \citep{Sako_2018, Zhong_2015, Zhong_2015-M, ramirez2020spectral, Maschmann_2020}. \citet{Zhong_2015} applied a template-fit method to identify and classify late-type K and M dwarfs from LAMOST. 2612 late-K and M dwarfs were identified which can help researchers to investigate the chemo-kinematics of the local Galactic disk and halo. \citet{Maschmann_2020} used two Guassian functions to fit the emission lines to find double-peak candidates and finally they found 5663 double-peak emission line galaxies at z < 0.34. Meanwhile, there is an important issue for rare object identification using template matching, that is, classifiers require sufficient high-quality spectral templates. In order to obtain ample qualified rare templates, researchers tried to construct new templates \citep{Kesseli_2017, 2014AJ....147..101W}. 

Template matching has been widely used in lots of surveys. However, some spectra are of low quality, template matching can not obtain precise results on redshift estimation, stellar parameters estimation and classification \citep{ article_interpolator}. Hence, for the inferior quality spectra, other machine learning algorithms like SVM based classification algorithms and artificial neural network (ANN) based classification algorithms are employed to get robust results. The other defect of template matching is that, for rare objects, we do not have enough samples to get representative template spectra. So rare objects are often misclassified.

\subsection{K-Nearest neighbour based classification algorithms}
\begin{figure}
\centering
\includegraphics[width=5.92cm,height=4.88cm]{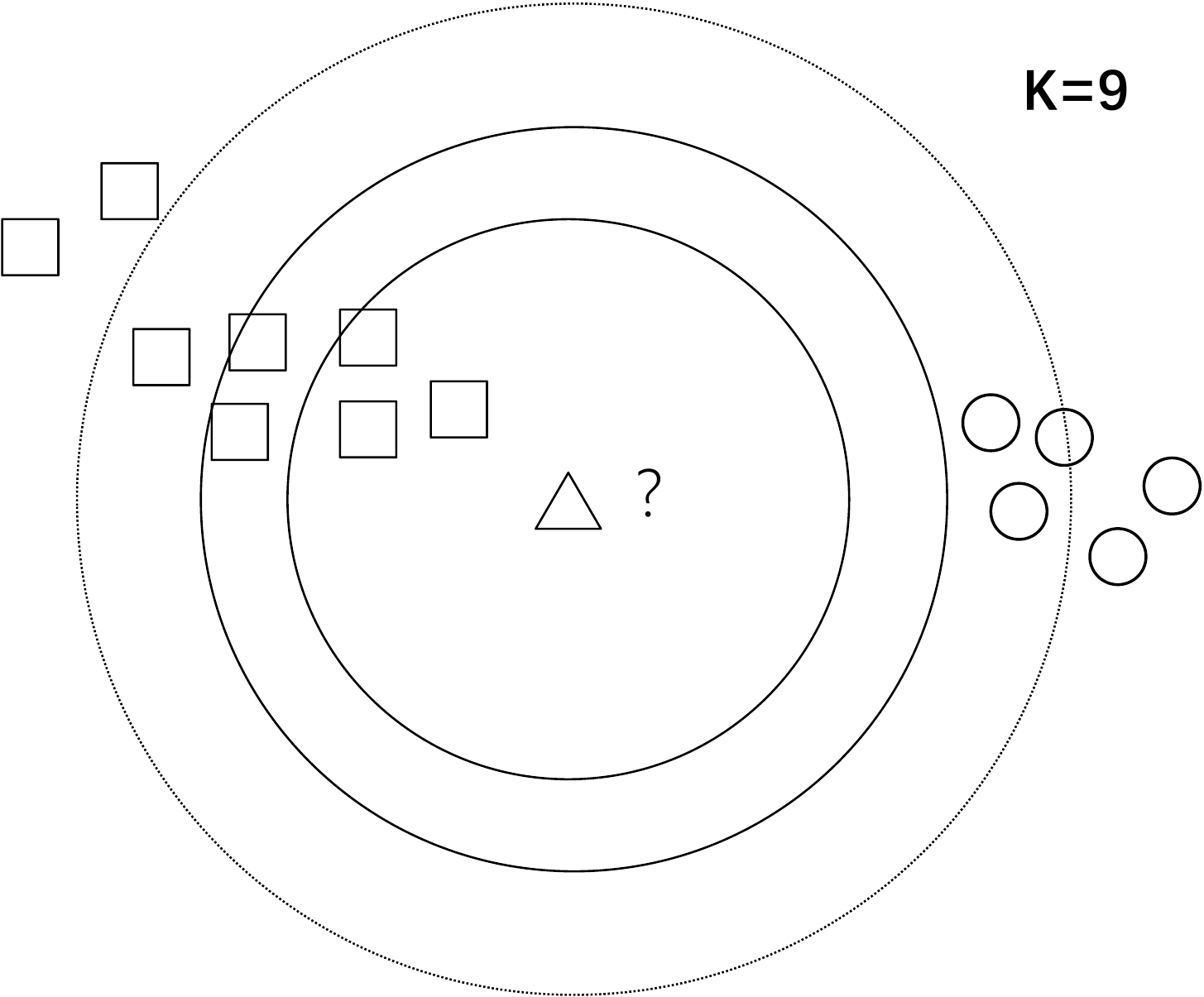}
\caption{Process of KNN. The middle triangle is the object needed to be predicted. Rectangles and circles are two types of known objects. The dashed circle needs to be enlarged to find k neighbours. K=9: three circles and six rectangles are the triangle's k neighbours.}

\label{fig:KNN}
\end{figure}

K-Nearest Neighbor (KNN) based classification algorithms assign labels to the target based on the majority labels of its K closest objects. More in depth explanations of KNN based classification algorithms can be found in \citet{ZHANG20072038, DENG2016143}. The main ideas are shown in Fig. \ref{fig:KNN}. They are intelligible and their time complexity is linear to the data volume. Taking these into consideration, KNN based classification algorithms have been used to classify astronomical spectra and combined with other methods to improve classification accuracy. Table \ref{table_KNN_related} displays the major astronomical applications of KNN based classification algorithms. 

\begin{table*}
\centering
\caption{Investigations of KNN based classification algorithms on astronomical spectra data}
\label{table_KNN_related}
\begin{threeparttable}
\resizebox{\linewidth}{!}{
\begin{tabular}{cccc} 
\hline
 Merits                                                                                                                                                                                                                                                                                                                                                                                                                                                                 & Caveats                                                                                                                                      & References                                                                                                                                                                                                          \\ 
\hline
 \multicolumn{1}{l}{\begin{tabular}[c]{@{}l@{}}Accurate and fast on proper features$^{1}$ \\ \\Combined with other methods$^{2}$ to improve accuracy\\ \\Applied to stellar spectra and subtypes classification$^{3}$\end{tabular}} & \multicolumn{1}{l}{\begin{tabular}[c]{@{}l@{}}Generally, RF\textgreater{}SVM\textgreater{}KNN\\ \\limited to large redshift objects\\ \\ Misclassification on F,G,K stars\end{tabular}} & \multicolumn{1}{l}{\begin{tabular}[c]{@{}l@{}}{\citet{2019AJ....158..188B}},\\{\citet{2020MNRAS.498.1750A}},\\{\citet{article_svm_knn}},\\{\citet{2019ApJ...886..128B}},\\{\citet{2019MNRAS.483.5077A}},\\{\citet{2021ChJPh..69..303X}},\\{\citet{2017A&A...605A.123P}},\\{\citet{sookmee2020globular}}\end{tabular}}  \\
\hline
\end{tabular}
}
 \begin{tablenotes}
    \footnotesize
    \item[1] : Features extracted by CNN; astronomical specific information.
    \item[2] : SVM, CNN, Decision tree, etc.
    \item[3] : MK classification, star/galaxy/quasar classification, Hot subdwarfs, symbiotic stars, Be stars, LSP/HSP, etc.
    \end{tablenotes}
    \end{threeparttable}

\end{table*}

KNN can be used for stellar classification. \citet{2019AJ....158..188B} used KNN and random forest (RF) for MK classification of stellar spectra. Considering high dimensional spectra data, they extracted absorption lines of spectra to reduce the time complexity. The results showed that KNN had a shorter training time but a longer testing time than RF. KNN could obtain the same accuracy as RF when using hybrid methods or oversampling balancing techniques. But for O-type stars which are few in the data sets, KNN performed poorly. This is a common phenomenon in most classification applications, that is, it is hard to get good classification results in unbalanced data sets.

For complex spectral classification tasks, it is not a good choice to only use the basic KNN based classification methods. Because from the comparison results of different classification methods, researchers found that good results were often produced by SVM or RF, rather than KNN \citep{2021ChJPh..69..303X, 2020MNRAS.498.1750A, 2017A&A...605A.123P}. To obtain better results, some improvements to KNN were also proposed, like KNN-DD to detect known outliers \citep{2012LNS...902..275B} and ML-KNN: a lazy learning approach to multi-label learning \citep{ZHANG20072038}. In addition, many researchers combined KNN with other methods to reduce the misclassification rate, like SVM+KNN to correct some prediction errors \citep{article_svm_knn}. And its classification accuracy of quasars reached 97.99\%.  

KNN based classification algorithms are arguably simple and efficient machine learning algorithms. And they have been demonstrated to be competitive methods because of the high accuracy under the premise of their simplicity and rapidness \citep{10.1007/s11222-009-9153-8, sookmee2020globular, guzman2018stellar}. They use Euclidean distance to measure the similarity of data and perform better on low dimensional data. After preprocessing high dimensional spectra, KNN based classification algorithms can also be applied in astronomy, such as star/galaxy/quasar classification and classification of small radial velocity objects. However, from the investigated researches, KNN based classification algorithms mainly suffer from three disadvantages: 1) the only hyper-parameter K is difficult to determine. 2) KNN based classification algorithms are ineffective for star classification because of the misclassification between adjacent classes. 3) Unbalanced data is another challenge for KNN based classification algorithms. Recently, some algorithms like Synthetic Minority Over-Sampling Technique (SMOTE) have been employed to adjust the data volume distributions to solve the third issue.

\subsection{Support vector machine based classification algorithms} 
Support vector machine (SVM) based classification algorithms are binary classifiers that learn a boundary from the training data to classify two types of data. And multiple binary SVM classifiers can be integrated into a multi-label classifier. Generally, the classification precision and robustness of SVM based classification algorithms are relatively superior to other single classifiers (non-ensemble algorithms). Table \ref{table_svm_related} shows the main astronomical researches of SVM based classification algorithms. 

\begin{figure}
\centering
\includegraphics[width=6cm,height=4.5cm]{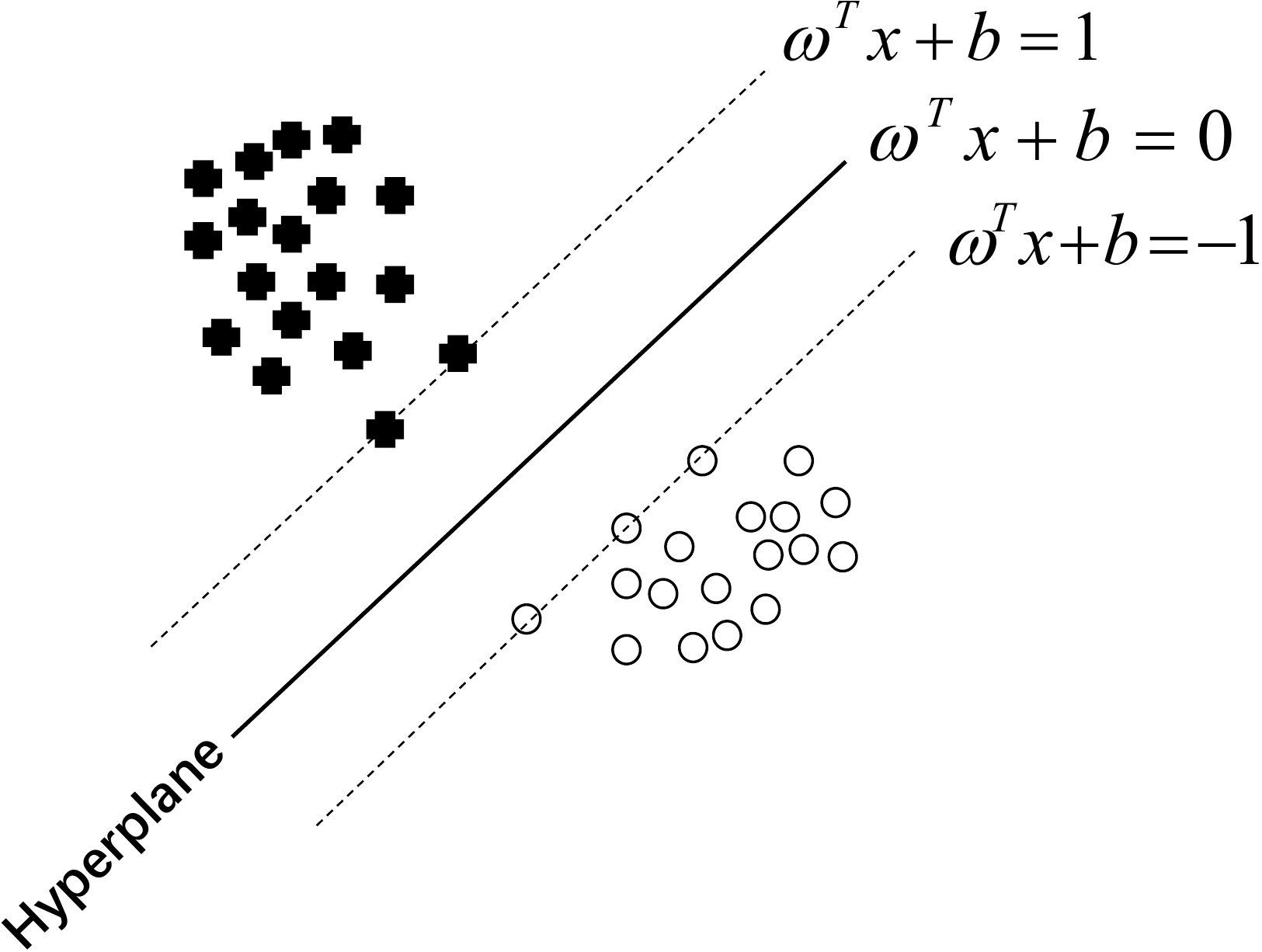}
\caption{Process of SVM. Black circles and white circles are two types of unknown groups. Solid line is the hyperplane to separate groups. The proper gap between paralleled dashed lines and solid lines can avoid overfitting and underfitting.}
\label{fig:SVM}

\end{figure}

\begin{table*}
\centering
\caption{Investigations of SVM based classification algorithms on astronomical spectra data}
\label{table_svm_related}
\begin{threeparttable}
\resizebox{\linewidth}{!}{
\begin{tabular}{clll} 
\hline
\multicolumn{1}{c}{Merits}                                                                                                                                                                                                                                                                                                                                                                                                                                                                                                                               & \multicolumn{1}{c}{Caveats}                                                                                                                                                                                   & \multicolumn{1}{c}{References}              \\ 
\hline
\multirow{11}{*}{\begin{tabular}[c]{@{}l@{}}Accurate and fast on proper features$^{1}$\\ \\ Optimizations of SVM to improve accuracy$^{2}$\\ \\Applied to unbalanced and large-scale data sets\\ \\Applied to stellar spectra and subtypes classification$^{3}$\end{tabular}} & \multirow{11}{*}{\begin{tabular}[c]{@{}l@{}}Stellar loci is better than SVM on MK classification\\ \\1D SCNN is better than SVM on stellar classification\\ \\limited to large redshift objects\\ \\ Need sufficient valid samples\end{tabular}} & {\citet{2015RAA....15.1137L}, \citet{guzman2018stellar}},\\ &                                                                                                                                                                                                                                                                                                                                                                                                                                                                                                                                                                                                                                                                                                                                                            & {\citet{2020MNRAS.498.1750A}, \citet{9079538}},                \\
                      &                                                                                                                                                                                                                                                                                                                                                                                                                                                                                                                                                                                                                                                                                                                                                                        & {\citet{2019MNRAS.483.4774L}, \citet{Govada_2015}},           \\
                      &                                                                                                                                                                                                                                                                                                                                                                                                                                                                                                                                                                                                                                                                                                                                                                        & {\citet{2014PASA...31....1F}, \citet{barrientos2020machine}},  \\
                      &                                                                                                                                                                                                                                                                                                                                                                                                                                                                                                                                                                                                                                                                                                                                                                        & {\citet{2012A&A...541A..50S}, \citet{2012A&A...537A..42T}},      \\
                      &                                                                                                                                                                                                                                                                                                                                                                                                                                                                                                                                                                                                                                                                                                                                                                        & {\citet{kou2020new}, \citet{liu2021stellar}},                  \\
                      &                                                                                                                                                                                                                                                                                                                                                                                                                                                                                                                                                                                                                                                                                                                                                                        & {\citet{2013A&A...557A..16M}, \citet{article_svm_knn}},       \\
                      &                                                                                                                                                                                                                                                                                                                                                                                                                                                                                                                                                                                                                                                                                                                                                                        & {\citet{2017A&A...606A..39S}, \citet{2017Ap&SS.362...98L}},      \\
                      &                                                                                                                                                                                                                                                                                                                                                                                                                                                                                                                                                                                                                                                                                                                                                                        & {\citet{2013PASA...30...24Y}, \citet{2021ChJPh..69..303X}},    \\
                      &                                                                                                                                                                                                                                                                                                                                                                                                                                                                                                                                                                                                                                                                                                                                                                        & {\citet{dong2020cascaded}, \citet{kong2018spectral}},          \\
                      &                                                                                                                                                                                                                                                                                                                                                                                                                                                                                                                                                                                                                                                                                                                                                                        & {\citet{2019ApJ...886..128B}, \citet{2016MNRAS.455.4289L}}    \\
\hline
\end{tabular}
}

\begin{tablenotes}
    \footnotesize
    \item[1] : Multi-frequency, color space, spectral lines, etc
    \item[2] : Within-Class Scatter and Between-Class Scatter (WBS-SVM), OCSVM, Twin Support Vector Machine(TWSVM).
    \item[3] : MK classification, LSP/HSP, K/F/G stars, Type IIP/IIL Supernovae, etc.
    \end{tablenotes}
    \end{threeparttable}

\end{table*}

Spectral classification is a common astronomical task for SVM based classification algorithms \citep{brice2019classification, barrientos2020machine, tao2018automated, guzman2018stellar, 2015RAA....15.1137L, liu2021stellar, liu2018classification}. \citet{2012A&A...541A..50S} used the infrared information to separate galaxies from stars and the accuracy reached 90\% for galaxies and 98\% for stars. \citet{2013A&A...557A..16M} trained a SVM classifier to classify stars, active galactic nucleus (AGNs) and galaxies using spectroscopically confirmed sources from the VIPERS and VVDS surveys. In the stellar spectral classification, A stars and G stars can be identified easily, while it was hard to identify O, B and K stars. Because the differences in the spectral features between late B type and early A type stars or between late G and early K type stars were very weak \citep{2015RAA....15.1137L}. \citet{dong2020cascaded} used SVM and cascaded dimensionality reduction techniques to classify spectra, which is better than principal component analysis (PCA) or t-distributed stochastic neighbor embedding (T-SNE).

In addition to classification, SVM based classification algorithms can also be used for peculiar spectra identification \citep{9079538}. More depth details of rare objects such as carbon stars and variable objects can be found in \citet{2012A&A...544A..95G, 2013ApJ...765...12G, 2021MNRAS.503.3828B, https://doi.org/10.48550/arxiv.2203.08125, kong2018spectral, kou2020new, 2017A&A...606A..39S, 9079538}. \citet{2020A&A...642A.103S} detected anomalous in the mid-infrared data using one-class SVM. Among the 36 identified anomalous, 53\% of them were low redshift galaxies, 33\% were particular quasi-stellar objects (QSOs), 3\% were galactic objects in dusty phases of their evolution and 11\% were unknown objects. The main problem in this task is that the number of some types of rare samples is far smaller than normal samples. So the classification model can not identify the rare classes well. There are also many approaches to solve this problem, like data augmentation, over-sampling, etc. \citet{2017Ap&SS.362...98L} proposed an entropy based methods for unbalanced spectral classification. And the performance was better than using KNN and SVM directly. 

SVM based classification algorithms are binary classifiers with rigorous mathematical theory. They try to find the optimal separating hyperplane to divide data into two categories (Fig. \ref{fig:SVM}). For multi-label classification, One-VS-One (OVO), One-VS-All (OVA) and Directed Acyclic Graph (DAG) are the main tactics to train different classifiers. There are two important tricks of SVM based classification algorithms. One is soft margin which uses a robust partition boundary to separate two types of data and tolerates the misclassification of some abnormal data. The other one is kernel function which can map linearly inseparable data into a linearly separable high-dimensional space. However, SVM based classification algorithms adopt kernel matrix to measure the similarity  of samples. So the computation time and space are two vital issues for classifiers on large amounts of data.

SVM based classification algorithms are promising classification methods that have a convincing theory and robust results. They have attracted a good deal of attention due to their high accuracy in multi-dimensional space and already have been applied to astronomical spectral classification, such as star/galaxy/quasar classification, stellar spectral classification and novelty detection. However, the time complexity of SVM based classification algorithms is exponentially related to the training size. So, it is indispensable to pre-process astronomical spectra to reduce the training time.

\subsection{Decision tree based classification algorithms}
\begin{figure}
\centering
\includegraphics[width=6.3cm,height=2.8cm]{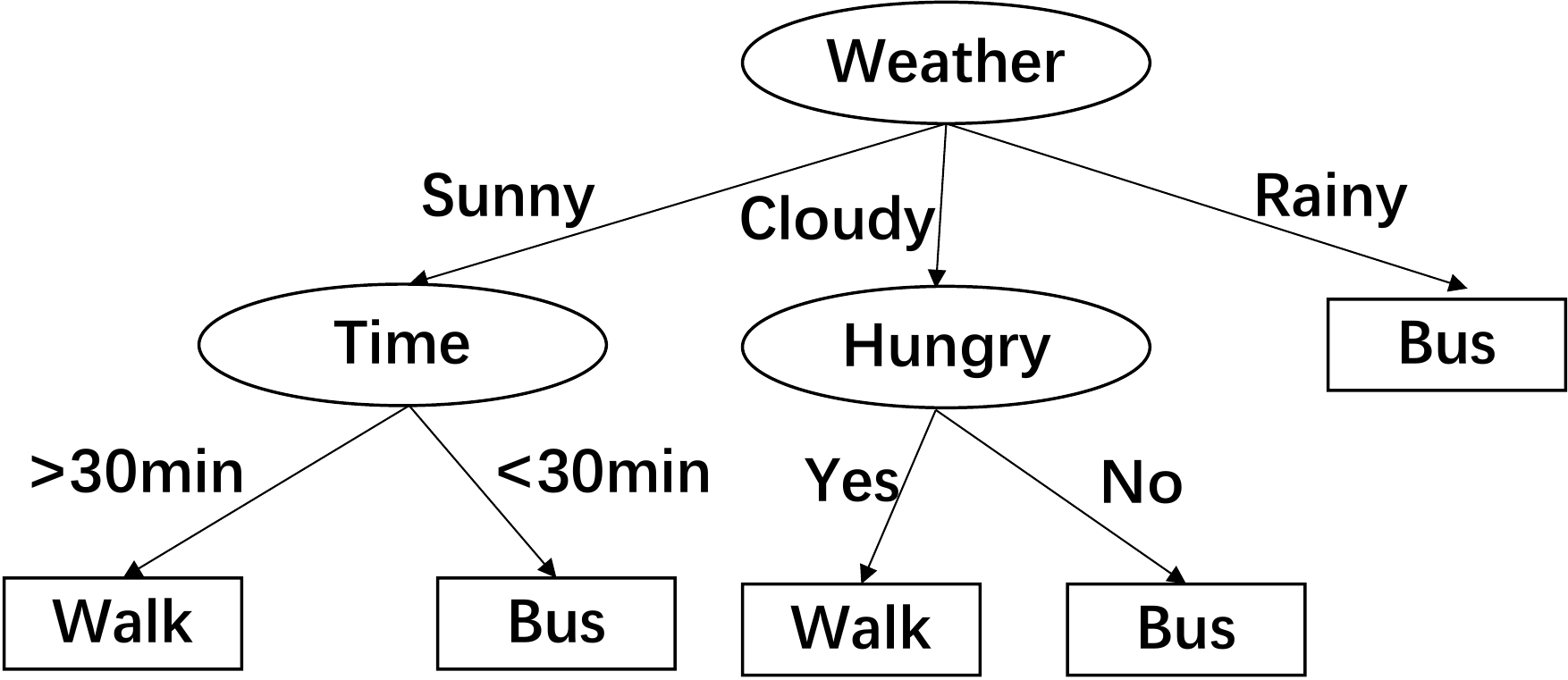}\caption{Main idea of decision tree. ID3, C4.5 and CART are three common decision trees, the principle of them are similar except for the different node-split tactics (Information Gain, Information Gain Ratio and Gini Index).  Ellipses represent the split nodes for classification, and rectangles represent the final multiple classification results. }
\label{fig:Decision Tree}
\end{figure}

Decision Tree (DT) based classification algorithms \citep{10.1145/234313.234346} are essential in machine learning algorithms. Their leaves represent classification results and internal nodes of branches are regarded as criteria for distinguishing objects. The graphical representation of decision tree is shown in Fig. \ref{fig:Decision Tree}. Decision tree and its variants have been applied in astronomy and many other fields \citep{bae2014clinical, li2005scalable, czajkowski2014multi, zhao2008comparison}. Table \ref{table_dt_ensemble_related} shows the main astronomical investigations of decision tree based classification algorithms.

\begin{table*}
\centering
\caption{Investigations of DT based classification algorithms and ensemble learning on astronomical spectra data}
\label{table_dt_ensemble_related}
\begin{threeparttable}
\resizebox{\linewidth}{!}{
\begin{tabular}{cccc} 
\hline
Merits                                                                                                                                                                                                                                                                                                                                                                                                                 & Caveats                                                                                                                                                                                                       & References                                                                                                                                                                                                                                                                                                                                                                                                                                                                                                                                                                                                                                                                                                                                                        \\ 
\hline
\multicolumn{1}{l}{\begin{tabular}[c]{@{}l@{}}Results interpretability and predicted probability\\  \\High accuracy on stellar and subtypes$^{1}$\\ \\Extract feature well for stellar and subtypes $^{2}$ \\ \\Classify spectra with missing values and noise\end{tabular}} & \multicolumn{1}{l}{\begin{tabular}[c]{@{}l@{}}Limited to large redshift objects\\ \\Misclassification on G,F,K stars\\ \\Adjust parameters manually \\ \\Poor performance on unbalanced data\end{tabular}} & \multicolumn{1}{l}{\begin{tabular}[c]{@{}l@{}}{\citet{2016ApJ...819...18P}, \citet{2019MNRAS.483.5077A}, \citet{2021arXiv210507110F}},\\{\citet{2021ChJPh..69..303X}, \citet{2011AJ....141..189V}},\\{\citet{2018MNRAS.481.4194M}, \citet{2020A&A...639A..84C}},\\{\citet{2021MNRAS.501.3457P}, \citet{2019RAA....19..111L}, \citet{2019AJ....157....9B}}, \\{\citet{2020MNRAS.498.1750A}, \citet{2019MNRAS.483.4774L}, \citet{2014AJ....147...33Y}}, \\{\citet{2020MNRAS.493.6050H}, \citet{2019RAA....19..111L}, \citet{baqui2021minijpas}}, \\{\citet{2018MNRAS.476.2117R}, \citet{2019AJ....158..188B}}, \\{\citet{2017A&A...605A.123P}, \citet{2021A&C....3600473I}, \citet{tao2018automated}}, \\{\citet{kyritsis2022new}, \citet{2022MNRAS.509.2674G}, \citet{brice2019classification}}, \\{\citet{https://doi.org/10.48550/arxiv.2203.08125}, \citet{hou2020spectroscopically}, \citet{ensembleuniverse7110438}},\\ {\citet{2021MNRAS.503.5263Z}, \citet{yue2021identify}, \citet{sookmee2020globular}} \end{tabular}} \\
\hline
\end{tabular}
}

\begin{tablenotes}
    \footnotesize
    \item[1] : Star/galaxy/quasar classification, MK classification, LSP/HSP, M star/others, etc.
    \item[2] : MK classification, stellar subtypes, M subtypes, etc.
    \end{tablenotes}
    \end{threeparttable}
\end{table*}

Decision tree based classification algorithms have been widely used for astronomical classification due to their good interpretability of classification results. Here are some examples of decision tree for astronomical classification. \citet{2018MNRAS.481.4194M} explored the classification boundaries of star and galaxy through decision tree. This visualized the classification process of the star-galaxy and helped astronomers understand the decision rules of celestial classification. \citet{6714666} used  parallel decision trees to classify different types of objects and evaluated the performance of classification results. \citet{2011AJ....141..189V} applied 13 different decision tree algorithms to analyse the classification performance of star/galaxy, and the functional tree algorithm yielded the best results.

According to the astronomical researches using decision tree based classification algorithms, there are three tips to improve the classification performance. Firstly, effectively pre-processing the raw observational spectra will assist and speed up the classification, such as noise reduction and data compression. Secondly, extracting valid features is also important. Most familiar approaches normalize and standard spectra data by prevalent methods without additional operations \citep{2016ApJ...819...18P, 2011AJ....141..189V}, yet these simple approaches will have high computational costs and could not improve classification accuracy effectively. So other valid features may be better to improve classification performance, such as line indices and astronomical-specific features. Thirdly, searching for appropriate methods is another vital approach to improving classification performance. Compared with other typical methods, RF performed best both on accuracy and time consuming in \citet{2021ChJPh..69..303X, 2019AJ....158..188B, 2021arXiv210507110F}, etc. Alternatively, integration of decision tree and other conventional classification methods can enhance the superiority of feature selection and results interpretation respectively \citep{2021A&C....3600473I}. However, heterogeneous data, large redshift objects, other stellar parameters regression and misclassification are still challenges for decision tree in astronomical research. These need to be solved in the future.

Iterative Dichotomiser 3 (ID3), C4.5 and Classification and Regression Trees (CART) are three widespread methods based on decision tree. ID3 adopts Information Gain (IG) as the node selection criterion for classification. While C4.5 chooses Information Gain Ratio (IGR) to alleviate the flaws of ID3 (IG: discrete data, incomplete attribution, overfitting, etc). Another upgraded method is CART which can be used for both classification and regression. It employs the Gini Index as a node selection standard instead of Information Entropy. The main advantage of decision tree based classification algorithms is interpretability of results, which is very helpful for astronomers to analyse the features of astronomical objects. And the disadvantage is that we often obtain a complex model which will be overfitting on the training data. So pruning parameters
is always required to reduce overfitting.

\subsection{Ensemble learning classification algorithms} 

Ensemble learning \citep{10.5555/645528.657623} combines multiple weak classifiers into a strong classifier to solve a task together. Generally, ensemble learning methods are sorted into bagging methods decreasing variance, boosting methods reducing deviation and stacking methods increasing prediction accuracy. Compared with the single decision tree, ensemble learning methods are more often used in astronomy.

\begin{figure}
\centering
\includegraphics[width=\columnwidth]{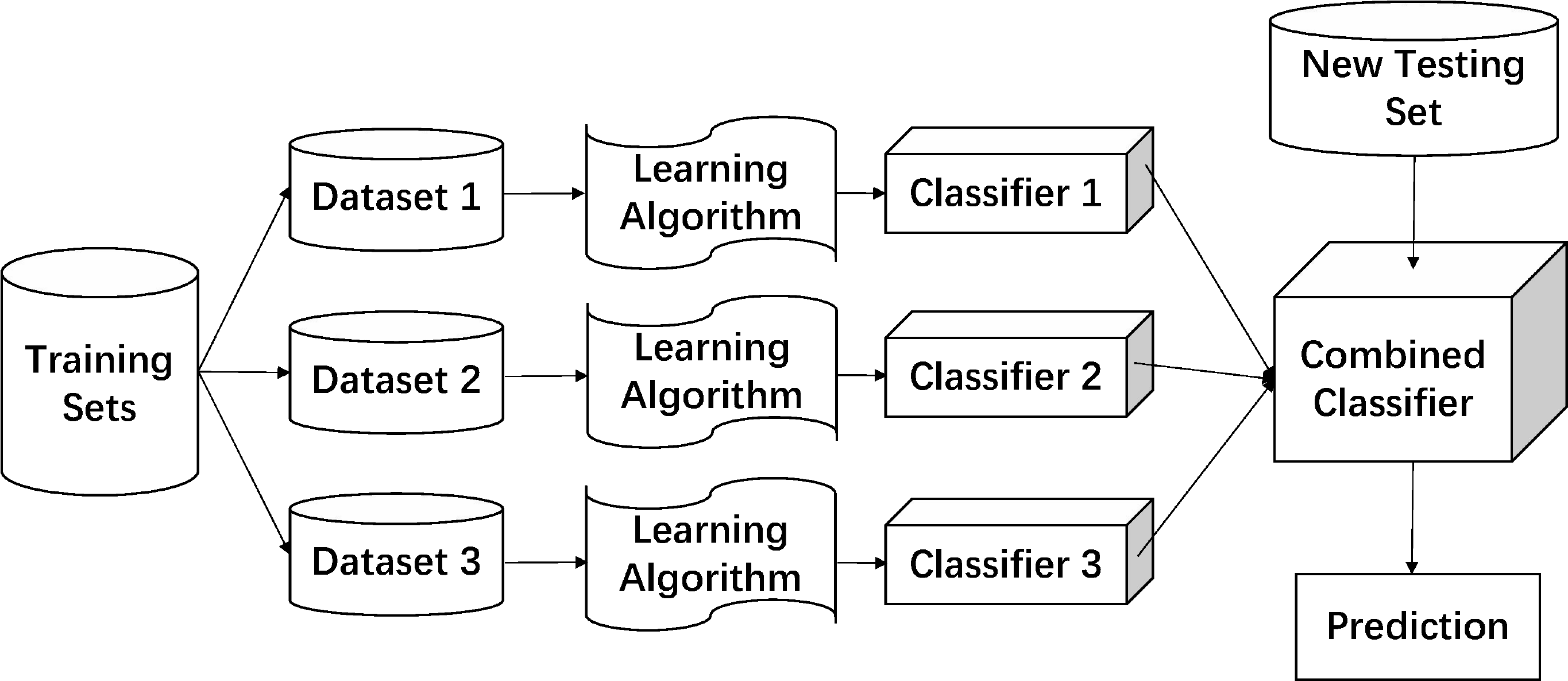}
\caption{Main idea of bagging. The left cylinders are training sets to train different base classifiers using learning algorithms. The upper cylinder is the new testing set to test the combined classifier which is made up of a collection of base classifiers. And the strong classifier is generated by voting for classifier i ( i=1, 2, 3 in Fig. \ref{fig:bagging}).}
\label{fig:bagging}
\end{figure}

\begin{figure}
\centering
\includegraphics[width=\columnwidth]{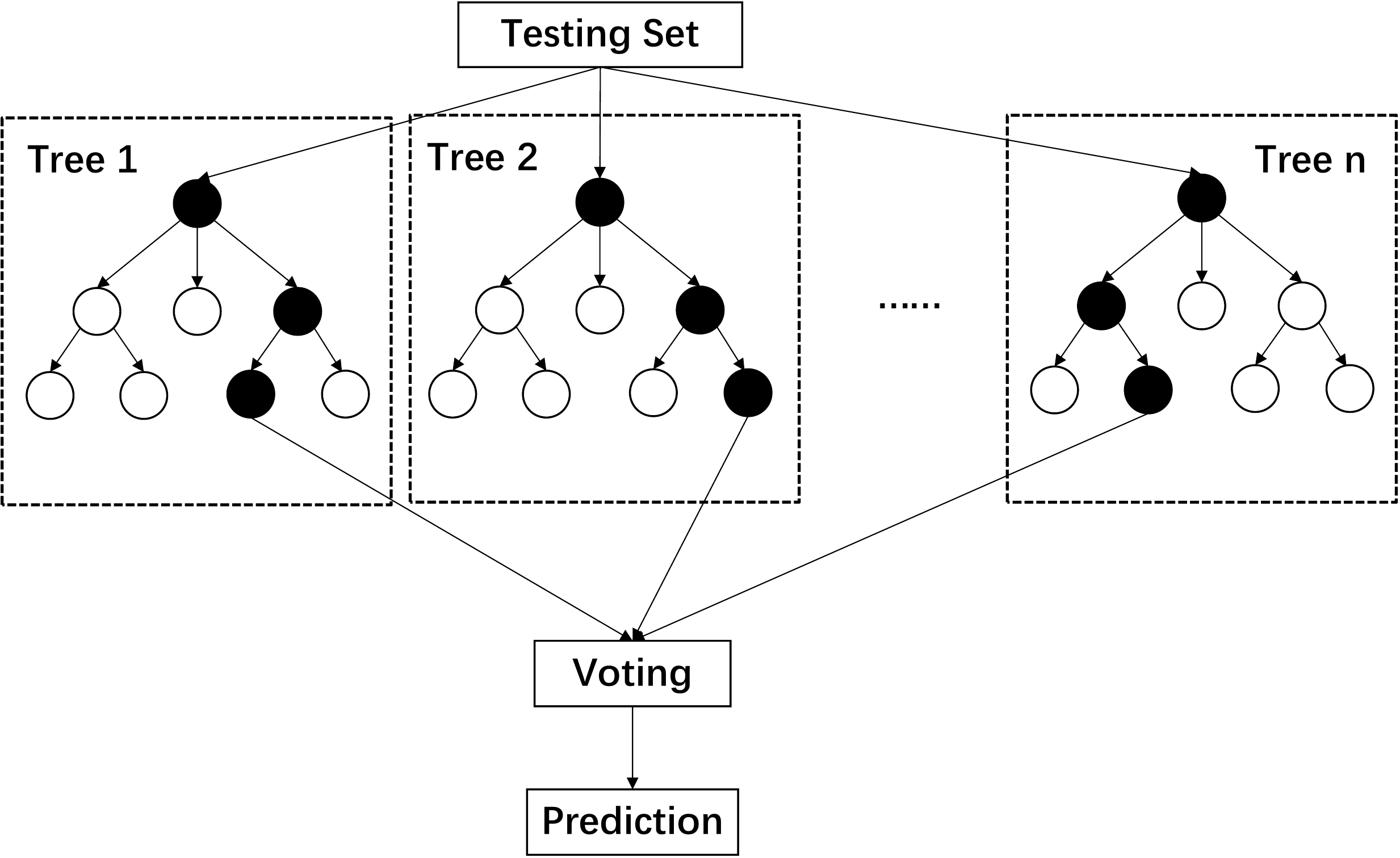}
\caption{Main idea of random forest. Tree i (i=1, 2, ......, n) are decision trees trained by certain strategies. Black circles are the basis nodes of classification. The voting methods adopt the majority or average results of tree i (i=1, 2, ......, n) as the final results.}
\label{fig:RF}
\end{figure}

Bagging usually trains different models with various training sets respectively and chooses one strategy to unify consequences. The principle of bagging is shown in Fig. \ref{fig:bagging}. Random Forest is the most notable bagging method which consists of several unrelated decision trees. Fig. \ref{fig:RF} describes the principle of Random Forest. RF can be applied to classification, clustering, regression and outlier detection due to its high accuracy and adaptability of high dimensional data sets.

RF is a robust classifier for spectral classification \citep{ brice2019classification, 2018MNRAS.481.4194M, 2019MNRAS.483.4774L, 2019RAA....19..111L,2019AJ....157....9B, 2014AJ....147...33Y, 2020MNRAS.493.6050H, biau2016random, 2019AJ....158..188B, baqui2021minijpas}. \citet{2020A&A...639A..84C} trained a RF classifier on 3.1 million labelled sources from Sloan Digital Sky Survey (SDSS) and applied this model on 111 million unlabelled sources. The result showed that the classification probabilities of stars were greater than 0.9 (about 0.99). Besides, RF performed well in the process of searching for rare objects \citep{hou2020spectroscopically, kyritsis2022new}. \citet{2021MNRAS.501.3457P} trained a random forest classifier to determine whether a black hole or a neutron star is hosted by a Low Mass X-ray binaries (LMXBs). It is difficult to accurately classify variable stars into their respective subtypes, hence \citet{2017A&A...605A.123P} proposed new robust feature sets and used RF to evaluate the classification performance. \citet{2019MNRAS.483.5077A} used classification tree for identifying symbiotic stars (SySts) from other H$\alpha$ emitters in photometric surveys. \citet{2022MNRAS.509.2674G} used random forest to identify white dwarfs in LAMOST DR5. \citet{2018MNRAS.476.2117R} used an unsupervised random forest to detect outliers on APO Galactic Evolution Experiment (APOGEE) stars. In addition, RF is often compared with other algorithms on classification tasks, and generally, it tends to be better than others \citep{2020MNRAS.498.1750A, 2019MNRAS.483.4774L, 2017A&A...605A.123P}.

\begin{figure}
\centering
\includegraphics[width=\columnwidth]{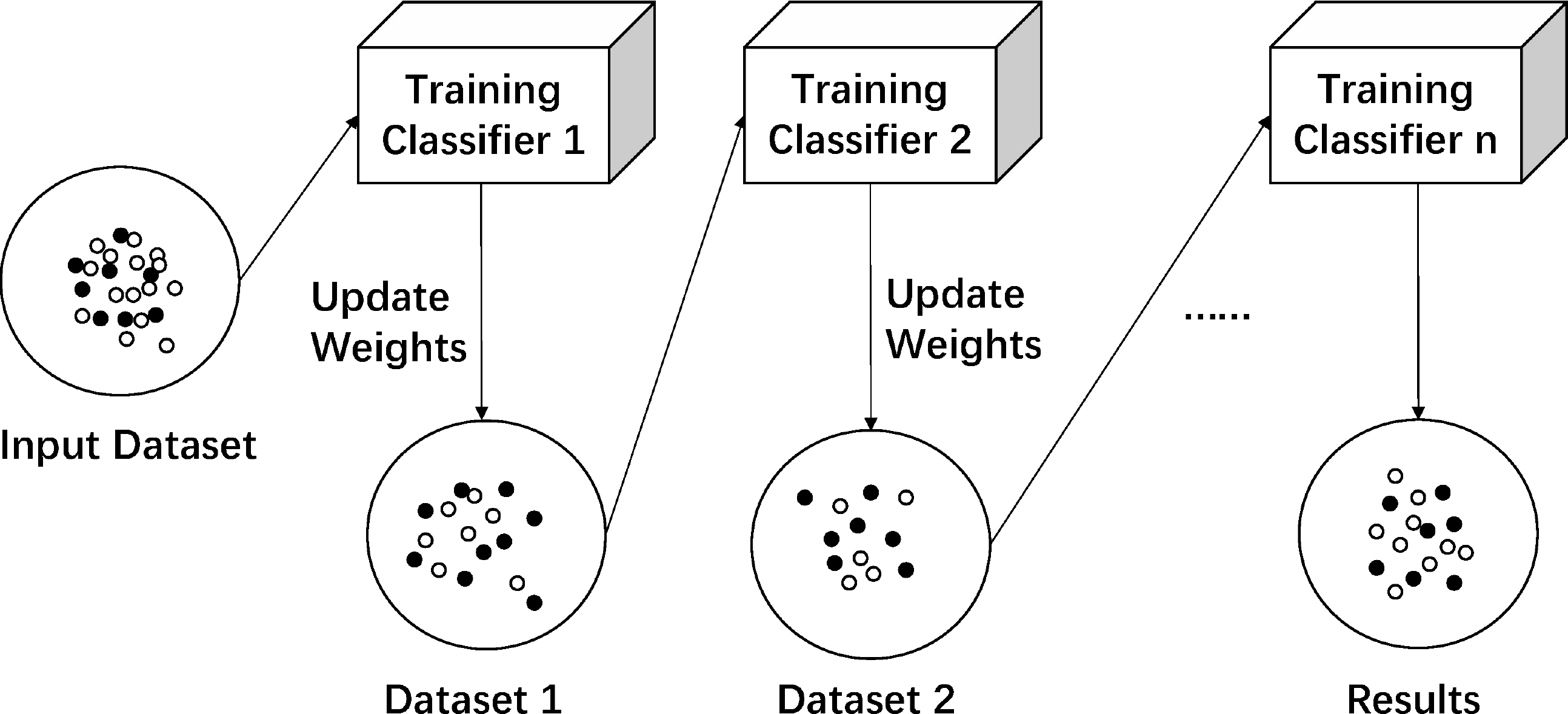}
\caption{ Main idea of boosting. The left data set is regarded as input to train classifier 1 (Training Classifier 1 in Fig. \ref{fig:boosting}). Data set 1: a new data set from training classifier 1 after updating its weights. Train the classifier and update the weights until the classification results converge.}
\label{fig:boosting}
\end{figure}

\begin{figure}
\centering
\includegraphics[width=\columnwidth]{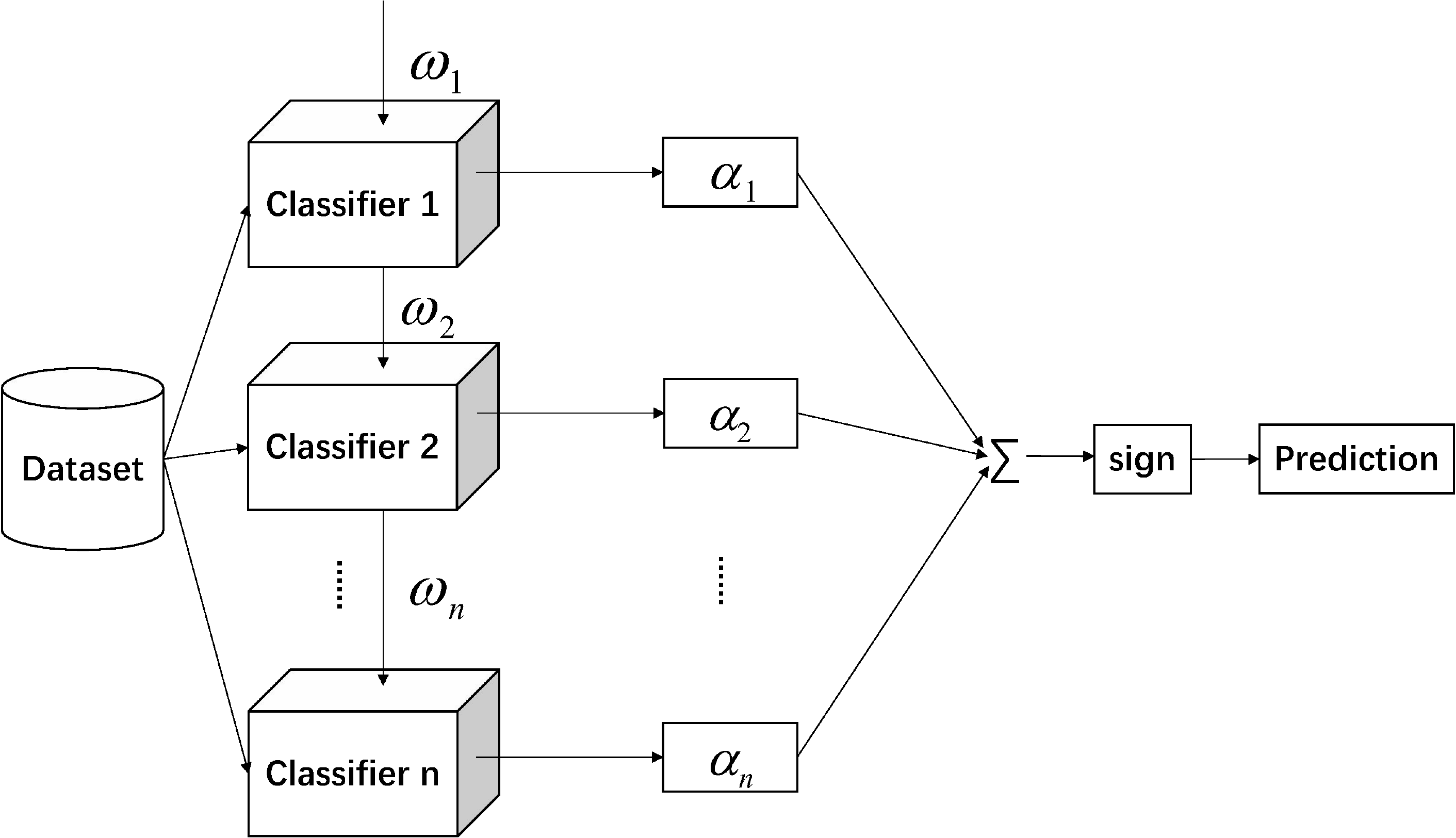}
\caption{Main idea of adaptive boosting (Adaboost). The left cylinder is input data used to train classifier i (i=1, 2, ......, n). 
${\mathop \omega \nolimits_i }
$ (${i}$=1, 2, ......,n) are the weights of data. 
${\mathop \alpha \nolimits_i }$ (${i}$=1, 2, ......,n) are the weights of classifiers. }
\label{fig:Adaboost}
\end{figure}

\begin{figure}
\centering
\includegraphics[width=6.3cm,height=4.9cm]{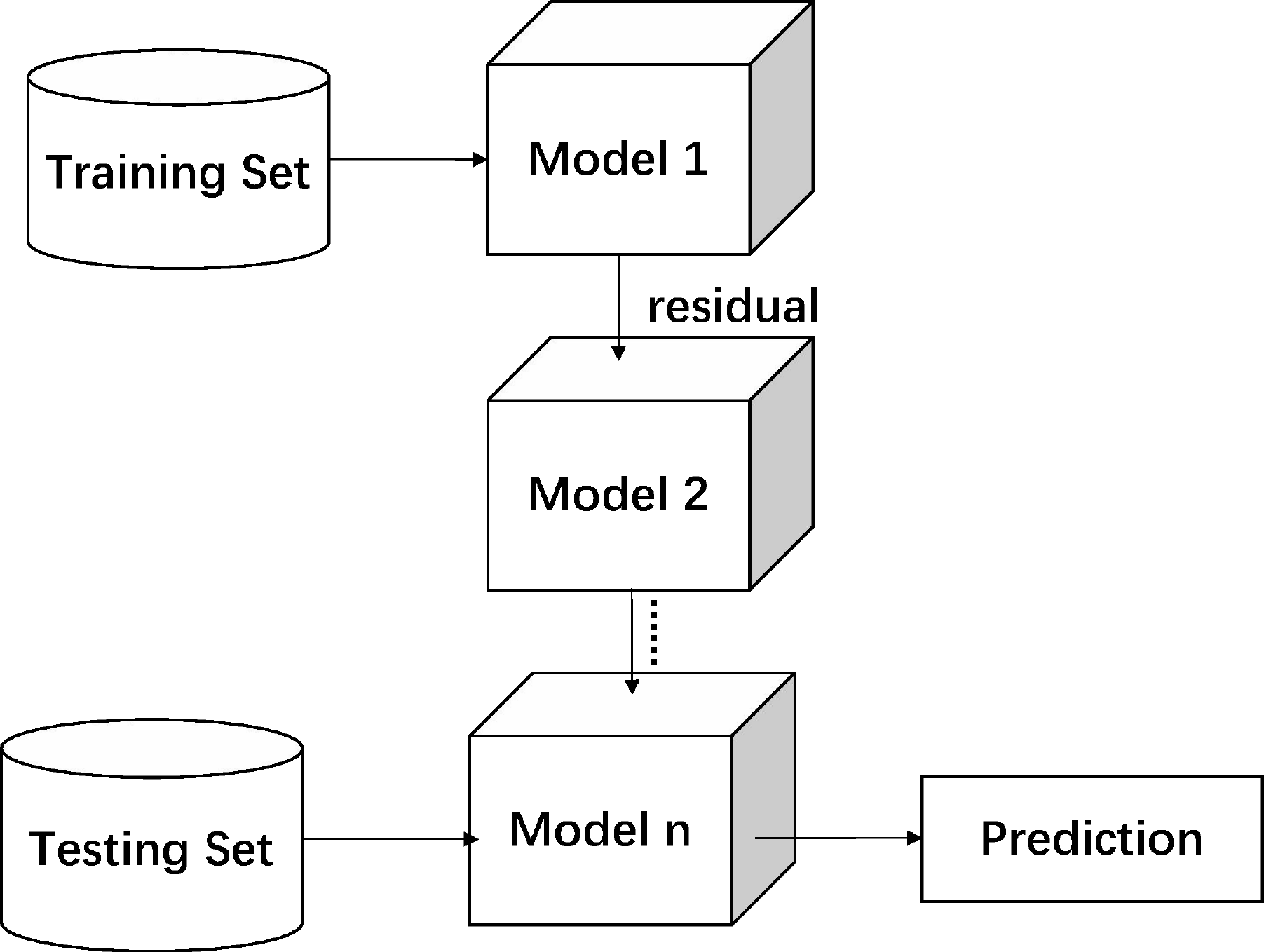}
\caption{Main idea of gradient boosting decision trees (GBDT). The residual of model i-1 is the input of model i (i=1,2,......,n). The goal of GBDT is to make residual as small as possible.}
\label{fig:GBDT}
\end{figure}

\begin{figure}
\centering
\includegraphics[width=\columnwidth]{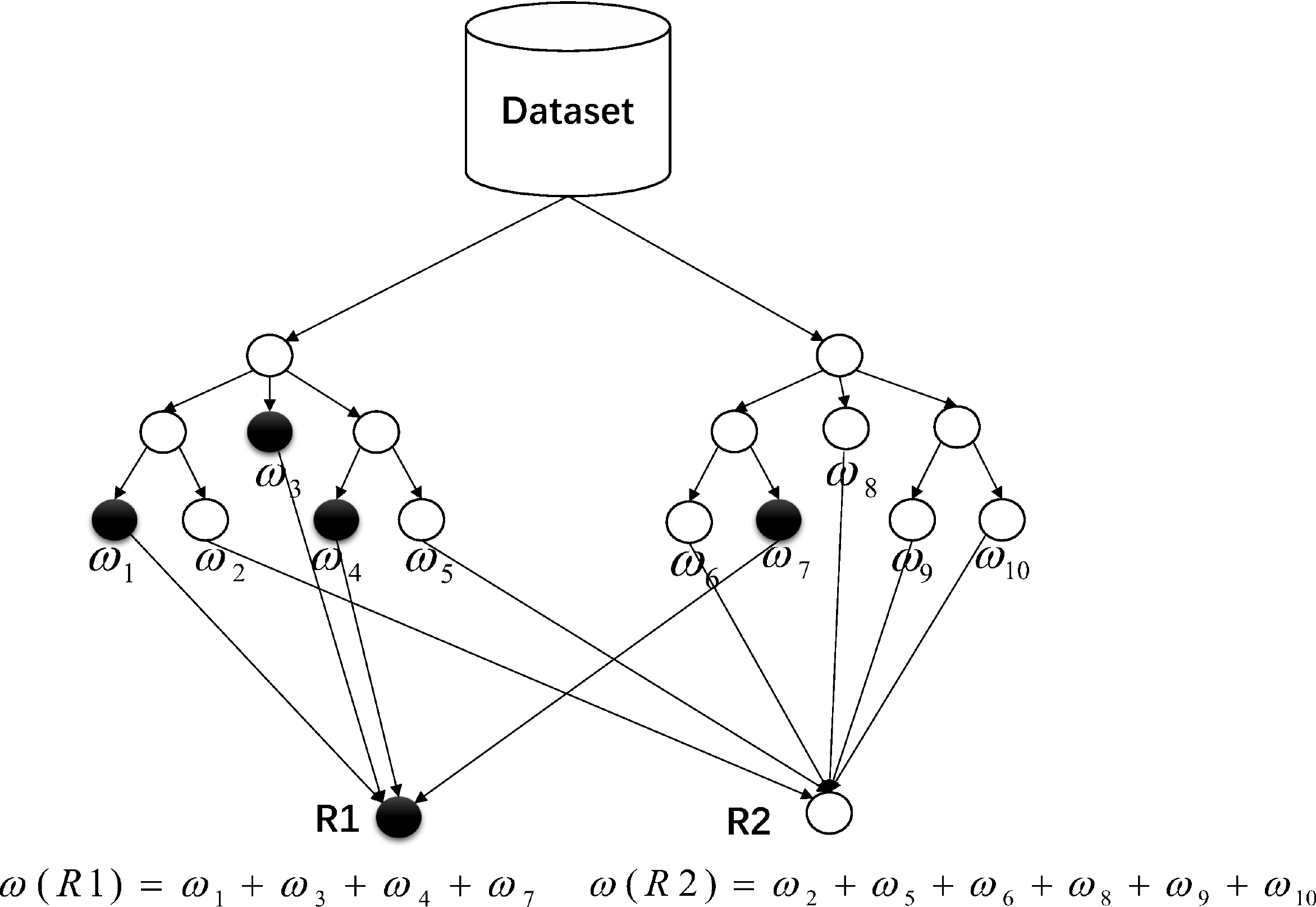}
\caption{Main idea of XGBoost. ${\mathop \omega \nolimits_i }$
 (${i}$=1, 2, ......, n) are the scores of leaves. R1 and R2 are the predicted labels which are the sum of ${\mathop \omega \nolimits_i }$. The black circles represent data. They are classified as R1 and the white circles are classified as R2. Bigger score between R1 and R2 is the final result. }
\label{fig:XGBOOST}
\end{figure}

Boosting trains models with adjusted data, that is, the weights of misclassified objects are augmented based on the former models. Fig. \ref{fig:boosting} is the principle of boosting. Gradient Boosting Decision Tree (GBDT), Adaptive boosting (Adaboost), extreme gradient boosting (XGBoost) and Light Gradient Boosting Machine (LightGBM) are prevalent boosting methods. Adaboost is a prominent boosting method that chooses single-layer decision trees as weak classifiers. In each iteration, it trains one weak classifier based on data weights generated in the last iteration. So Adaboost pays more attention on misclassified data. The other essential parameters are weights of each classifier. They are computed based on classification accuracy of every classifier. And the final results are obtained after inputting the sum of each weak classifier into a sign function. Fig. \ref{fig:Adaboost} is the main principle of Adaboost. GBDT \citep{2018MNRAS.481.4194M, 2017A&A...605A.123P}, another typical boosting algorithm, can also be regarded as an optimized version of Adaboost. GBDT chooses the residual from the previous iteration as input to train the next classifier till the residual is close to zero. Besides, GBDT can take more objective functions and train models using negative gradient, whereas Adaboost only sets data weights automatically. Fig. \ref{fig:GBDT} shows the main principle of GBDT. XGBoost optimized GBDT by supporting different meta classifiers, adding regularization to limit model complexity, adapting to different data samplings and so on. Fig. \ref{fig:XGBOOST} shows the main principle of XGBoost. 

GBDT and XGBoost are two powerful ensemble classifiers \citep{friedman2001greedy, chen2016xgboost} and have been applied to spectral classification and rare object identification. \citet{CHAO2019539} used XGBoost to classify star and galaxy on dark sources of SDSS photometric data sets and the results showed that XGBoost outperformed other methods. \citet{ensembleuniverse7110438} searched for Cataclysmic Variables (CVs) in LAMOST-DR7 using LightGBM which is based on the ensemble tree model. They found 225 CV candidates including four new CV candidates which were verified by SIMBAD and published in catalogs. \citet{yue2021identify} also identified M sub-dwarfs using XGBoost. In order to get better classification results, many new ensemble algorithms have been proposed in recent years \citep{CHAO2020345_2, zhao2022automated, 10.1007/978-3-030-81007-8_56}.

\begin{figure}
\centering
\includegraphics[width=\columnwidth]{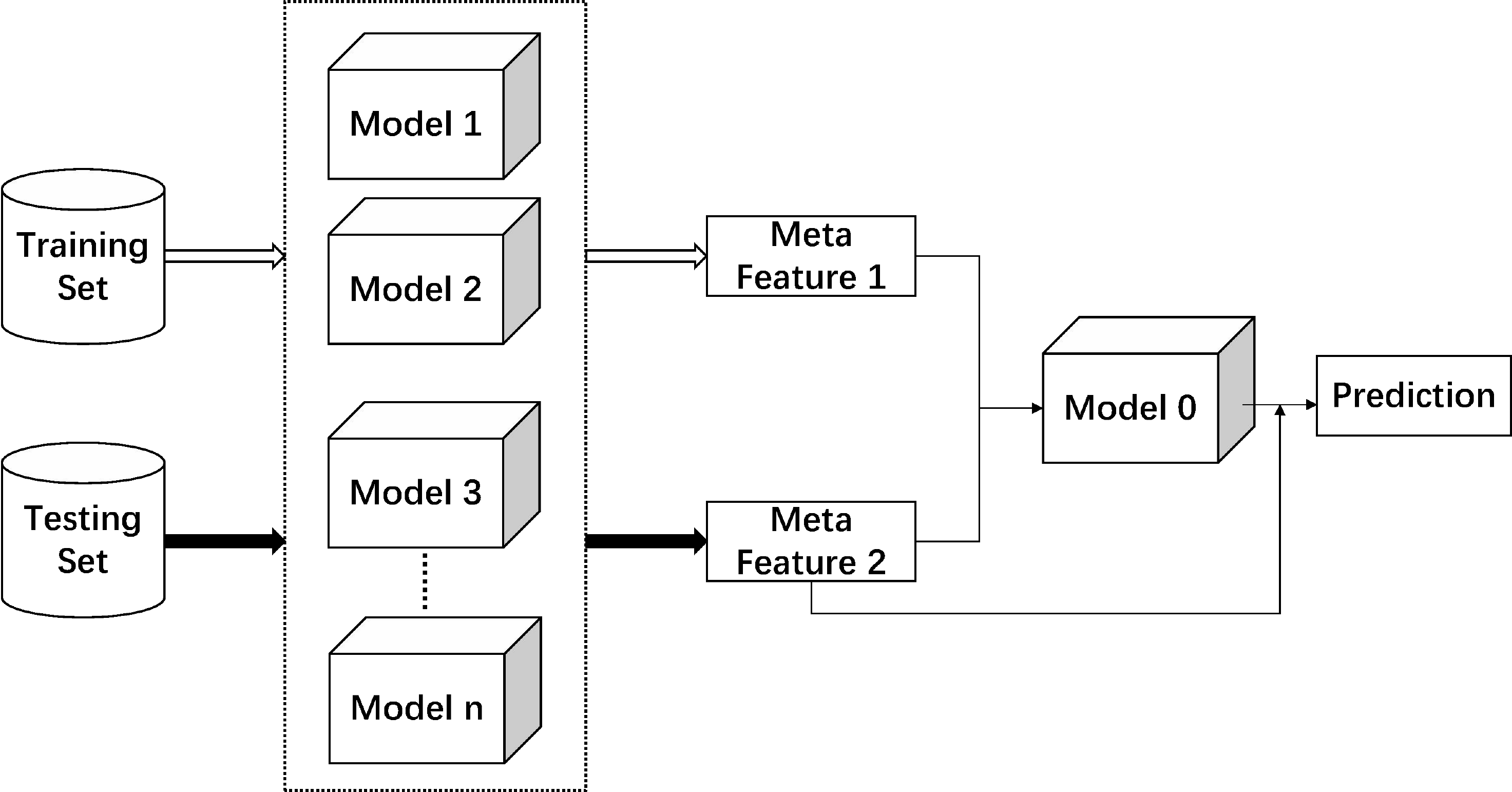}
\caption{Main idea of stacking. The left cylinders are training sets and test sets. Meta feature 1 and meta feature 2 are  the results of model i (i=1,2,.....,n) on training sets and test sets respectively. Model 0 is trained by meta feature 1 and meta feature 2. Then meta feature 2 is input into model 0 to predict results.}
\label{fig:stacking}
\end{figure}

Stacking uses a new model to fit meta features which are obtained by multi-predictors on training sets and testing sets. And this new model will be validated with the following meta features. Fig. \ref{fig:stacking} introduces the principle of stacking. 

Ensemble learning has obtained desirable results in astronomical spectral analysis. And random forest is the most frequently used ensemble method in astronomy. Because it has good generalization performance on large scale high-dimensional data sets. It is good at probabilistic prediction and is insensitive to noise. However, multi-value attribute still troubles RF. In addition, ensemble learning methods are also limited to heterogeneous data, unbalanced data and optimal parameters (the number of decision tree, weak classifiers).

\subsection{Neural network based classification algorithms} 
\begin{figure}
\centering
\includegraphics[width=\columnwidth]{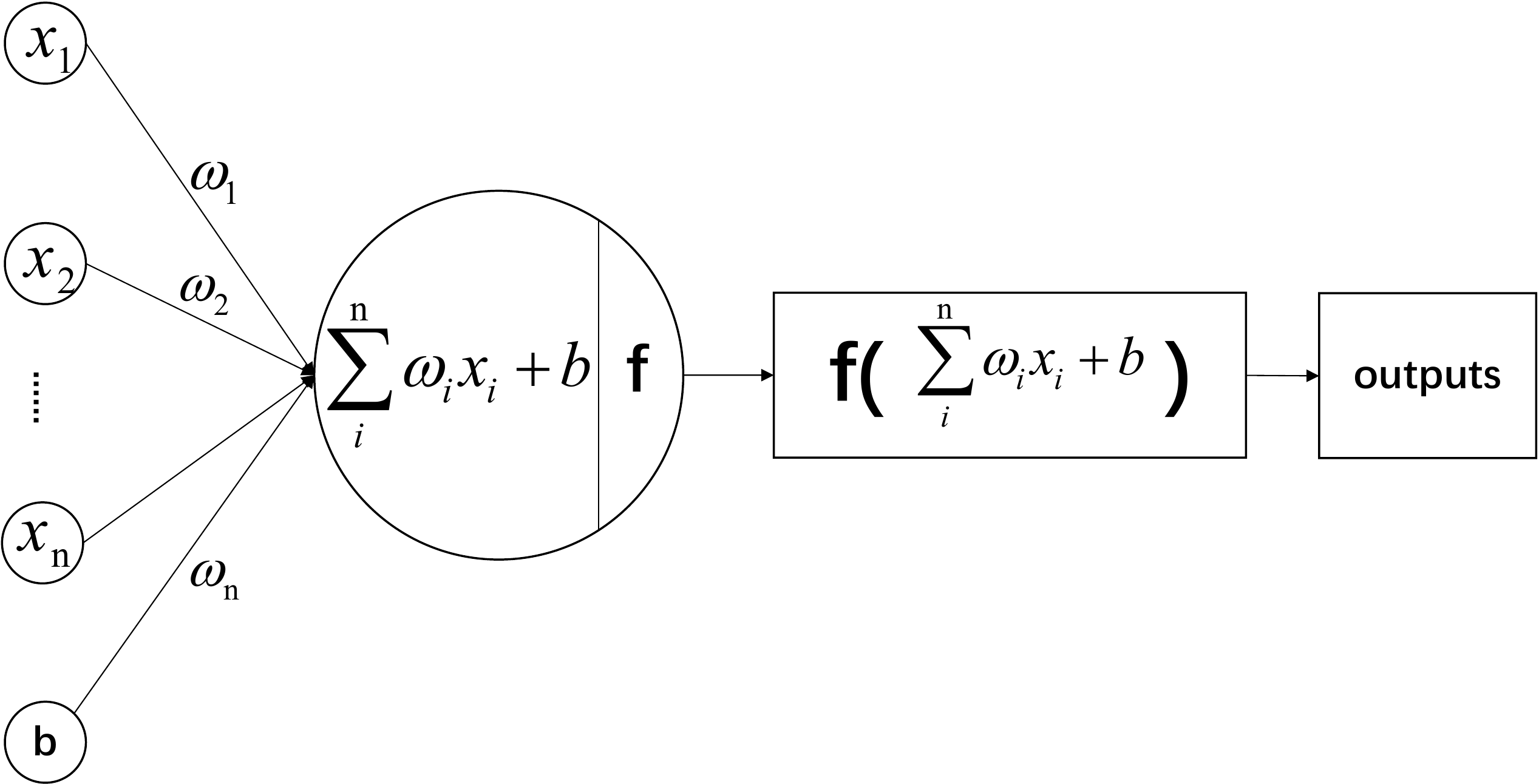}
\caption{Main idea of neural unit. ${\mathop x\nolimits_i }$
 (${i}$=1, 2, ......, n) is input data. ${\mathop \omega \nolimits_i }$ (${i}$=1, 2, ......, n): weights of ${\mathop x\nolimits_i }$. b: (biases) is also the input of neural units. Big circle in the middle contains a linear combination of input and an activation function f.
 }
\label{fig:neural unit}
\end{figure}

\begin{figure}
\centering
\includegraphics[width=\columnwidth]{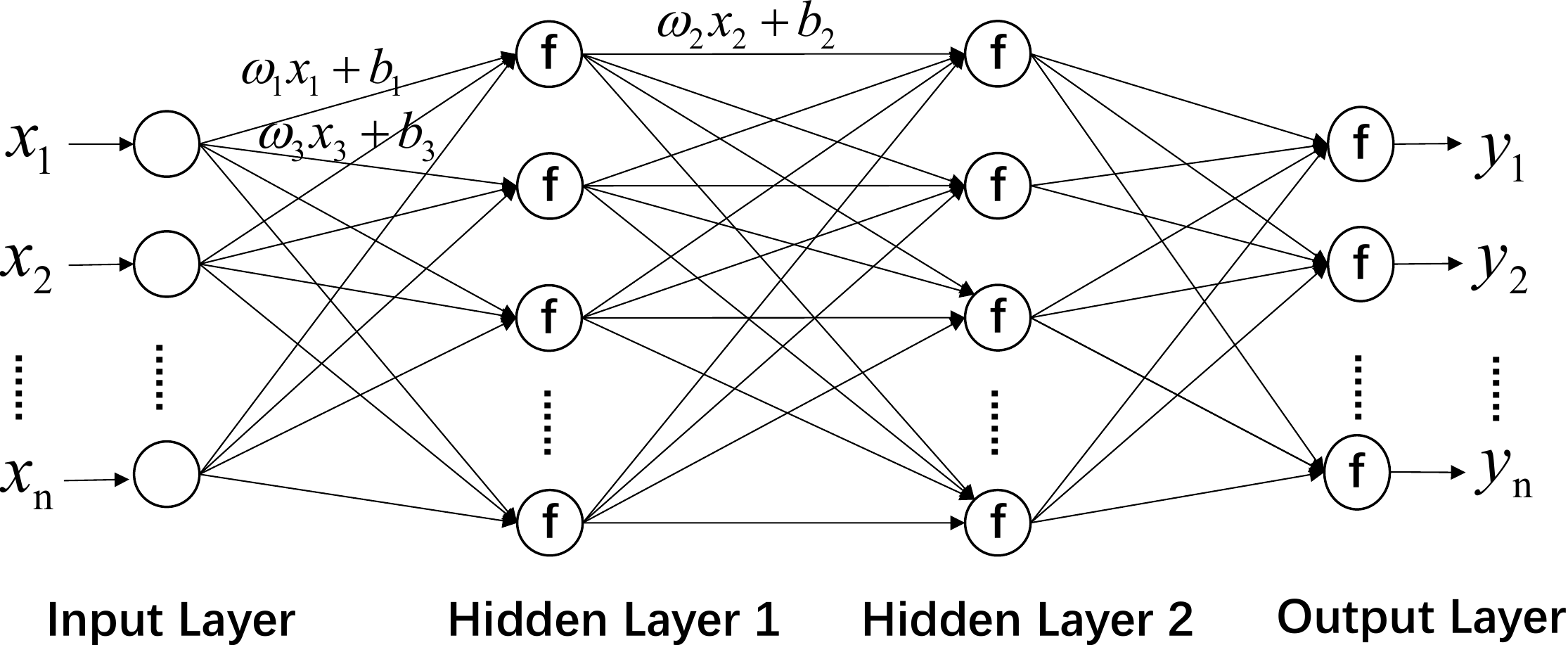}
\caption{Main idea of artificial neural network. ANN contains multiple fully connected neural units . ${\mathop \omega \nolimits_i }$ (${i}$=1, 2, ......, n)
 and ${\mathop b\nolimits_i }$ (${i}$=1, 2, ......, n)
are updated during iterations. f is the activation function. ${\mathop y\nolimits_i }$ (${i}$=1, 2, ......, n)
is the final result.}
\label{fig:ANN}
\end{figure}

Artificial Neural Network, also known as Multi Layer Perception (MLP), is a machine learning method that imitates the signal transmission mechanism in the brain. It consists of an input layer, multiple hidden layers and an output layer. The neural unit in each hidden layer tackles input data and sends results to the next fully connected layer. The output layer generates the final consequences. Fig. \ref{fig:neural unit} is the principle of a neural unit. And Fig. \ref{fig:ANN} is the main principle of ANN. Particularly, Pseudo Inverse Learning (PIL) is a classic neural network. It can get globally optimal results and is faster than Back-propagation(BP) algorithm. Besides, it does not require manual tuning of parameters. So it has been used for some simple tasks. However, for complicated tasks, optimal versions of neural network are necessary. Deep learning (DL) is an essential extension of ANN, and it contains more hidden layers and complex network structures \citep{Bergen2019earth}. Convolutional Neural Network (CNN), Auto Decoder (AE) and Deep Belief Networks (DBN) are three chief methods of DL. Moreover, other variant versions of neural network have been proposed to adapt to different data formats, like Visual Geometry Group (VGG), Residual Networks (ResNet), Recurrent Neural Network (RNN), Generative Adversarial Networks (GAN) and others. Moreover, pre-trained models, attention blocks, transfer learning and many other tricks have been used to improve the deep learning performance effectively.

\begin{figure}
\centering
\includegraphics[width=\columnwidth]{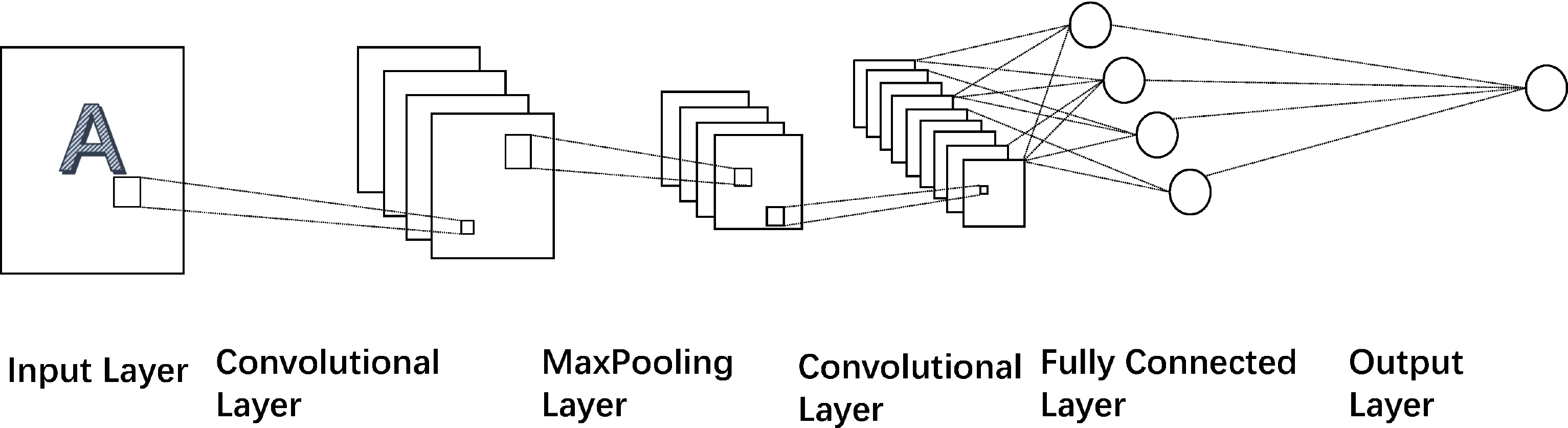}
\caption{Main idea of convolutional neural network. Convolutional layers are used to learn features from different layers. MaxPooling layer can reduce dimensionality. The role of the fully connected layer is equivalent to the classifier. The output layer is designed to represent the classification results according to the concrete classification task.}
\label{fig:CNN}   
\end{figure}

Convolutional Neural Network (CNN) consists of convolutional layers that extract image features, pooling layers that reduce dimensionality and fully connected layers that generate results. CNN automatically extracts features without destroying them. So it can get better accuracy and cope with high dimensional data. But its vanishing gradient problem and local optimal phenomenon still annoyed us. Fig. \ref{fig:CNN} shows the main principle of CNN.

\begin{figure}
\centering
\includegraphics[width=\columnwidth]{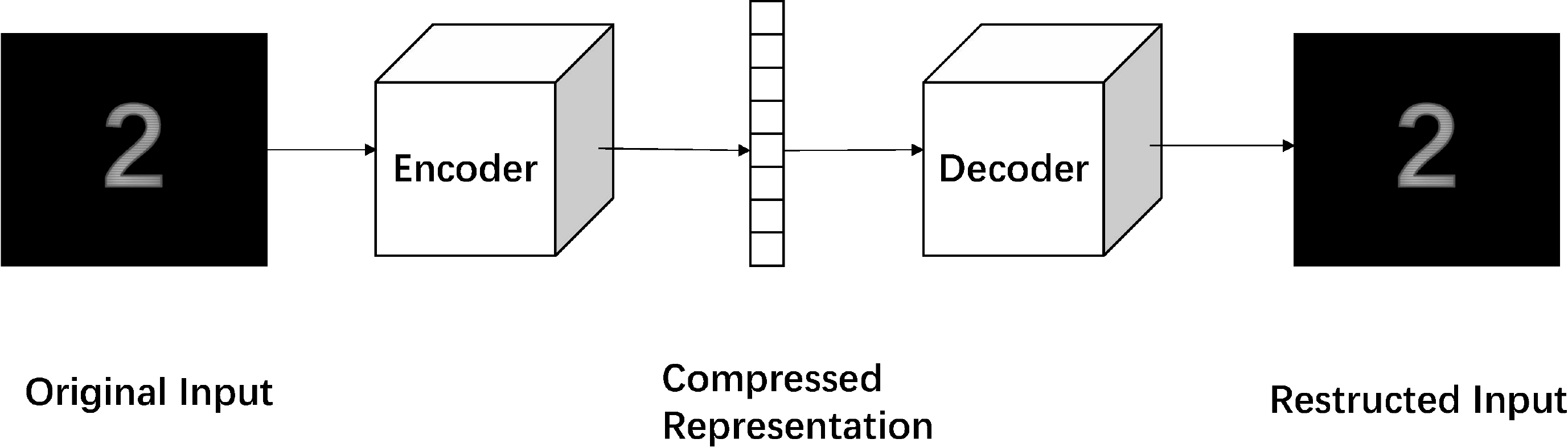}
\caption{Main idea of auto encoder. Compressed representation of original input is achieved by encoder model. Another vital model is decoder transforming compressed representation into the restructured input.}
\label{fig:AE}   
\end{figure}

Auto Decoder (AE) is a neural network whose input equals its output and its main idea is sparse code. It restructures the input using an encoder and a decoder. And it has been widely used for noise and dimensionality reduction to visualize data. Fig. \ref{fig:AE} is the principle of AE.

DBN is a probabilistic generative model. Its generative model builds a joint distribution between observations and labels. DBNs consist of multiple layers of Restricted Boltzmann Machines which is a probabilistic graphical model with stochastic neural network. The output states of each neural unit are activation and deactivation. 

Different examples of astrophysical research projects exploiting neural network are listed in Table \ref{table_neural_related} and summarized in the next parts.

\begin{table*}
\centering
\caption{Investigations of neural network based classification algorithms on astronomical spectra data}
\label{table_neural_related}
\begin{threeparttable}
\resizebox{\linewidth}{!}{
\begin{tabular}{cccc} 
\hline
Merits                                                                                                                                                                                                                                                                                                                                                                                                                                                                                                                                                                                                                                                                               & Caveats                                                                                                                                                                                                                                                         & References                                                                                                                                                                                                                                                                                                                                                                                                                                                                                                                                                                                                                                                                                                          \\ 
\hline
\multicolumn{1}{l}{\begin{tabular}[c]{@{}l@{}}High accuracy on stellar spectra and specific spectra$^{1}$\\ \\Better than other methods$^{2}$ \\ \\Generate synthetic data\\ \\Redshift estimation for quasar, SNe Ia/others\\ \\Tackle different types of input data sets$^{3}$\\ \\Need less additional information\\ \\Provide vital supplements to categories$^{4}$\end{tabular}} & \multicolumn{1}{l}{\begin{tabular}[c]{@{}l@{}}Limited to unbalanced data sets\\ \\Results are affected by noise\\ \\Poor performance on large redshift objects\\ \\Bad performance on weak features\\ \\Overfitting\\ \\ Misclassification on K/F stars\\ \\High computation time\end{tabular}} & \multicolumn{1}{l}{\begin{tabular}[c]{@{}l@{}}{\citet{2019MNRAS.483..529C}, \citet{2014A&A...562A..36S}},\\{\citet{2020PASP..132d4503Z}, \citet{2020JCAP...11..015F}},\\{\citet{2017MNRAS.465.4311W}, \citet{2019MNRAS.483.4774L}},\\{\citet{davison2022stag}, \citet{2019ApJ...886..128B}},\\{\citet{2015A&A...580A.138A}, \citet{2014PASA...31....1F}},\\{\citet{2020MNRAS.498.1750A}, \citet{2019MNRAS.485.2167G}},\\{\citet{Rastegarnia_2022}, \citet{sharma2020application}},\\{\citet{2019ApJ...879...72G}, \citet{fremling2021sniascore}},\\{\citet{chen2021classifying}, \citet{2021arXiv210507110F}, \citet{zou2019classification}},\\{\citet{zheng2020generalization}, \citet{Tan_2022}},\\{\citet{lu2020study}, \citet{astsatryan2021astronomical}},\\{\citet{jingyi2018deep}, \citet{jiang2021spectral}},\\{\citet{jing2020new}, \citet{jiang2020automated}},\\{\citet{vskoda2020active}, \citet{kerby2021multiwavelength}},\\{\citet{luo2008automated}, \citet{zheng2020classification}},\\{\citet{vilavicencio2020application}} \end{tabular}}  \\
\hline
\end{tabular}
}
\begin{tablenotes}
    \footnotesize
    \item[1] : M stars/others, BAL quasars/others, Pulsars/blazars, etc.
    \item[2] : RF, template matching, KNN, etc.
    \item[3] : Spectra, image, photometric data, etc.
    \item[4] : Quasar, star, double-lined spectroscopic binaries, etc.
    \end{tablenotes}
    \end{threeparttable}
\end{table*}

Astronomical spectral classification is a typical task for neural network. \citet{2019MNRAS.483..529C} used CNN to classify star and galaxy on low-resolution spectra from narrow-band photometry with accuracy over 98\%. \citet{jingyi2018deep, astsatryan2021astronomical} used deep CNN to classify quasar and galaxy. Many new improvements of neural network emerged in recent years have been proven to be effective, like residual structures and attention mechanisms \citep{2020PASP..132d4503Z}. A multi-task residual neural network was applied to classify M-type star spectra. It reduced the number of parameters in spectral classification and improved the model efficiency \citep{lu2020study}. Compared to other methods, neural network always worked best on the complex data \citep{2015A&A...580A.138A, chen2021classifying, kerby2021multiwavelength, sharma2020application, vilavicencio2020application, 2019ApJ...879...72G}. 

Rare object identification is another vital task of neural network \citep{2019ApJ...885...85M, jiang2020automated, kou2020new, luo2008automated, vskoda2020active, Tan_2022, zhang2022spectroscopic, zheng2020classification, zou2019classification, 2020MNRAS.496.2346M, Tan_2022,2019MNRAS.485.2167G}. \citet{2014A&A...562A..36S} searched for metal-poor galaxy (MPG) in large surveys and achieved an MPGs acquisition rate about 96\%. \citet{zheng2020generalization} used 1D CNN to search for O stars. \citet{2019ApJ...885...85M, fremling2021sniascore, davison2022stag} proposed a software package that used deep learning models to classify the type, age, redshift, and host galaxy of supernova spectra.
\citet{9079538} identified spectrum J152238.11+333136.1 from LAMOST DR5 and discussed the rare features of P-Cygni profiles.

Neural network based classification algorithms can be used to extract spectral features by different layers (i.e. hidden layers in Fig. \ref{fig:ANN}, convolutional layers in Fig. \ref{fig:CNN}). These layers can automatically learn rich and complex relationships between data. So neural network based algorithms can obtain high accuracy \citep{2019MNRAS.485.2167G, 2020PASP..132d4503Z, 2019MNRAS.483.4774L, 2014PASA...31....1F, 2017MNRAS.465.4311W, jiang2021spectral, jing2020new, MORAES2013621, 2020AJ....160...45P, 2017MNRAS.465.4311W, 2020AJ....160...45P}. Furthermore, neural network could also handle input features well even without color or morphological information \citep{2019ApJ...886..128B, 2019MNRAS.483..529C} which greatly expanded the size and formats of input data sets.

In short, neural network can learn deep features of data, which will provide subtle differences for classification. More importantly, with the introduction of tricks (i.e., residuals and attention blocks), ANN pays more attention on the valid features. In addition, ANN increases its depth to handle complex and high dimensional data. So it has been widely used in astronomy, such as star/galaxy/quasar classification, MPGs/MRGs classification, rare object identification and spectral feature selection, etc \citep{Rastegarnia_2022}. Although neural network model can produce good results, it is a black box that is difficult to interpret results. Compared with decision tree, the results of neural network are difficult for astronomers to analyse the characteristics of celestial objects.

\begin{table*}
\centering
\caption{Investigations of statistics and ranking on astronomical spectra data}
\label{table_statistic_ranking_related}
\resizebox{\linewidth}{!}{
\begin{tabular}{ccc} 
\hline
Merits                                                                                                                                 & Caveats                                                                                                                               & References                                                                                                                                                                                                                                                                                                                                                                                                                                                                                                                                                                               \\ 
\hline
\multicolumn{1}{l}{\begin{tabular}[c]{@{}l@{}} Ranking methods can identify rare objects efficiently  \\ \\ CRC-WPLS are used on non-linear unbalanced data \end{tabular}} & \multicolumn{1}{l}{\begin{tabular}[c]{@{}l@{}}CRC-WPLS is not a prevalent method\\ \\Ranking methods also require ample data\end{tabular}} & \multicolumn{1}{l}{\begin{tabular}[c]{@{}l@{}}{\citet{doi:10.1146/annurev.astro.36.1.369}, \citet{2010JApA...31..177L}},\\{\citet{2015RAA....15.1671S}, \citet{2018ApJS..234...31L}},\\{\citet{2015MNRAS.452.4183H}, \citet{Kang_2021}},\\{\citet{2016PASP..128c4502D}, \citet{2011AJ....142..203D}},\\{\citet{SONG201879}, \citet{2021MNRAS.503..484P}},\\{\citet{tao2018automated}, \citet{2020MNRAS.498.1750A}},\\ {\citet{pruzhinskaya2019anomaly}, \citet{
luo2008automated}}\end{tabular}}  \\
\hline
\end{tabular}
}
\end{table*}

\subsection{Gaussian naive Bayes based classification algorithms} 
Assuming that features are independent, Gaussian naive Bayes based classification algorithms simplify the Bayesian algorithm. They prefer to deal with features in a Gaussian distribution and the maximum posterior probability is the final results. Eq \ref{eq_gaussian_logistic_target_function} is the objective of Gaussian Naive Bayes based classification algorithms and Eq \ref{eq_guassian_logistic_guassian_probability} is Gaussian probabilities. Table \ref{table_statistic_ranking_related} represents astronomical studies of Gaussian naive Bayes based classification algorithms.

\begin{equation}\label{eq_gaussian_logistic_target_function}
y = \mathop {{\mathop{\rm argmax}\nolimits} }\limits_{{c_k}} P\left( {Y = {c_k}} \right)\prod\limits_j P \left( {{X_j} = {x_j}\mid Y = {c_k}} \right)
\end{equation}

where

\begin{equation}\label{eq_guassian_logistic_guassian_probability}
P\left( {{x_i}\mid y} \right) = \frac{1}{{\sqrt {2\pi \sigma _y^2} }}\exp \left( { - \frac{{{{\left( {{x_i} - {\mu _y}} \right)}^2}}}{{2\sigma _y^2}}} \right)
\end{equation}

${\mathop \delta \nolimits_y }$ is variance of ${\mathop x\nolimits_i}$ (${i}$=1, 2, ......, n) and ${\mathop \mu \nolimits_y }$
is average of 
${\mathop x\nolimits_i}$
 in Eq \ref{eq_guassian_logistic_guassian_probability}.

Guassian Naive Bayes based classification algorithms are good at dealing with continuous small data generated from Gaussian distribution. Under the assumption of reliable and sufficient prior spectral information, they could identify rare objects from a large number of spectra data, such as carbon stars \citep{doi:10.1146/annurev.astro.36.1.369, 2010JApA...31..177L, 2020MNRAS.498.1750A, pruzhinskaya2019anomaly, 2015MNRAS.452.4183H}. And they were good at reducing noise of stellar spectra, which increased classification accuracy \citep{Kang_2021}.   

\subsection{Logistic regression based classification algorithms} 
\begin{figure}
\centering
\includegraphics[width=\columnwidth]{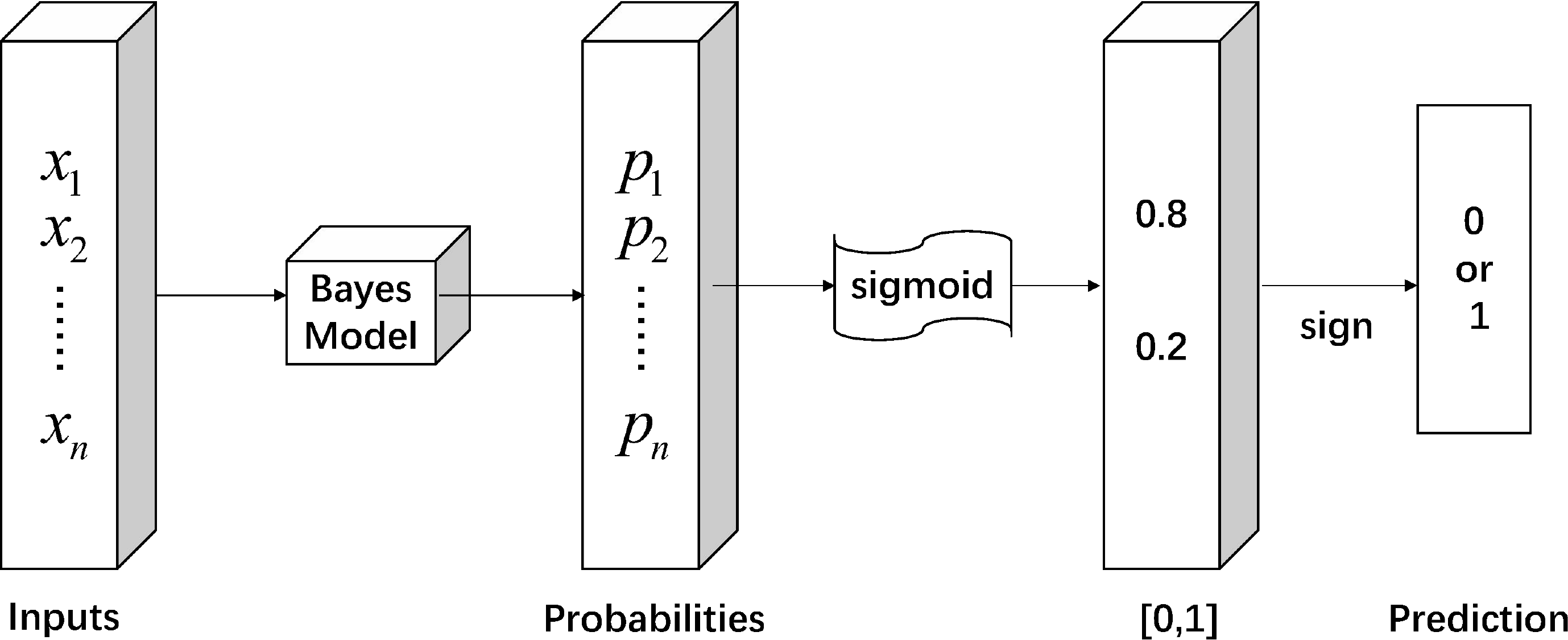}
\caption{Main idea of Bayesian logistic regression. ${\mathop p\nolimits_i }$
 (${i}$=1, 2, ......, n) are probabilities of ${\mathop x\nolimits_i }$ (${i}$
=1, 2, ......, n) and are generated by Bayes model. Sigmoid function activates ${\mathop p\nolimits_i }$ (${i}$=1, 2, ......, n) into value between 0 and 1. Sign function transforms probabilities into label 0 or 1.}
\label{fig:Bayesian Logistic Regression}
\end{figure}

Bayesian Logistic Regression (LR) based classification algorithms obtain posterior probability distributions from linear regression models. And we can get classification results through the sigmoid function. The main researches of LR based classification algorithms are shown in Table \ref{table_statistic_ranking_related}.
Fig. \ref{fig:Bayesian Logistic Regression} is the principle of Bayesian Logistic Regression based classification algorithms.

LR based classification algorithms can be used for quick regression. However, they can not get desirable accuracy due to underfitting, bipartition data and linear data in small feature spaces.
In astronomy, logistic regression based classification algorithms were often combined with other techniques to predict physical parameters and classify celestial objects  \citep{2021MNRAS.503..484P, tao2018automated, luo2008automated}.

\subsection{Collaborative representation based classifier and partial least squares discriminant analysis} 

Partial Least Squares Discriminant Analysis (PLS-DA) belongs to the discriminant analysis of multi-variance data analysis techniques and can be used for classification and discrimination. It handles data in the same cluster rather than data in different clusters. Data in the same group varies widely. And data volumes between groups differ a lot. It extracts principle components of the independent variable X and the controlled variable Y, and finds the relationship between principle components in a two high-dimensional space. Table \ref{table_statistic_ranking_related} displays the main astronomical researches of CRC-PLS based classification algorithms. 

CRC is a novel machine learning algorithm that represents a query by a linear
integral of training samples. And CRC classifies the above queries based on the representation \citep{2011AJ....142..203D}. It has the ability to handle unbalanced, non-linear and multi-label data.

CRC-PLS reaps the merits of PLS regression and CRC. So it can classify the high-dimensional spectra data \citep{SONG201879}.

\subsection{Ranking based classification algorithms} 
Ranking based  positive-unlabelled(PU) learning algorithms have been frequently used in astrophysical object retrieval. Graph based ranking methods successfully identify carbon stars from massive astronomical spectra data, such as manifold algorithm and efficient manifold algorithm \citep{2015RAA....15.1671S}, Locally linear embedding.  The bipartite ranking is another typical method to improve ranking performance and it has been introduced to search for carbon stars \citep{2016PASP..128c4502D}. Alternatively, bagging is a popular method to obtain better performance by integrating different classifiers. The idea of bagging has been well applied in rare object retrieval wonderfully \citep{2018ApJS..234...31L, 2016PASP..128c4502D}.

The core idea of ranking based classification methods is to learn a ranking based model which usually ranks data sets by predefined evaluation methods. They have two goals: 1) positive samples are ranked ahead of negative samples. 2) the scores of related samples tend to be similar. Many optimal ranking methods have emerged to improve classification performance and reduce time consumption, such as efficient manifold algorithms and bagging TopPush. And these methods have already discovered carbon stars from extensive spectra data which is a significant supplement to the catalogs of carbon stars (Table \ref{table_statistic_ranking_related} 
).
\section{Experiment Analysis}
Recently, lots of basic or improved classification algorithms have been successfully applied to various astronomical data analyses. However, due to the diversity of classification tasks and classification data, it is difficult to assess the advantages and disadvantages of these methods from the current literature. So, in this section, we construct unified experimental spectral data sets from LAMOST survey and  SDSS survey to evaluate the commonly used methods.
\subsection{Experimental data introduction}
In the experimental design, we construct several groups of data sets using the spectra data from LAMOST \citep{2015RAA....15.1095L} and SDSS.

LAMOST (The Large Sky Area Multi-Object Fiber Spectroscopic Telescope, also known as Guo Shou Jing Telescope) is a special reflective Schmidt telescope with an effective aperture of 3.6–4.9 meters and a field of view of 5$\degree$. It is equipped with 4000 fibers, a spectra resolution of R $\approx$ 1800 and a wavelength ranging from 3800 to 9000 $\mathop A\limits^ \circ $
(\url{http://www.lamost.org/public/?locale=en}). Its scientific goal is to make a 20000 square degrees spectroscopic survey (DEC: -10$\degree$ $\sim$ +90$\degree$).  After seven years of surveying, LAMOST has observed tens of millions of low-resolution spectra data, providing important data for astronomical statistical research.  

The Sloan Digital Sky Survey (SDSS) is an international collaboration of scientists to build the most detailed Three-Dimensional Imagery of the Universe. It uses a wide-field telescope with a diameter of 2.5 meters and a field of view of 3$\degree$. The photometric system is matched with five filters in u, g, r, i and z bands to photograph celestial objects. It covers 7500 square degrees of the sky around the South Galactic Pole and records data on nearly 2 million celestial objects. 

Experimental data are selected from LAMOST DR8 and SDSS DR16. The LAMOST DR8 data sets include a total of 17.23 million released spectra. The number of high-quality spectra of DR8 (that is, the S/N > 10) reaches 13.28 million and DR8 includes a catalog of about 7.75 million groups of stellar spectral parameters. The SDSS DR16 covers more than one-third of the sky and contains about 5,789,200 total spectra and 4,846,156 useful spectra. And DR16 contains new optical and infrared spectra, including the first infrared spectra observed by Las Campanas Observatory in Chile.

\begin{table}
\centering
\caption{Data preprocessing}
\label{table:dataprepocess}
\begin{tabular}{ll} 
\hline
                 & Data selection and preprocessing                                    \\ 
\hline
Data release     & LAMOST DR8, SDSS DR16                                 \\
Extinction      & 1D spectra from LAMOST(l > 45°)                          \\
Redshift        & Rest wavelength frame spectra for star/galaxy/quasar  \\
Flux calibration & Relative flux calibration: cut off 5700$\mathop A\limits^ \circ $-5900$\mathop A\limits^ \circ $       \\
\hline
\end{tabular}
\end{table}

We select and preprocess the spectra from four aspects. These are shown in Table \ref{table:dataprepocess}.

(1) Data release. We select spectra from LAMOST DR8 and SDSS DR16. 

(2) Extinction problem. In order to decrease the influence of reddening on classification performance, 1D spectra in data sets are selected from LAMOST (45$\degree$< l) \citep{2022mnras}. 

(3) Flux calibration. LAMOST uses relative flux calibration. We cut off the
overlapping region (5700\AA $<\lambda<$ 5900\AA) known to have calibration issues to minimize their effect on our classification.

(4) Redshift. For star/galaxy/quasar classification, we convert original spectra into the rest-frame wavelengths by applying the redshift values from LAMOST and compare the performance of classification on the rest wavelength frame spectra and original spectra. Because the radial velocity of stellar spectra are small under the current resolution of LAMOST, which has little influence on classification results. Spectra for stellar classification are left in the observed-frame wavelengths.

We determine three classification tasks among multiple astronomical researches, including A/F/G/K stars classification, star/galaxy/quasar classification and rare object identification. Rare objects includes carbon stars \citep{doi:10.1146/annurev.astro.36.1.369, 2012A&A...544A..95G, 2010JApA...31..177L}, double stars, artefacts: bad merging of red and blue segments (A common phenomenon that occurs in the spectra of LAMOST). 

We design six groups of data sets for the above tasks. Data sets 1 - data sets 3 are constructed for A/F/G/K stars classifications. They are divided by data characteristics, S/Ns and data volumes, and each data sets contain three or four sub-datasets. Data sets 4 are used to evaluate the classification performance of star/galaxy/quasar on original spectra and rest wavelength frame spectra. Data set 5 is used to identify rare objects: carbon stars, double stars and artefacts. And the classifier is trained on 200 rare objects and 19900 other non-rare objects. Non-rare objects include 10000 normal stars, 6500 galaxies and 3400 quasars. We analyse the results of rare object identification by accuracy, precision, recall and F1 score.
Spectra of the first five groups of data sets are selected from LAMOST. Because the sources of LAMOST have considerable overlaps with SDSS, we construct the matching data sets (data sets 6 in Table \ref{dataset_6}) from SDSS and LAMOST to compare the classification performances on them. The analyses of experimental results on data sets 6 are elucidated in Section 3.2.1. 

The composition of testing sets in all data sets is the same as their training sets. The ratio of training sets and testing sets for data sets 1, data sets 2, data sets 3, data sets 4 and data sets 6 is 8:2 and the ratio of training sets and testing sets for data set 5 is 1:1. Details of data sets are shown in Table \ref{dataset_6}.

\begin{table*}
\centering
\caption{Data sets of spectral classification}
\label{dataset_6}
\begin{threeparttable}
\resizebox{\linewidth}{!}{
\begin{tabular}{lllll} 
\hline
                            & Data sets introduction$^{1}$                                                             & Data components$^{2}$                                                                                                     & S/N                                                                                         & Characteristics              \\ 
\hline
\\ 
\multirow{4}{*}{Data sets 1} & \multirow{3}{*}{\begin{tabular}[l]{@{}l@{}}A/F/G/K stars classifications\\ on four characteristics\end{tabular}} & A : F : G : K Stars = 5000 : 5000 : 5000 : 5000                                                     & $>$10                                                                      & 1D Spectra                             \\
                            &                                                                                  &                                                                                                                    A : F : G : K Stars = 5000 : 5000 : 5000 : 5000 &                                                                                            $>$10                                                                      & PCA (100 dimensions)         \\
                            &                                                                                  &                                                                                                                    A : F : G : K Stars = 5000 : 5000 : 5000 : 5000 &                                                                                            $>$10                                                                      & Line Indices                 
                                                                                                          \\ \\

\multirow{3}{*}{Data sets 2} & \multirow{3}{*}{\begin{tabular}[l]{@{}l@{}}A/F/G/K stars classifications\\ on three S/Ns\end{tabular}}                             & A : F : G : K Stars = 5000 : 5000 : 5000 : 5000                                                    & $<$10                                                                                       & 1D Spectra  \\
                            &                                                                                  &                                                                                                                    A : F : G : K Stars = 5000 : 5000 : 5000 : 5000                                                    & 10-30                                                                                       & 1D Spectra                             \\
                            &                                                                                  &                                                                                                                    A : F : G : K Stars = 5000 : 5000 : 5000 : 5000                                                    & $>$30                                                                                       & 1D Spectra                             \\ \\

\multirow{4}{*}{Data sets 3} & \multirow{4}{*}{\begin{tabular}[l]{@{}l@{}}A/F/G/K stars classifications\\ on four volumes\end{tabular}}                           & A : F : G : K Stars = 2000 : 2000 : 2000 : 2000                                                                      & $>$10                                                                      & 1D Spectra  \\                           &                                                                                  & A : F : G : K Stars = 5000 : 5000 : 5000 : 5000                                                                      &                                                                                            $>$10                                                                      &                             1D Spectra  \\
                            &                                                                                  & A : F : G : K Stars = 10000 : 10000 : 10000 : 10000                                                                  &                                                                                            $>$10                                                                      &                             1D Spectra  \\
                            &                                                                                  & A : F : G : K Stars = 20000 : 20000 : 20000 : 20000                                                                  &                                                                                        $>$10                                                                      &                             1D Spectra  \\ \\
\multirow{2}{*}{Data sets 4}                   &\multirow{2}{*}{Star/galaxy/quasar classifications                                              }  & star : galaxy : quasar = 5000 : 5000 : 5000                                                                         &\multirow{2}{*}{\begin{tabular}[l]{@{}l@{}} stars : $>$10,\\galaxies , quasars : all \end{tabular}}                           
 & Original Spectra \\&& star : galaxy : quasar = 1000 : 1000 : 1000                                                                                        &&Rest Wavelength Frame Spectra \\ \\

Data set 5                   & Search for rare objects                                                          $^{3}$ & \begin{tabular}[l]{@{}l@{}}rare objects : normal stars : galaxies : quasars\\=200 : 10000 : 6500 : 3400\end{tabular} & \begin{tabular}[c]{@{}c@{}}normal stars : $>$10,\\galaxies , quasars : all\end{tabular} & 1D Spectra                   \\ \\
\multirow{2}{*}{Data sets 6}                   &\multirow{2}{*}{\begin{tabular}[l]{@{}l@{}}A/F/G/K stars classifications\\ on LAMOST and SDSS\end{tabular}}& A : F : G : K = 5824 : 5380 : 4151 : 6240(LAMOST) & \multirow{2}{*}{$>$10                                                                                     } & \multirow{2}{*}{1D Spectra}  \\&&A : F : G : K = 5797 : 5355 : 4144 : 6229(SDSS)&&\\ \\
\hline
                            &                                                                                  &                                                                                                                     &                                                                                             &                             
\end{tabular}
}

\begin{tablenotes}
    \footnotesize
    \item[1] : Spectra of data sets 1 - data set 5 are selected from LAMOST. Spectra of data sets 6 are selected from LAMOST and SDSS.
    \item[2] : The values of the data componences in this table are the actual data volume.
    \item[3] : Rare objects : carbon stars, double stars, artefacts : bad merging of red and blue segments.
     
    \end{tablenotes}
    \end{threeparttable}
\end{table*}

\subsection{Result analysis}
In this section, nine basic methods including K-Nearest Neighbour, Support Vector Machine, Decision Tree, Random Forest, Gradient Boosting Decision Tree, Logistic Regression, Pseudo Inverse Learning and Convolutional Neural Network are tested on astronomical spectra data and we fairly evaluate the classification performance. 

Our experiments use grid search \citep{Syarif2016SVMPO} to identify the optimal parameters of each algorithm. And we take the average accuracy of 5-fold cross validation \citep{10.1007/s11222-009-9153-8} as the final accuracy to avoid the influence of sample selection.

\subsubsection{Performance analysis on 1D spectra, PCA and line indices} 

\begin{figure*}
\centering
\includegraphics[width=12.2cm,height=5.12cm]{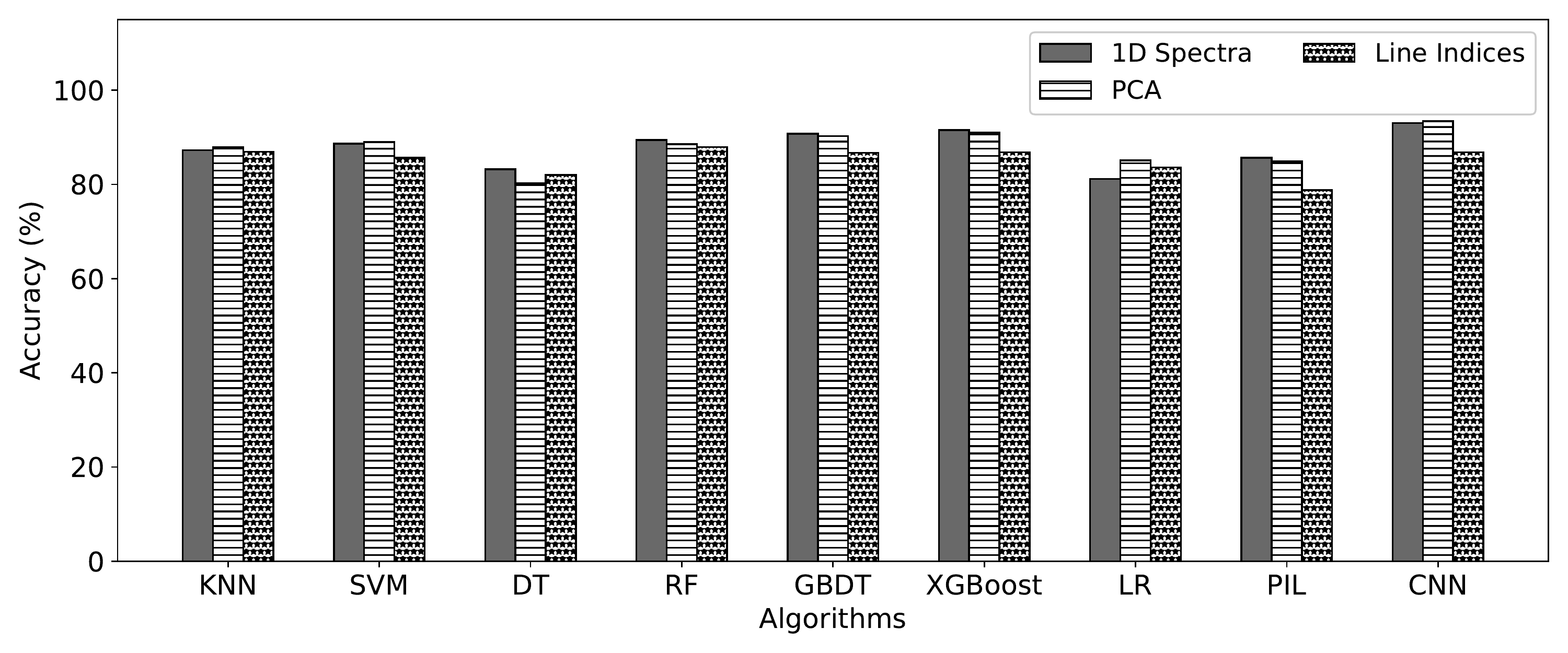}
\caption{Accuracy of algorithms on different data characteristics. Three different types of bars stand for various data characteristics.}
\label{fig:AFGK_accuracy_characteristic_2w_10-_bar}   
\end{figure*}

Fig. \ref{fig:AFGK_accuracy_characteristic_2w_10-_bar} represents the accuracy of nine basic algorithms on three data characteristics (1D spectra, PCA, line indices).

In the classification on 1D spectra, CNN achieves the highest accuracy. Because it can extract complex features through different layers. However, CNN still suffers from two unavoidable drawbacks. One is that it has to spend a long time to obtain the optimal model. The other is overfitting which can not be easily eliminated even by L2 regularization or dropout method. In order to reduce the training time, we can extract features by PCA and classify the pre-processed spectra. Because accuracy on PCA features is equal to that on 1D spectra and the training time is shorter.

\begin{figure*}
\centering
\includegraphics[width=11.6cm,height=9.79cm]{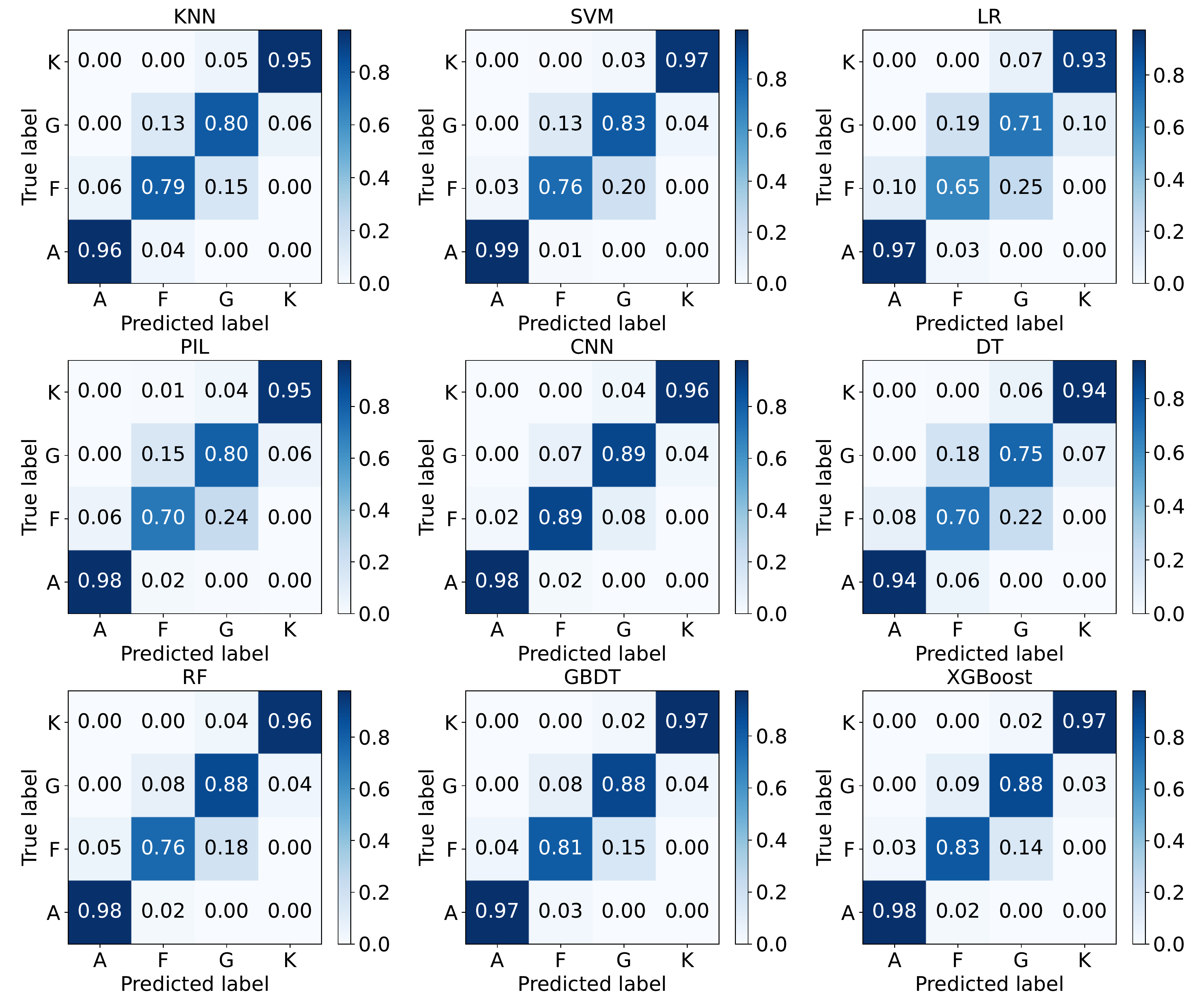}
\caption{Accuracy of algorithms on 1D spectra of A/F/G/K stars. x axis represents predicted labels conducted by experiments. y axis represents true labels of spectra. Figures in the grids are the consistent probabilities between predicted labels and true labels. Algorithm names are presented on the above of each confusion matrix.}
\label{fig:AFGK_raw_confusion}   
\end{figure*}

In Fig. \ref{fig:AFGK_raw_confusion}, A stars and K stars can be distinguished admirably whereas F stars and G stars have 
disappointing accuracy. Because F stars and G stars are more similar than A stars and K stars in the global shape of 1D spectra. Stellar rotation might become another reason for the misclassification because it broadens spectral lines and might cause the global shape of 1D spectra if lines are blended because of insufficient spectral resolution. So it is necessary to alleviate the influence of stellar rotation on classification. Moreover, researchers can use other spectra characteristics to avoid the caveats of 1D spectra. Results also show that LR, Pseudo Inverse Learning (PIL), DT can not get desirable results on F stars and G stars due to the weak spectral shapes. While strong classifiers (CNN, ensemble methods, SVM) show superiority. 

\begin{figure*}
\centering
\includegraphics[width=11.6cm,height=9.79cm]{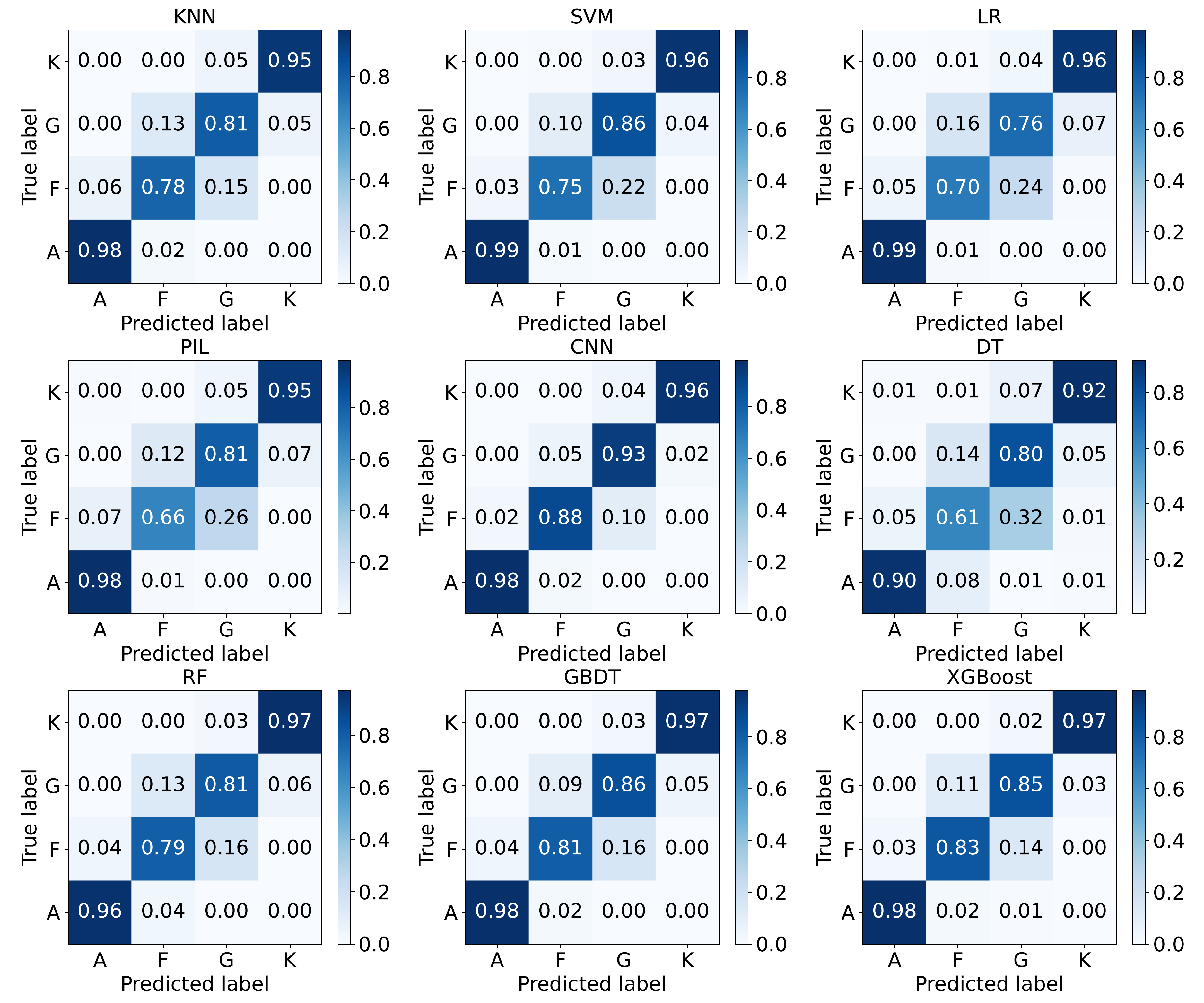}
\caption{Accuracy of algorithms on PCA of A/F/G/K stars. x axis represents predicted labels conducted by experiments. y axis represents true labels of spectra. Figures in the grids are the consistent probabilities between predicted labels and true labels. Algorithm names are presented on the above of each confusion matrix.}
\label{fig:AFGK_PCA_confusion}   
\end{figure*}

PCA is a useful dimensionality reduction tool in many fields. Technically, it extracts principle components of spectra. And the principle components preserve the main information of spectra as much as possible. So accuracy shows little difference with 1D spectra (Figs. \ref{fig:AFGK_accuracy_characteristic_2w_10-_bar}-\ref{fig:AFGK_PCA_confusion}). However, the consistent results can not be explained well because linear PCA may be misleading to tackle the non linear spectral lines. This phenomenon has also confused researchers \citep{tao2018automated}. And spectra preprocessed by PCA are a linear sum of different dimensional characteristics from 1D spectra which lacks concrete (astro)physical meaning. These problems need to be explored in the future. The main merit of PCA is that the spectra preprocessed by PCA can reduce the number of features and the computation time. So it has been widely used in astronomical tasks.

\begin{figure*}
\centering
\includegraphics[width=11.6cm,height=9.79cm]{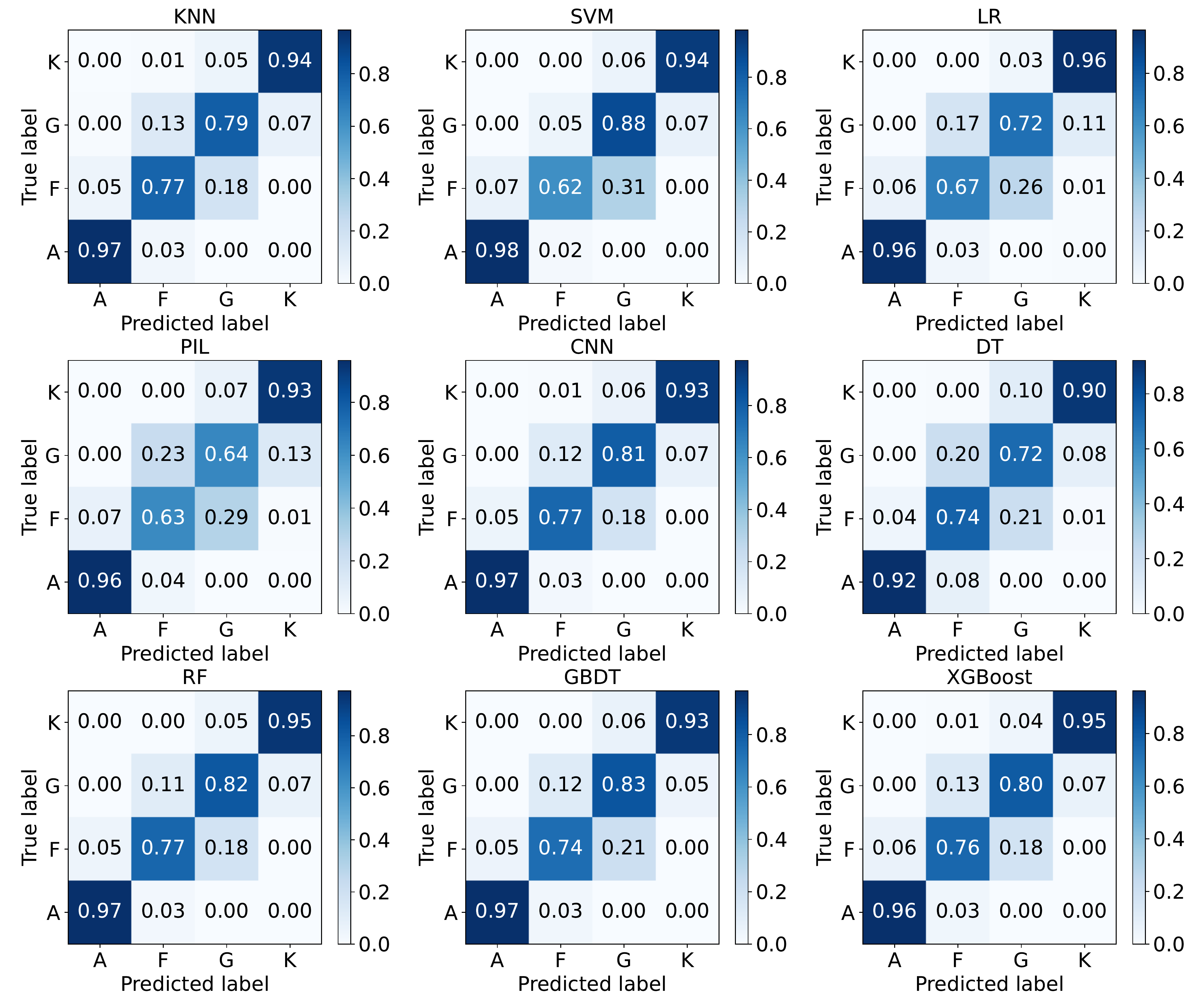}
\caption{Accuracy of algorithms on line indices of A/F/G/K stars. x axis represents predicted labels conducted by experiments. y axis represents true labels of spectra. Figures in the grids are the consistent probabilities between predicted labels and true labels. Algorithm names are presented on the above of each confusion matrix.}
\label{fig:AFGK_line_index_confusion}   
\end{figure*}

Line indices are vital features for spectral analysis. They refer to the relative intensity of absorption or emission lines produced by certain elements. And stellar absorption lines can be used to distinguish stars. Fig. \ref{fig:AFGK_accuracy_characteristic_2w_10-_bar} illustrates the results of nine basic algorithms on line indices. Overall, nine basic classification algorithms performed similarly.
Compared with the results on 1D spectra, simple KNN is superior to CNN in the low dimensional space of line indices. Because the powerful feature selection of CNN tends to show advantages in high dimensional space.   
Fig. \ref{fig:AFGK_line_index_confusion} show more misclassifications between A stars and F stars. Misclassification between F stars and G stars has decreased a little. And we can clearly see that F stars can be distinguished better than other stars.

\begin{figure*}
\centering
\includegraphics[width=12.8cm,height=5.4cm]{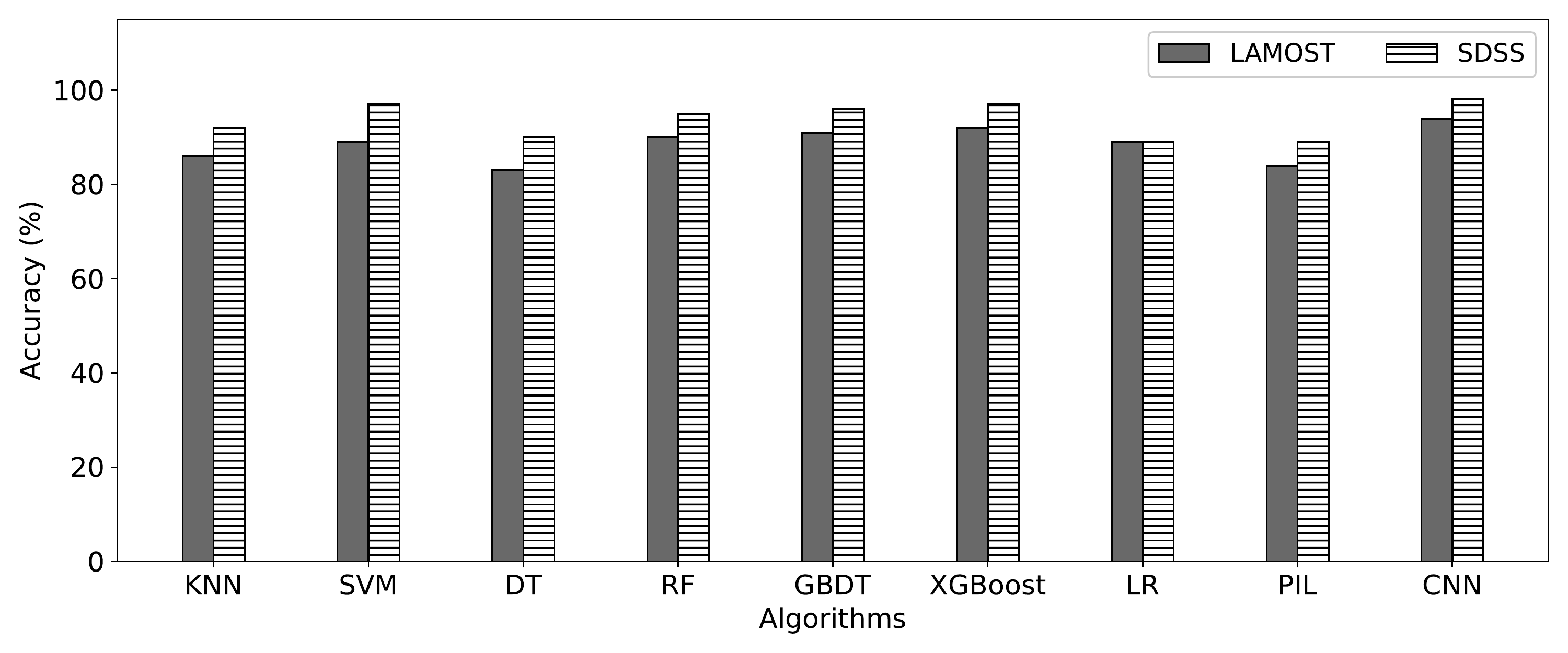}
\caption{Accuracy of algorithms on A/F/G/K stars of LAMOST and SDSS. Two different bars represent the spectra from LAMOST and SDSS.}
\label{fig:AFGK_LAMOST_SDSS_bar}   
\end{figure*}

\begin{figure*}
\centering
\includegraphics[width=11.6cm,height=9.79cm]{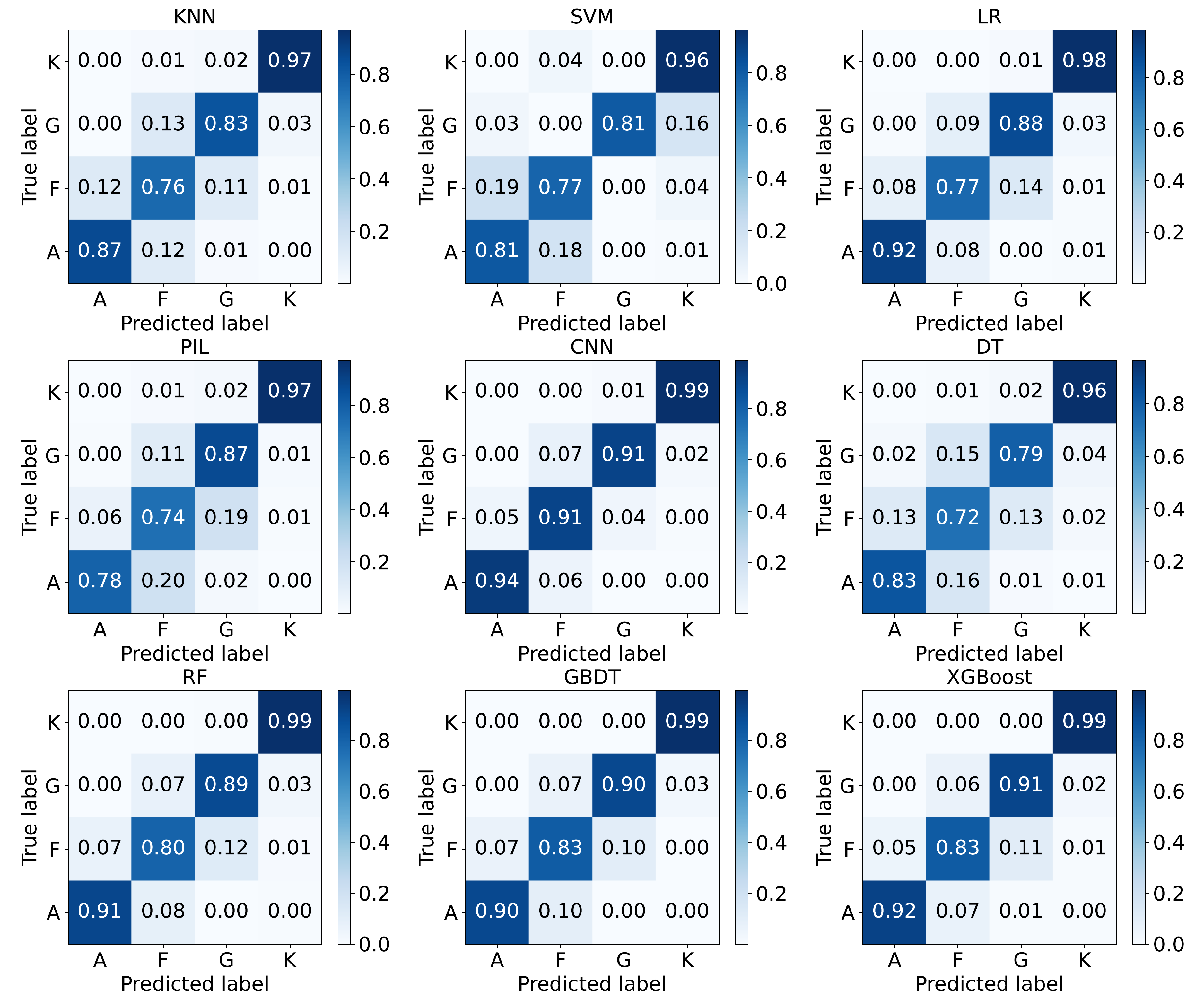}
\caption{Confusion matrices of algorithms on A/F/G/K stars of LAMOST. x axis represents predicted labels conducted by experiments. y axis represents true labels of spectra. Figures in the grids are the consistent probabilities between predicted labels and true labels. Algorithm names are presented on the above of each confusion matrix.}
\label{fig:AFGK_LAMOST_confusion}   
\end{figure*}

\begin{figure*}
\centering
\includegraphics[width=11.6cm,height=9.79cm]{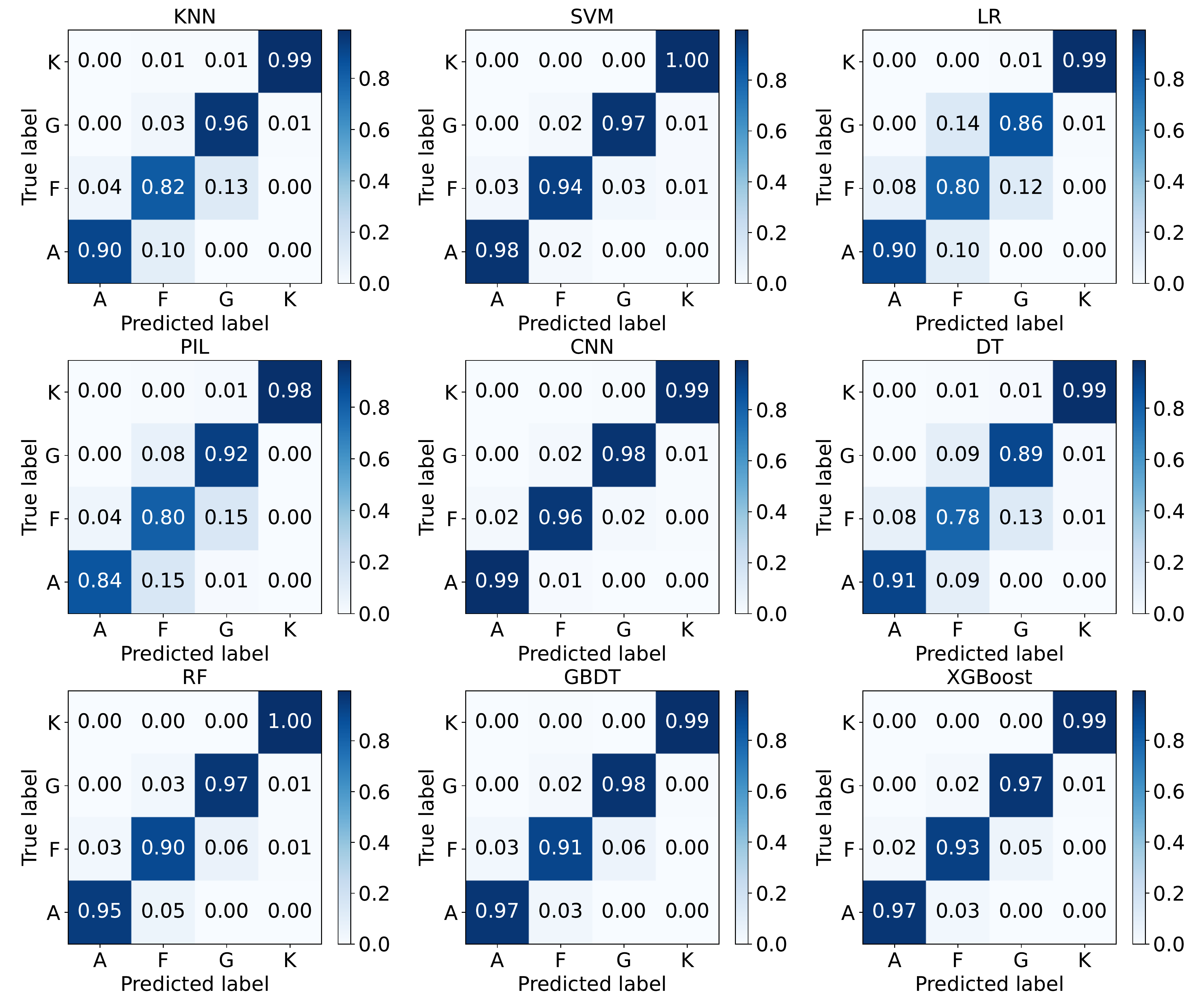}
\caption{Confusion matrices of algorithms on A/F/G/K stars of SDSS. x axis represents predicted labels conducted by experiments. y axis represents true labels of spectra. Figures in the grids are the consistent probabilities between predicted labels and true labels. Algorithm names are presented on the above of each confusion matrix.}
\label{fig:AFGK_SDSS_confusion}   
\end{figure*}

\textit{Comparative results analysis of LAMOST and SDSS.} As can be seen from Fig. \ref{fig:AFGK_LAMOST_SDSS_bar}, the classification algorithms perform better on SDSS instead of LAMOST. The reason may be that the calibration quality of LAMOST will be influenced by fiber-to-fiber sensitivity variations, further causing the slight differences on classification results. As shown in Fig. \ref{fig:AFGK_LAMOST_confusion}, all classification algorithms perform best on K-type stars from LAMOST. But they perform poorly on F-type stars from LAMOST. Similarly, the performance of classification algorithms on F-type stars from SDSS is bad (Fig. \ref{fig:AFGK_SDSS_confusion}). And it can be clearly seen that the performance of classification algorithms on A, G, K stars from SDSS is similar, but slight better than that from LAMOST.

\subsubsection{Performance analysis on spectra qualities}
\begin{figure*}
\centering
\includegraphics[width=12.8cm,height=5.376cm]{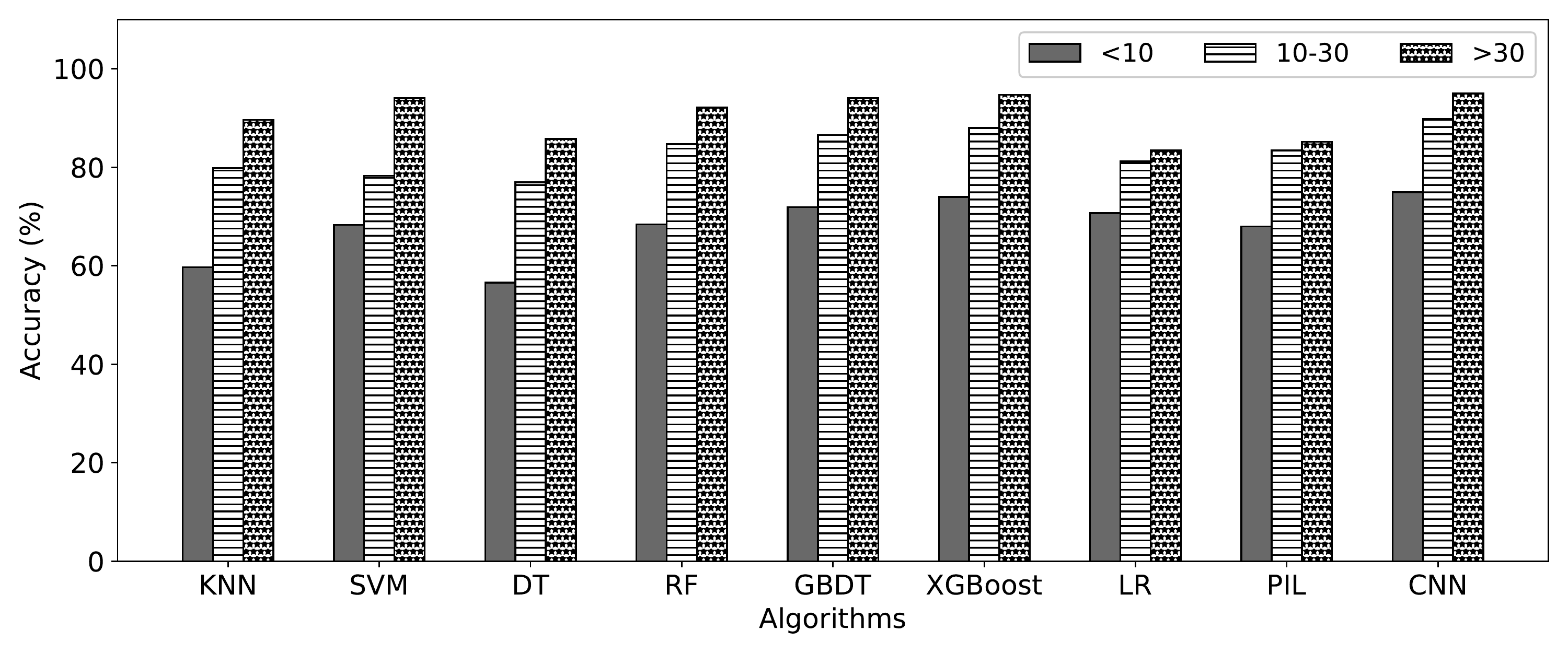}
\caption{Accuracy of algorithms on different S/Ns. Different color bars stand for different S/Ns.}
\label{fig:AFGK_accuracy_snr_2w_raw_bar}   
\end{figure*}

On the whole, the accuracy is in direct proportion to S/N (Fig. \ref{fig:AFGK_accuracy_snr_2w_raw_bar}). 
Paying more attention on S/N$>$30, we can draw a conclusion that SVM, ensemble methods and CNN can achieve better results than KNN. And, the classification performance of PIL is better than that of LR. Because PIL can extract complicated features through a simple three-layer neural network while LR fails in high-dimensional space.

The accuracy of classification on S/N: 10-30 drops completely because spectral data on S/N: 10-30 are always mixed with noise. CNN continues to remain top of the nine basic algorithms because it has added regularization and dropout methods to alleviate overfitting. 

\begin{figure*}
\centering
\includegraphics[width=11.6cm,height=9.79cm]{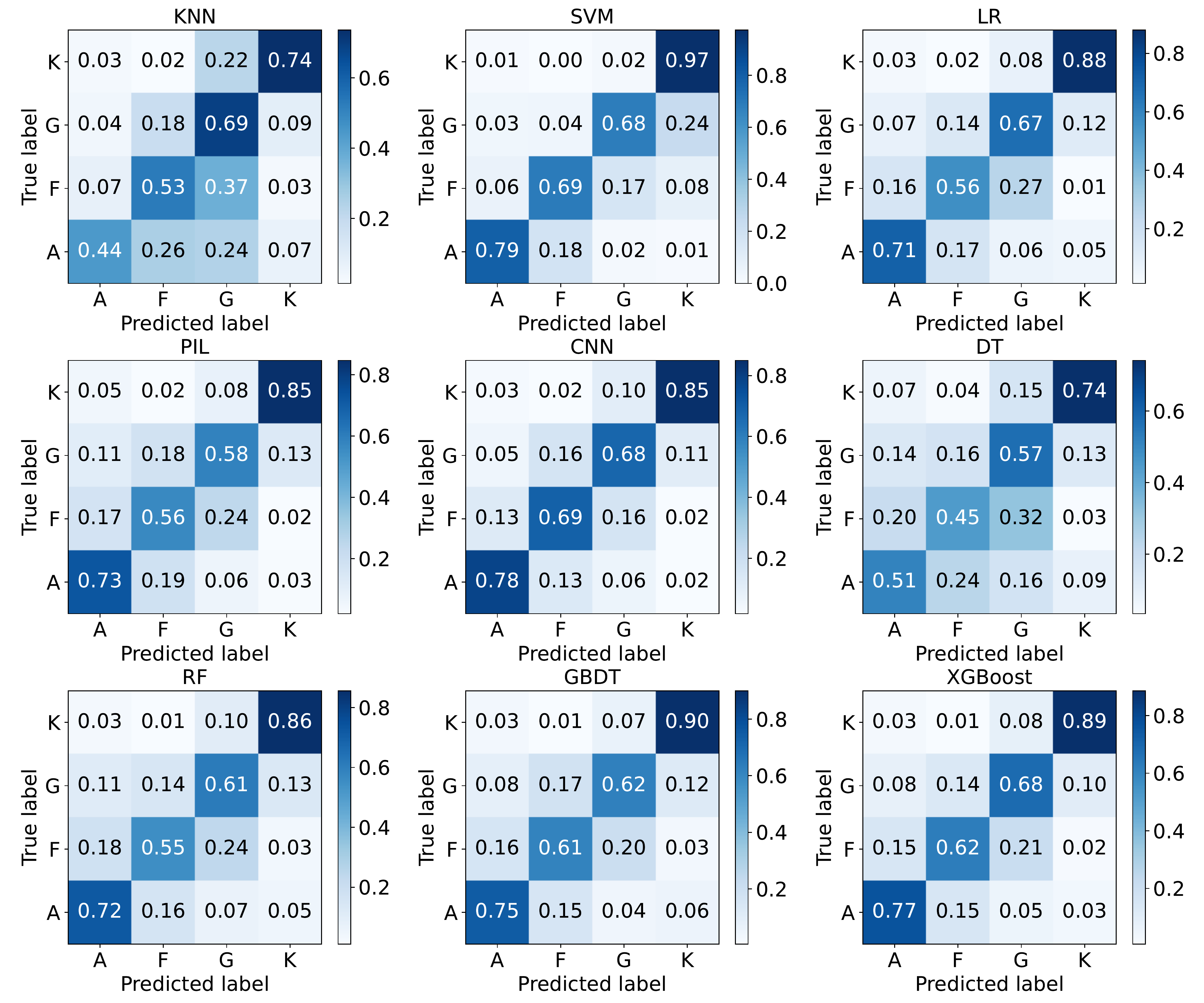}
\caption{Accuracy of algorithms on S/N$<$10 of A/F/G/K stars. x axis represents predicted labels conducted by experiments. y axis represents true labels of spectra. Figures in the grids are the consistent probabilities between predicted labels and true labels. Algorithm names are presented on the above of each confusion matrix.}
\label{fig:AFGK_-10_confusion}   
\end{figure*}

\begin{figure*}
\centering
\includegraphics[width=11.6cm,height=9.79cm]{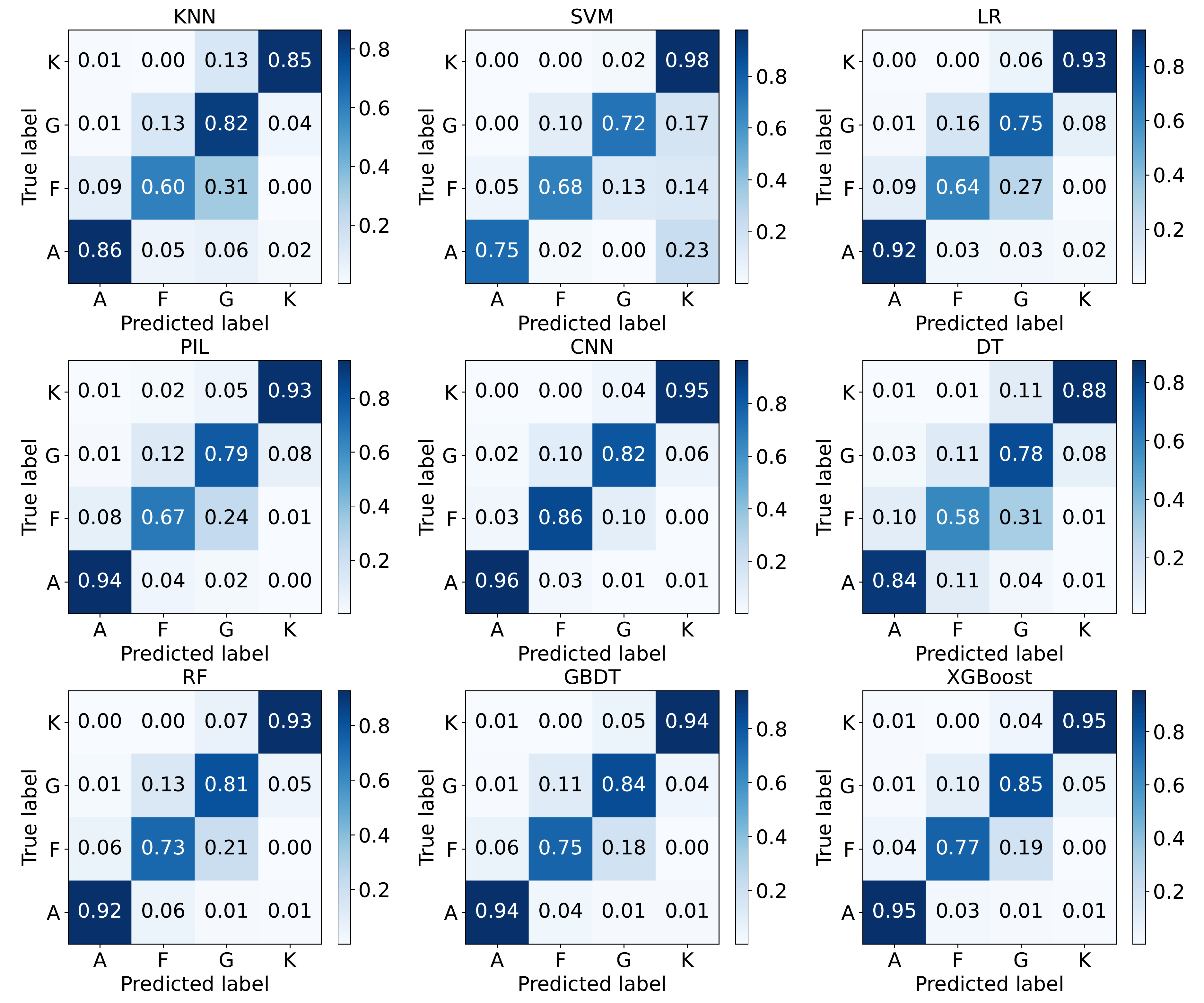}
\caption{Accuracy of algorithms on S/N: 10-30 of A/F/G/K stars. x axis represents predicted labels conducted by experiments. y axis represents true labels of spectra. Figures in the grids are the consistent probabilities between predicted labels and true labels. Algorithm names are presented on the above of each confusion matrix.}
\label{fig:AFGK_10-30_confusion}   
\end{figure*}

\begin{figure*}
\centering
\includegraphics[width=11.6cm,height=9.79cm]{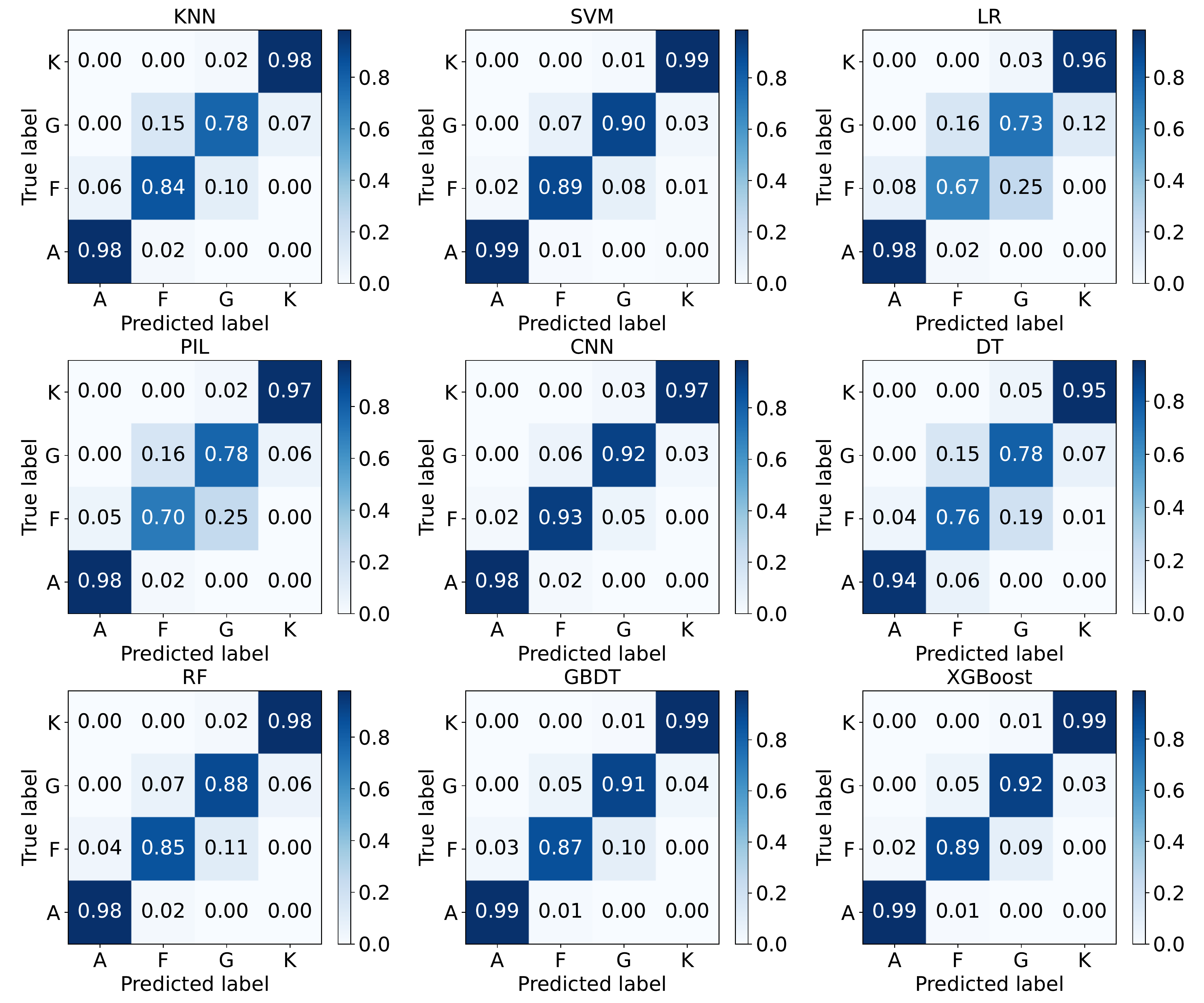}
\caption{Accuracy of algorithms on S/N$>$30 of A/F/G/K stars. x axis represents predicted labels conducted by experiments. y axis represents true labels of spectra. Figures in the grids are the consistent probabilities between predicted labels and true labels. Algorithm names are presented on the above of each confusion matrix.}
\label{fig:AFGK_30-_confusion}   
\end{figure*}

It is difficult to mine information from spectra on S/N$<$10 which are often regarded as unqualified spectra. As a result, it is prevalent to obtain low accuracy on spectra with S/N$<$10. We divide algorithms into three parts according to their classification accuracy. Obviously, SVM, CNN, ensemble methods and LR are the leading echelons followed by PIL. SVM shows robustness on S/N$<$10. Because the soft margin of SVM guarantees that most spectra are classified correctly even for some misclassified samples. Likewise, CNN gains 70\% accuracy depending on the strong ability of feature selection. GBDT and XGBoost adopt gradient boosting methods to reduce errors and attain higher accuracy than RF which only merges different decision trees.
KNN and decision tree can not satisfy us. They perform worst. Because KNN uses Euclidean distance as a distance metric. So it is susceptible to noise. Likewise, decision tree can not find proper splitting features because of noise. We can find misclassification of spectra on S/N$<10$ from Figs. \ref{fig:AFGK_-10_confusion}, \ref{fig:AFGK_10-30_confusion}, \ref{fig:AFGK_30-_confusion}, such as  the poor performance of PIL and decision tree methods on F stars . 

\subsubsection{Performance analysis on different data volumes}
\begin{figure*}
\centering
\includegraphics[width=12.2cm,height=5.08cm]{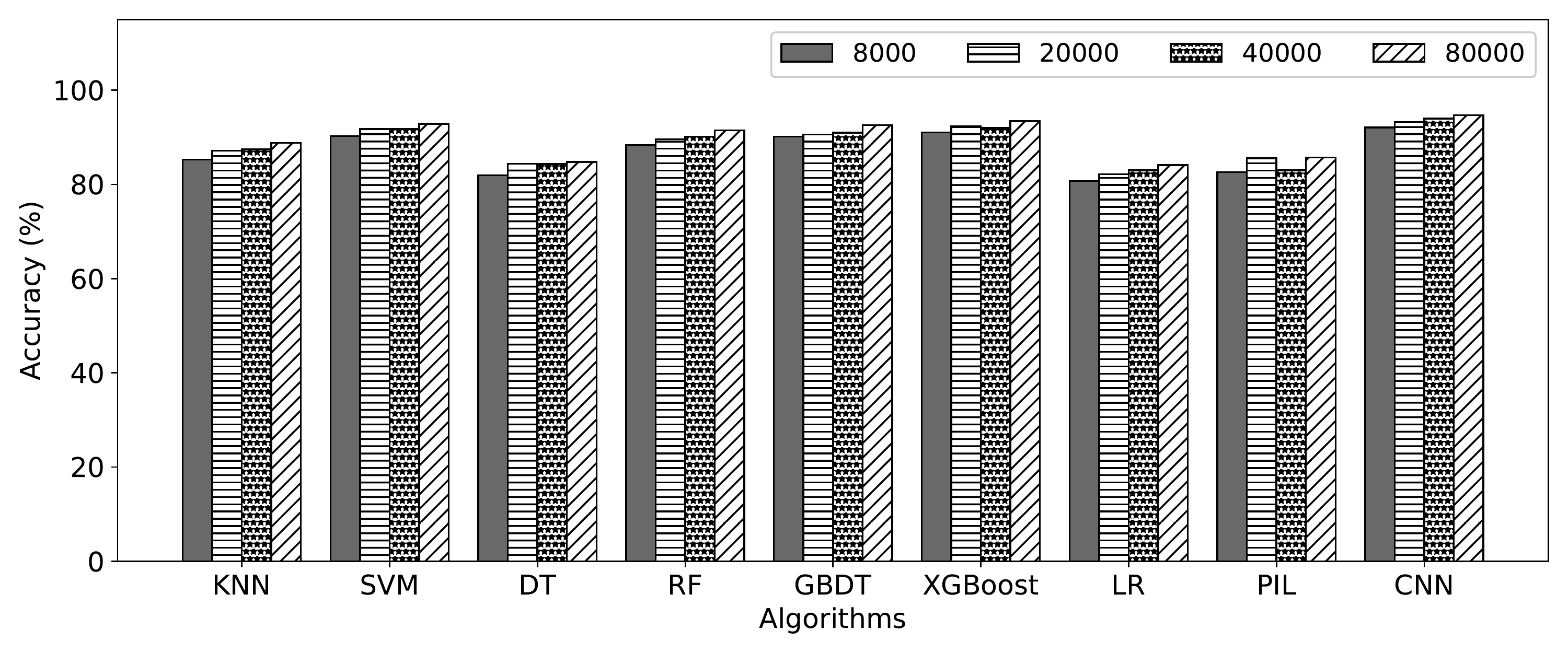}
\caption{Accuracy of algorithms on different data volumes. Four different bars stand for four types of data volumes.}
\label{fig:AFGK.accuracy_volume_raw_10-_bar}   
 
\end{figure*}

\begin{figure}
\centering
\includegraphics[width=\columnwidth]{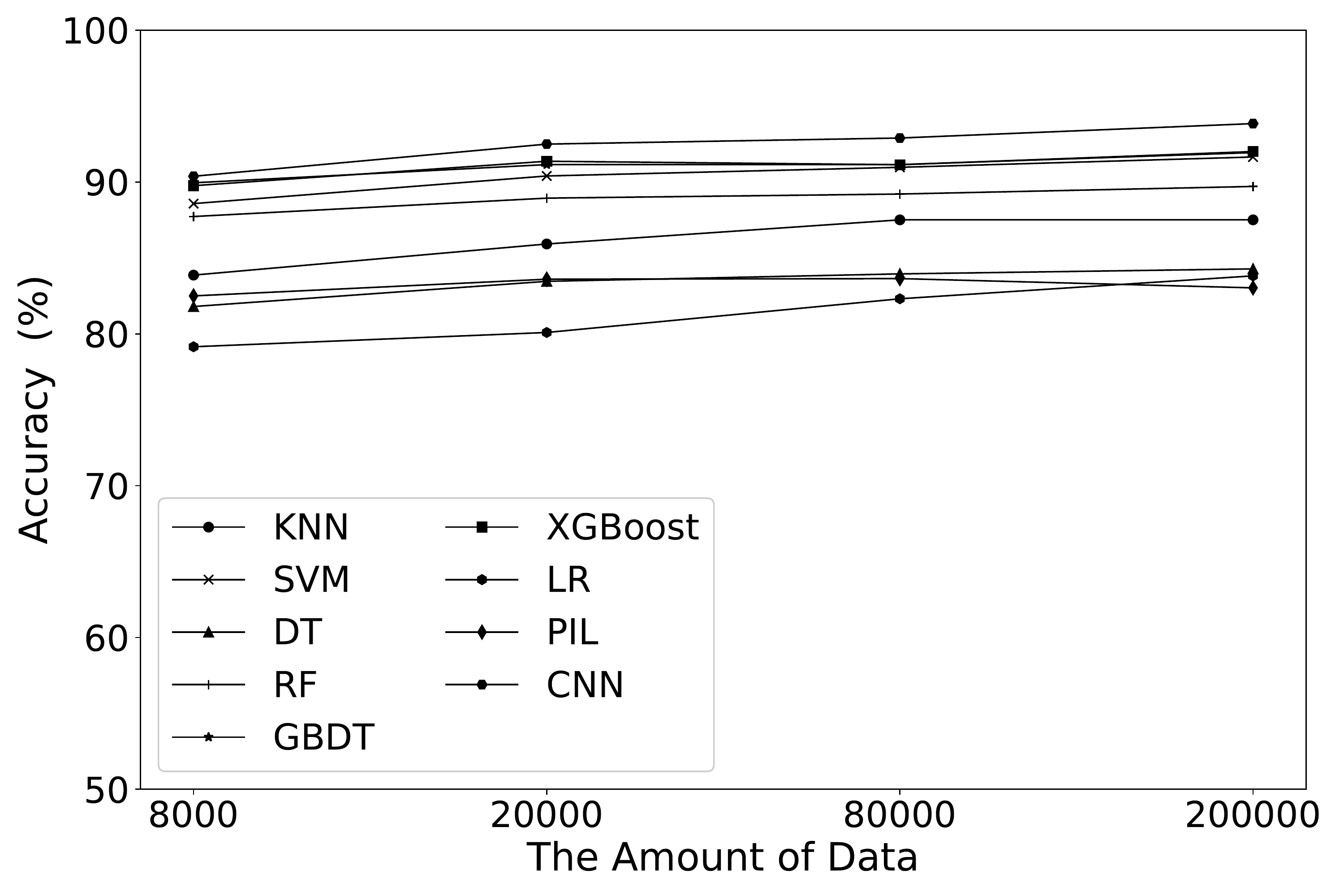}
\caption{Accuracy of algorithms on different data volumes. Different shapes in lines represent different classification algorithms.}
\label{fig:AFGK_accuracy_amount_linechart}   
\end{figure}

\begin{figure}
\centering
\includegraphics[width=\columnwidth]{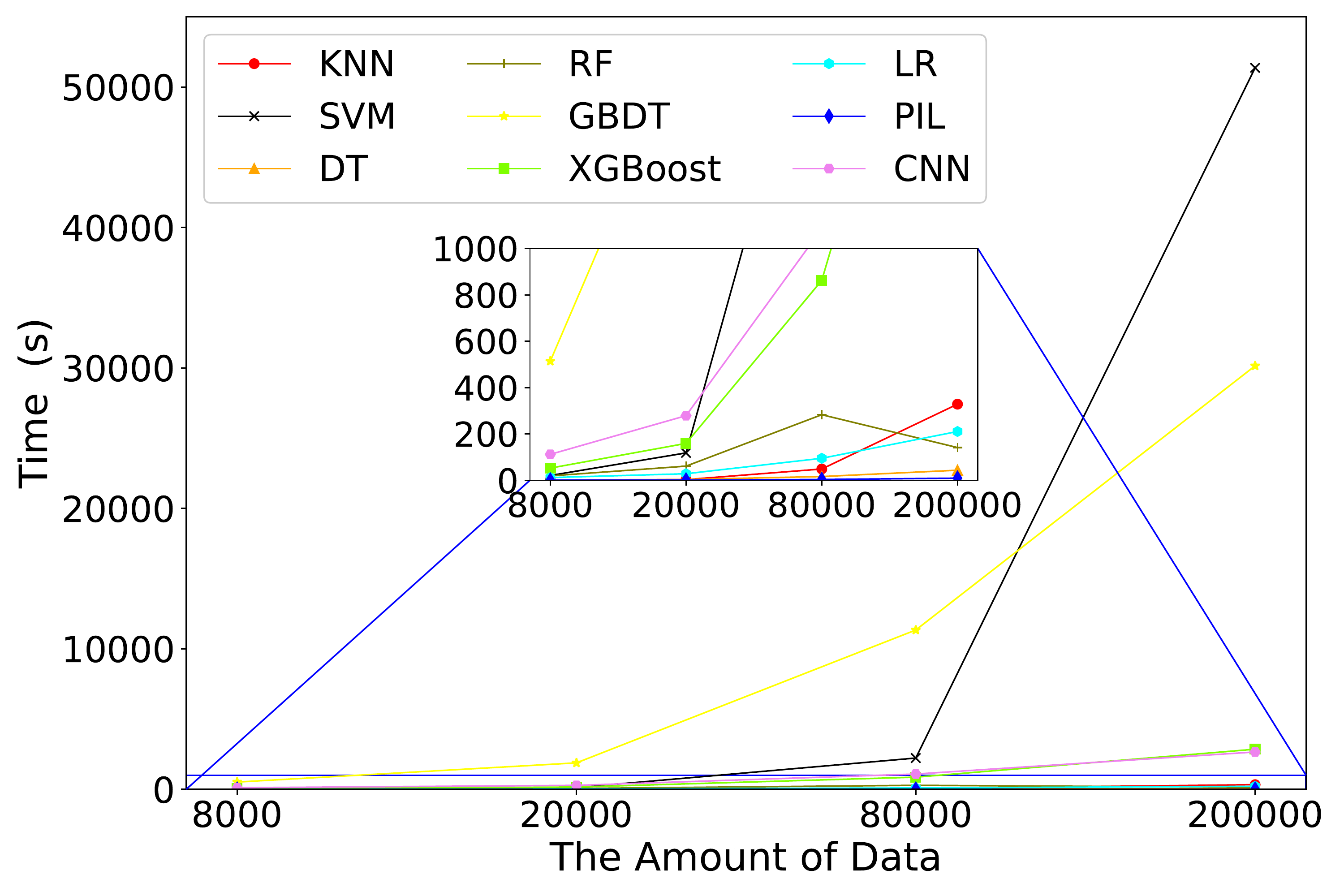}
\caption{Time of algorithms on different data volumes. Different line colors mean various algorithms. Time between 0 and 1000 is clearly shown in the middle rectangle.}
\label{fig:AFGK_time_amount_linechart}   
\end{figure}
Figs. \ref{fig:AFGK.accuracy_volume_raw_10-_bar} and \ref{fig:AFGK_accuracy_amount_linechart} show the performance of nine basic classification algorithms on the four different data volumes.

There is a slight improvement in the accuracy with the increase of data volumes. Because the large number of spectral data will provide more information to obtain better classifiers. 

Fig. \ref{fig:AFGK_time_amount_linechart} shows the computation time of nine basic classification algorithms on four different data volumes. Compared with other algorithms, SVM and CNN spend more time on classification. Besides, the computation time of SVM, CNN and LR  increases rapidly as the data volumes increase.
\begin{figure*}
\centering
\includegraphics[width=11.6cm,height=9.79cm]{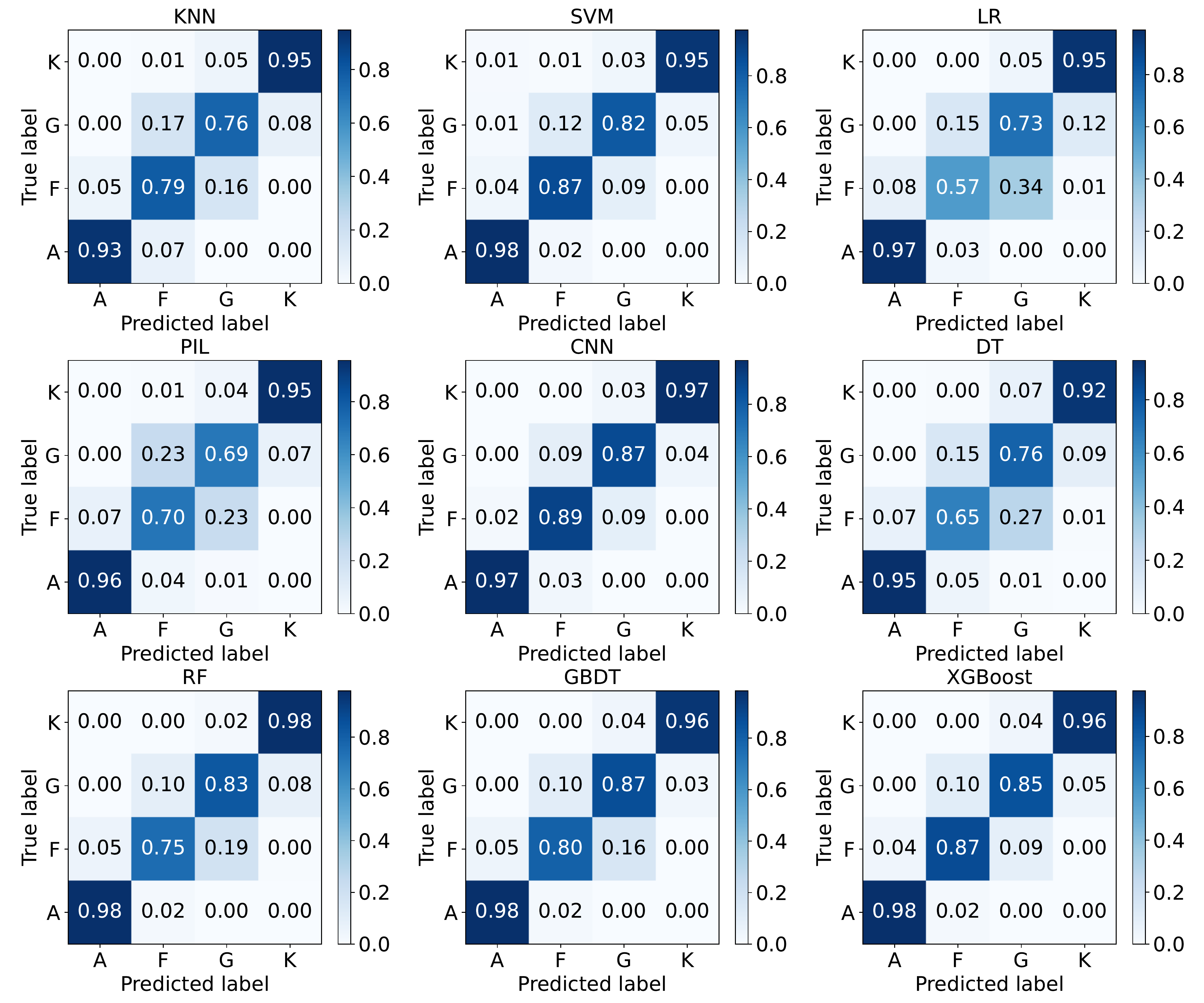}
\caption{Accuracy of algorithms on data volume of 8000 for A/F/G/K stars. x axis represents predicted labels conducted by experiments. y axis represents true labels of spectra. Figures in the grids are the consistent probabilities between predicted labels and true labels. Algorithm names are presented on the above of each confusion matrix.}
\label{fig:AFGK_8000_confusion}   
\end{figure*}

\begin{figure*}

\centering
\includegraphics[width=11.6cm,height=9.79cm]{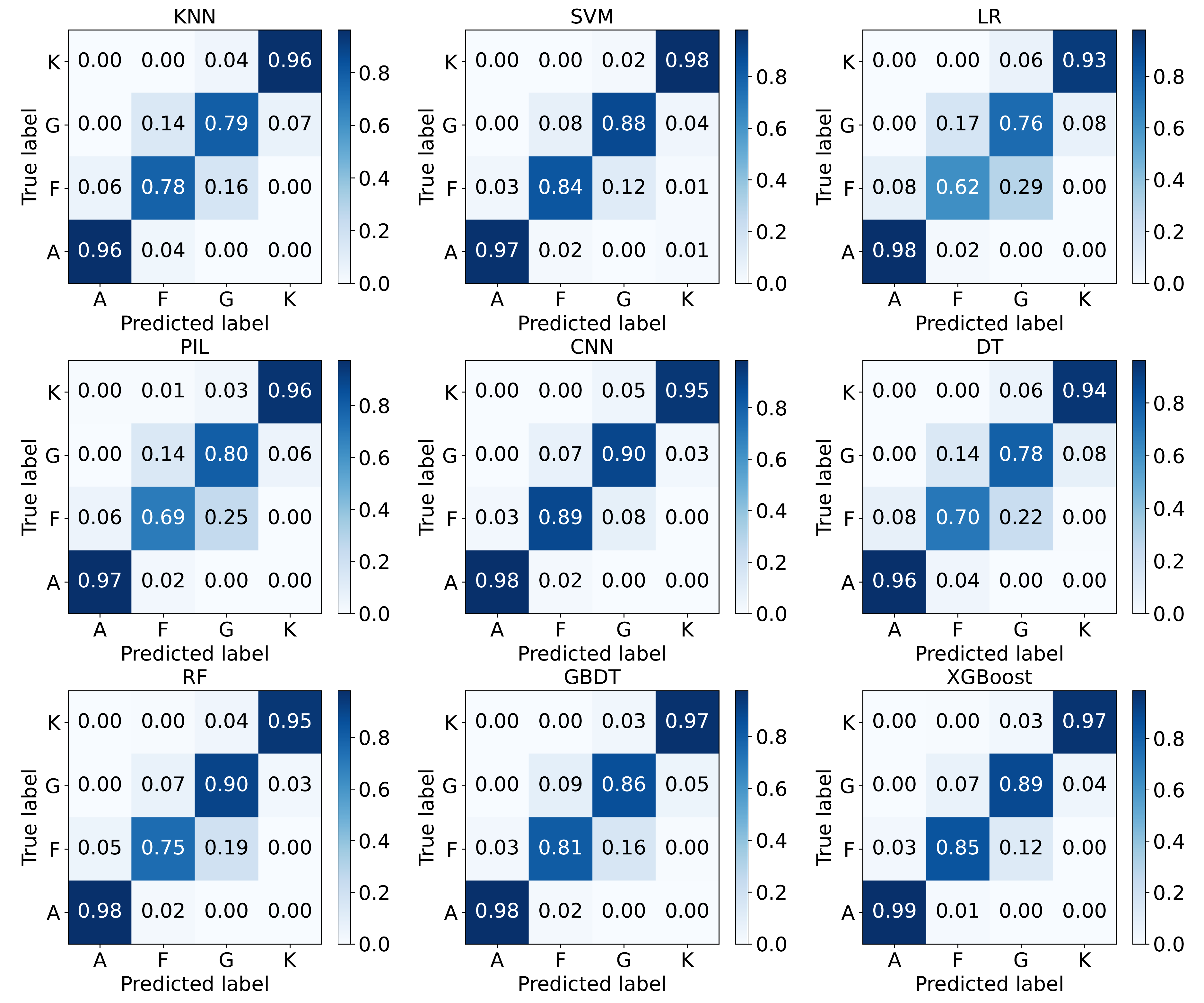}
\caption{Accuracy of algorithms on data volume of 20000 for A/F/G/K stars. x axis represents predicted labels conducted by experiments. y axis represents true labels of spectra. Figures in the grids are the consistent probabilities between predicted labels and true labels. Algorithm names are presented on the above of each confusion matrix.}
\label{fig:AFGK_20000_confusion}   
\end{figure*}
\begin{figure*}
\centering
\includegraphics[width=11.6cm,height=9.79cm]{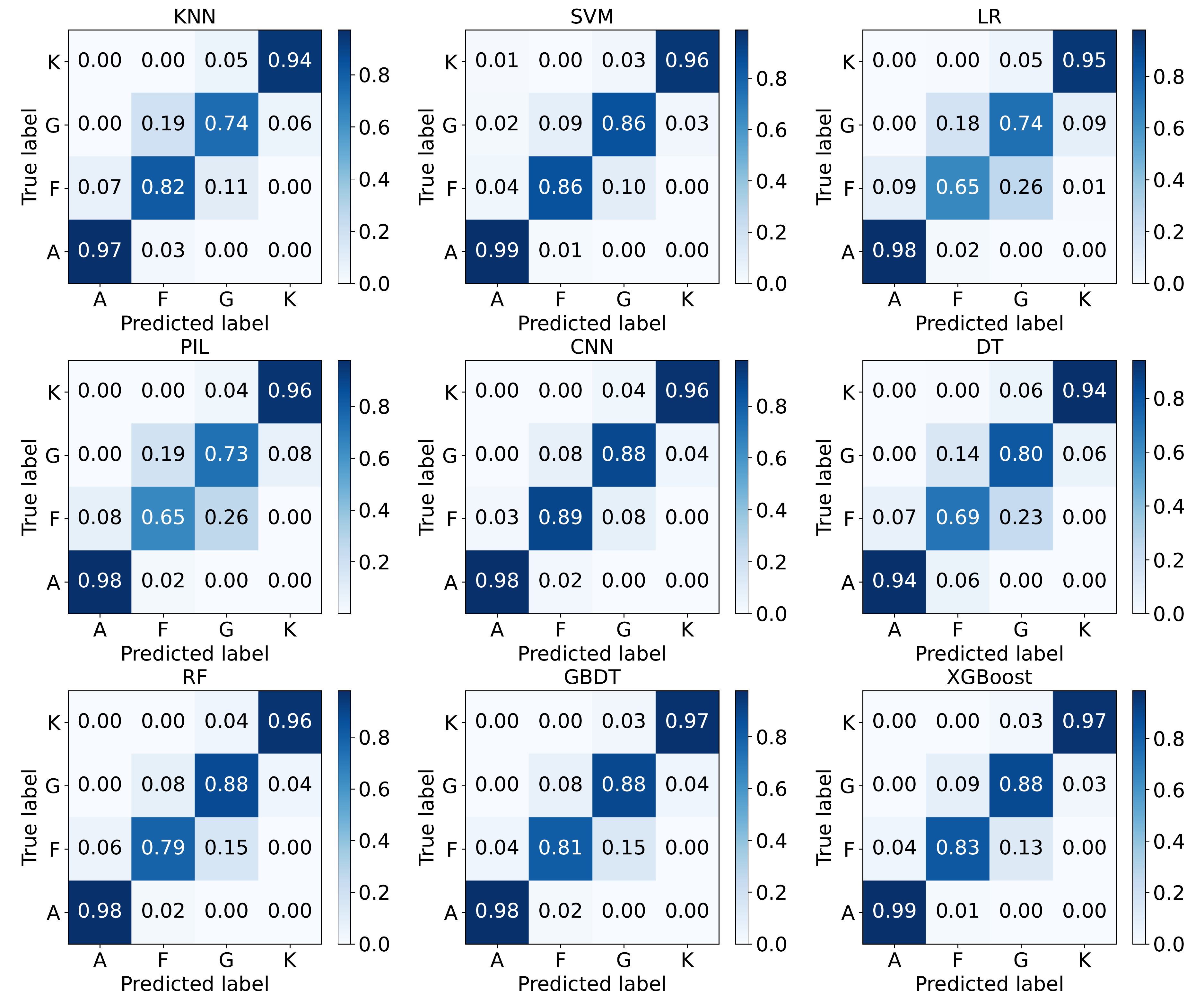}
\caption{Accuracy of algorithms on data volume of 40000 for A/F/G/K stars. x axis represents predicted labels conducted by experiments. y axis represents true labels of spectra. Figures in the grids are the consistent probabilities between predicted labels and true labels. Algorithm names are presented on the above of each confusion matrix.}
\label{fig:AFGK_40000_confusion}   
\end{figure*}
\begin{figure*}
\centering
\includegraphics[width=11.6cm,height=9.79cm]{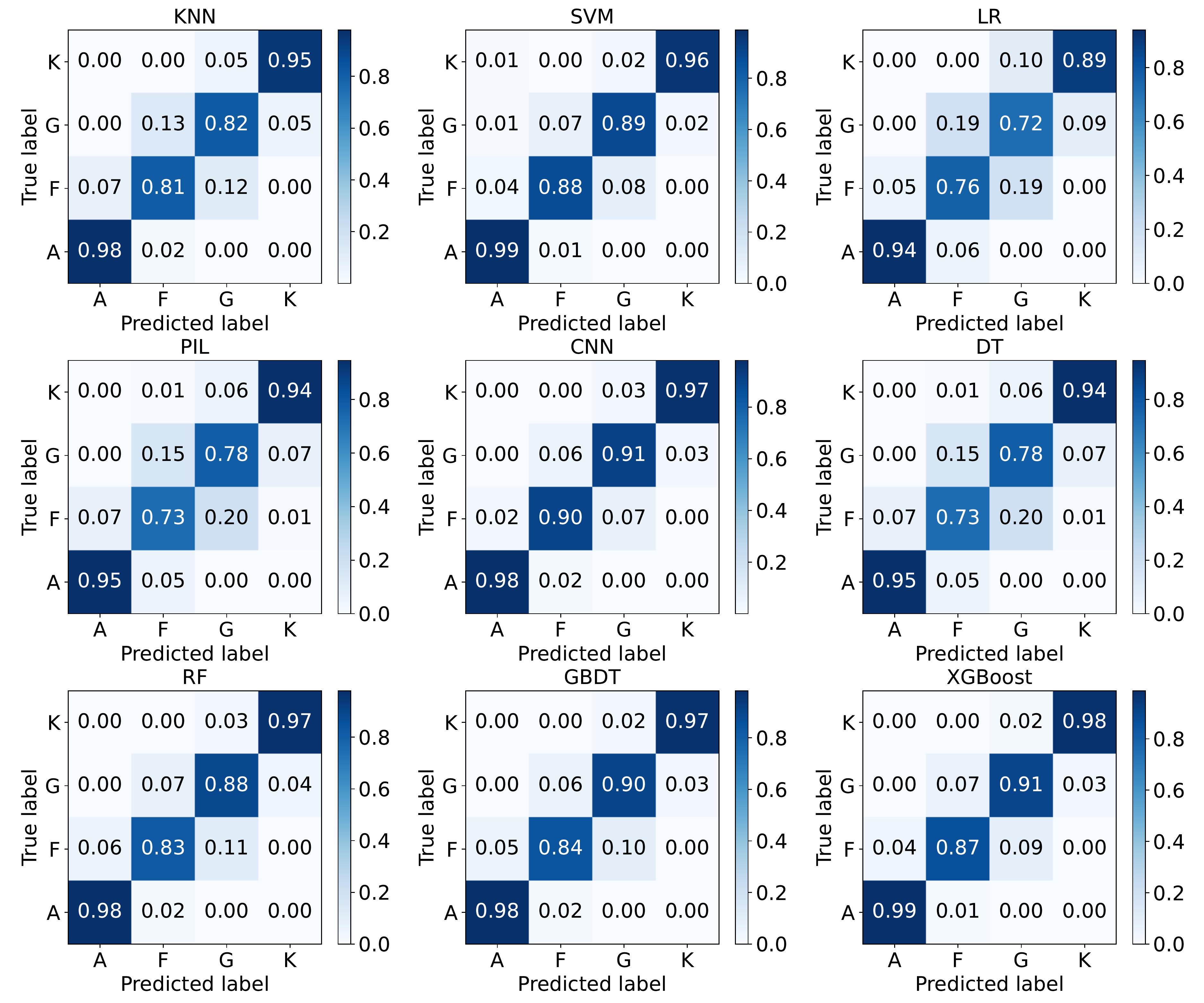}
\caption{Accuracy of algorithms on data volume of 80000 for A/F/G/K stars. x axis represents predicted labels conducted by experiments. y axis represents true labels of spectra. Figures in the grids are the consistent probabilities between predicted labels and true labels. Algorithm names are presented on the above of each confusion matrix.}
\label{fig:AFGK_80000_confusion}   
\end{figure*}
There is little difference in the confusion matrices of different data volumes. And the main misclassification exists between F stars and G stars in Figs. \ref{fig:AFGK_8000_confusion}-\ref{fig:AFGK_80000_confusion}.
\subsubsection{Performance analysis of star, galaxy and quasar classification}
\begin{figure*}
\centering
\includegraphics[width=12.2cm,height=5.08cm]{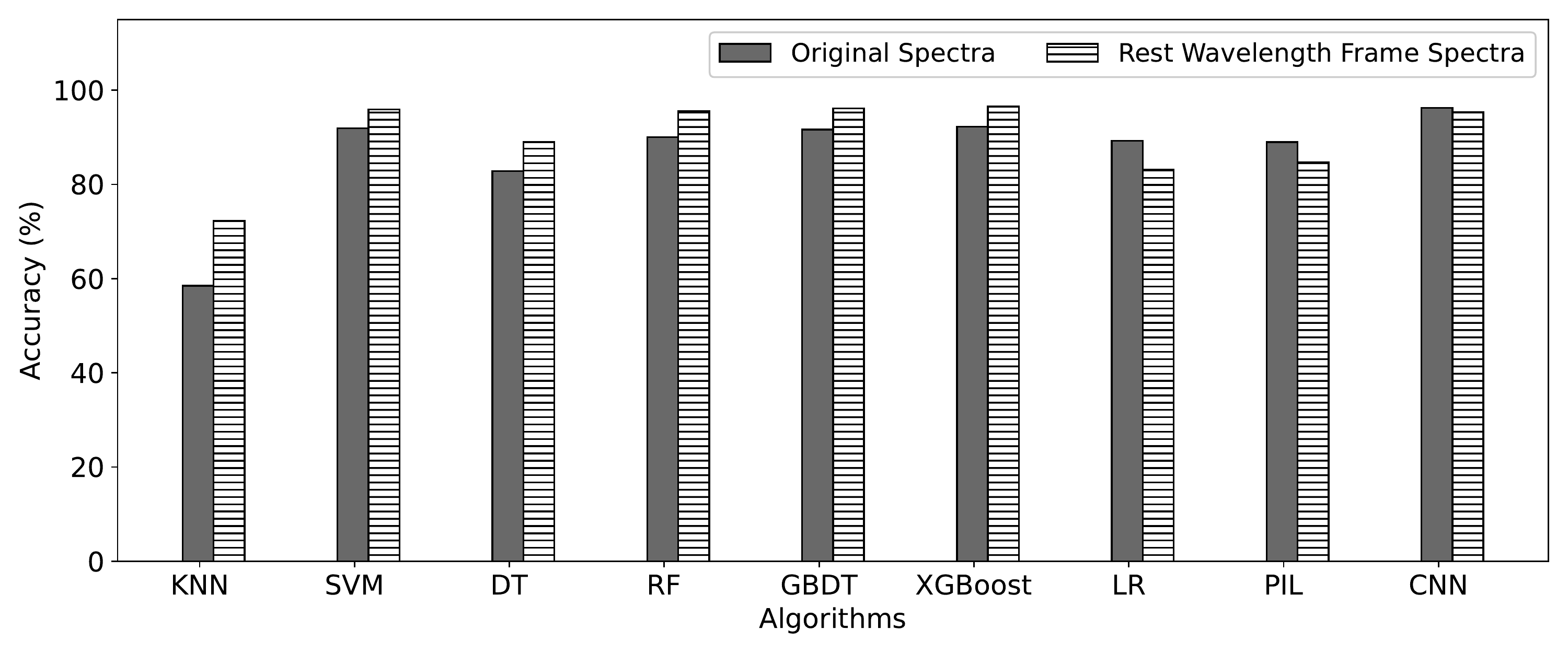}
\caption{Accuracy of algorithms on star/galaxy/quasar on original spectra and rest wavelength frame spectra. Two bars represent two spectra characteristics.}
\label{fig:SGQ_accuracy_15k_10-_raw_bar}   
\end{figure*}

As can be seen from Fig. \ref{fig:SGQ_accuracy_15k_10-_raw_bar}, most classification algorithms perform better on the rest wavelength frame spectra than on the original spectra. Because redshift causes feature shift problems, overlapping phenomena between nearby galaxies and high-velocity stars. These problems affect the performance of classification algorithms on original spectra. We also found that LR, PIL and CNN algorithms perform better on original spectra. Because LR fits more complex polynomials to classify spectra and the other two methods learn deep features for better classification.
So the above issues caused by redshift make little influence on the classification performance of these methods. In addition, the dimensionality of rest wavelength frame spectra is reduced and some information will be lost, which will also lead to poor classification performances of LR, PIL and CNN.

\begin{figure*}
\centering
\includegraphics[width=11.66cm,height=8.9cm]{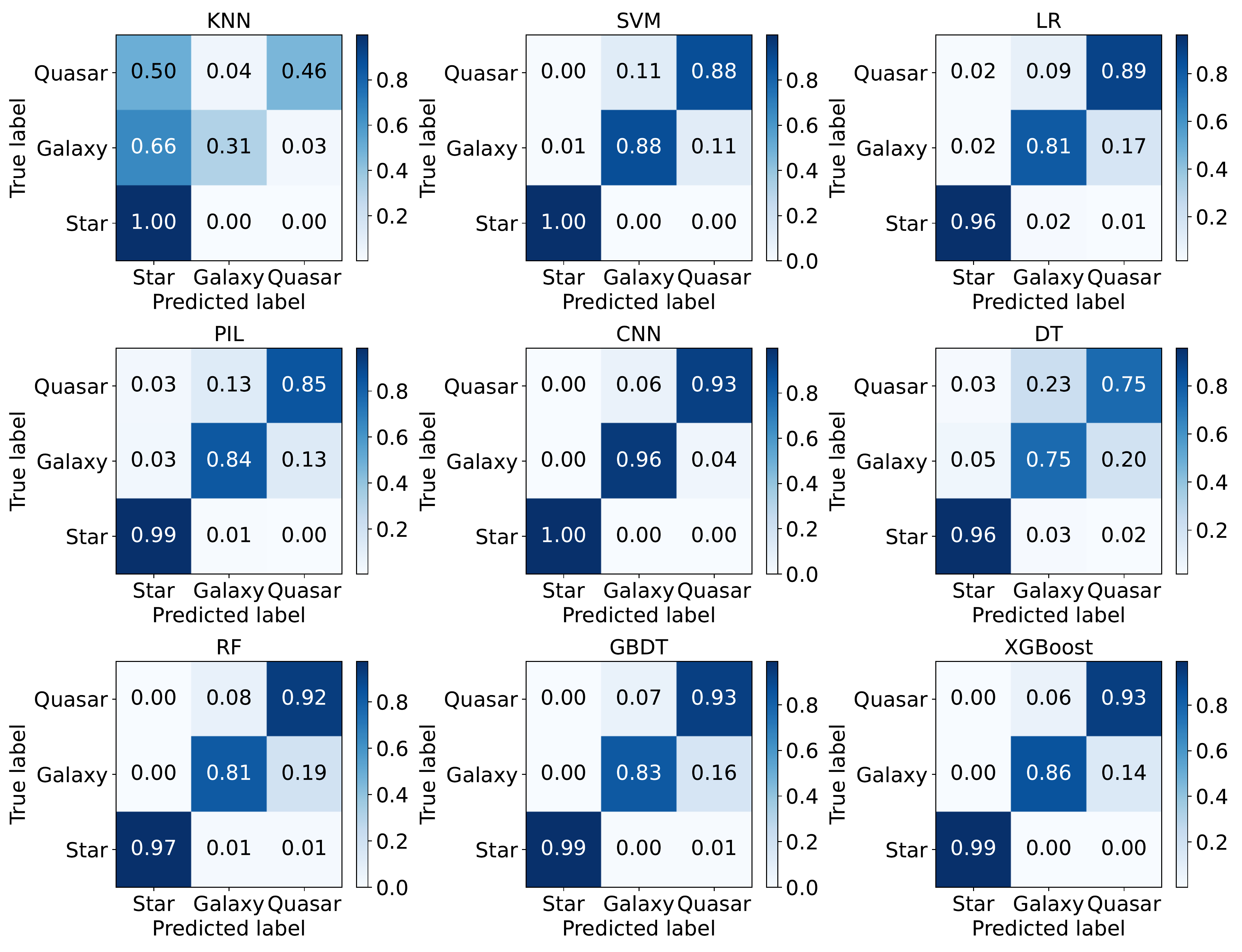}
\caption{Accuracy of algorithms on star/galaxy/quasar on original spectra. x axis represents predicted labels conducted by experiments. y axis represents true labels of spectra. Figures in the grids are the consistent probabilities between predicted labels and true labels. Algorithm names are presented on the above of each confusion matrix.}
\label{fig:SGQ_confusion_1d}   
\end{figure*}

\begin{figure*}
\centering
\includegraphics[width=11.66cm,height=8.9cm]{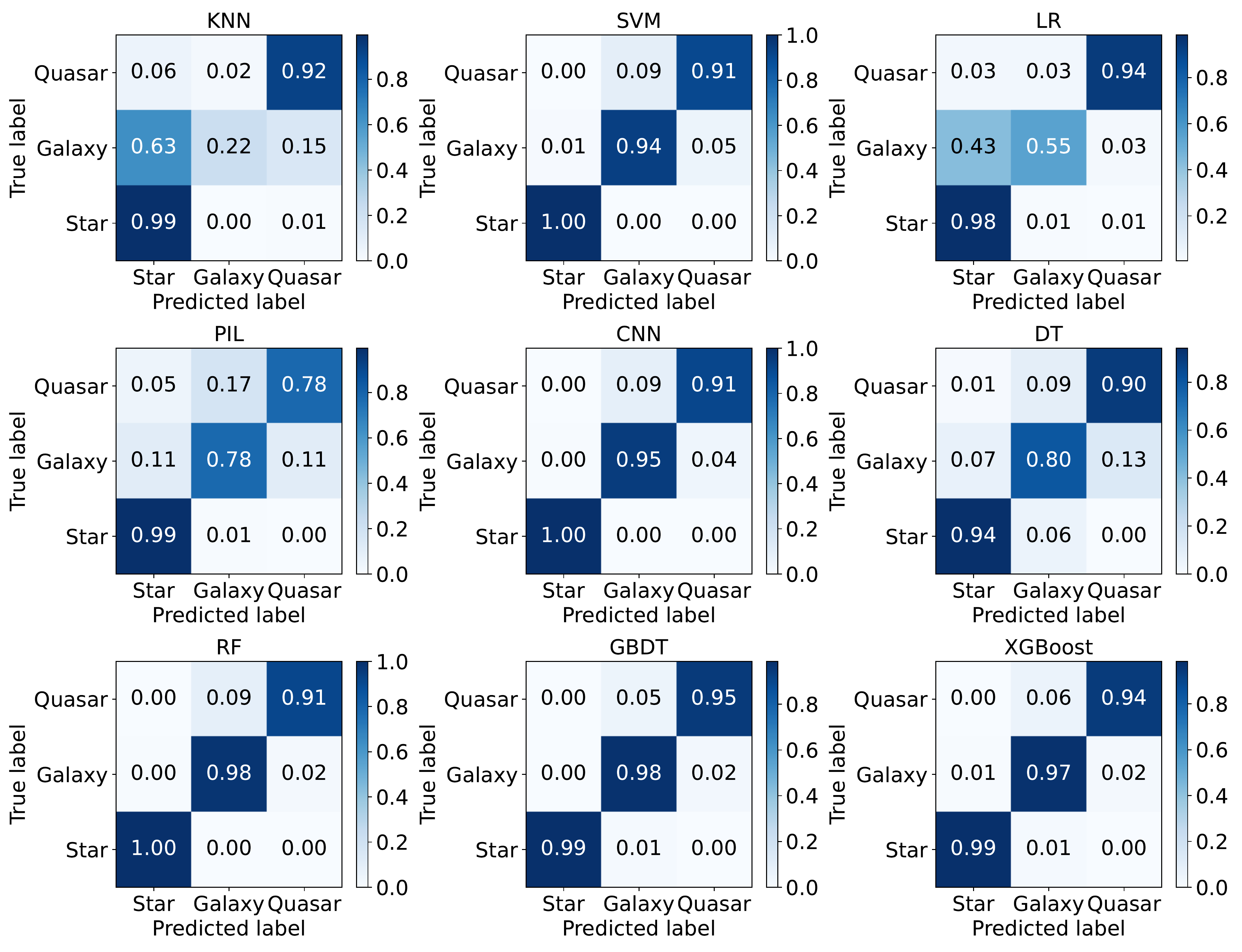}
\caption{Accuracy of algorithms on star/galaxy/quasar on rest wavelength frame spectra. x axis represents predicted labels conducted by experiments. y axis represents true labels of spectra. Figures in the grids are the consistent probabilities between predicted labels and true labels. Algorithm names are presented on the above of each confusion matrix.}
\label{fig:SGQ_confusion_restframe}   
\end{figure*}

Pay more attention on the classification algorithms in Fig. \ref{fig:SGQ_accuracy_15k_10-_raw_bar}, they can be divided into three parts: CNN, SVM, RF, GBDT, XGBoost; DT, LR, PIL; KNN. CNN performed better than others for its powerful ability of feature selection. The classical classifier SVM can also find a suitable hyperplane to separate the rest wavelength frame spectra. Methods such as RF, GBDT, XGBoost can classify rest wavelength frame spectra well due to their integration. Decision tree and random forest can not choose the split nodes well because of the inconsistent features. KNN can not classify galaxy and quasar well. Because the feature lines are inconsistent on spectra shape and position due to redshift. Misclassification can also be found in Figs. \ref{fig:SGQ_confusion_1d} and \ref{fig:SGQ_confusion_restframe}. 


\subsubsection{Performance analysis on rare targets}

\begin{figure*}

\centering
\includegraphics[height=5cm]{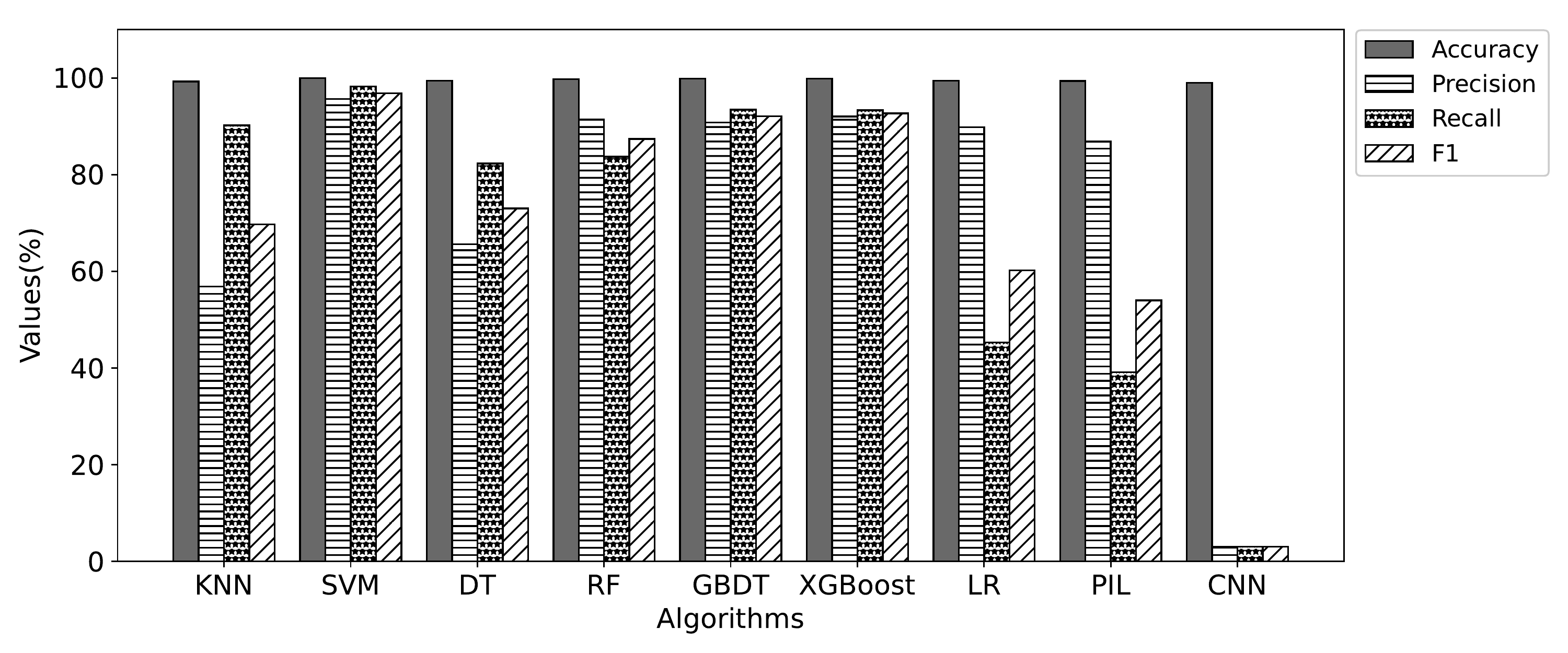}
\caption{Results of algorithms on carbon stars. Four bars are four evaluation criteria of classification results.}

\label{fig:carbon_accuracy_bar_bar}   
\end{figure*}

\begin{figure*}

\centering
\includegraphics[height=5cm]{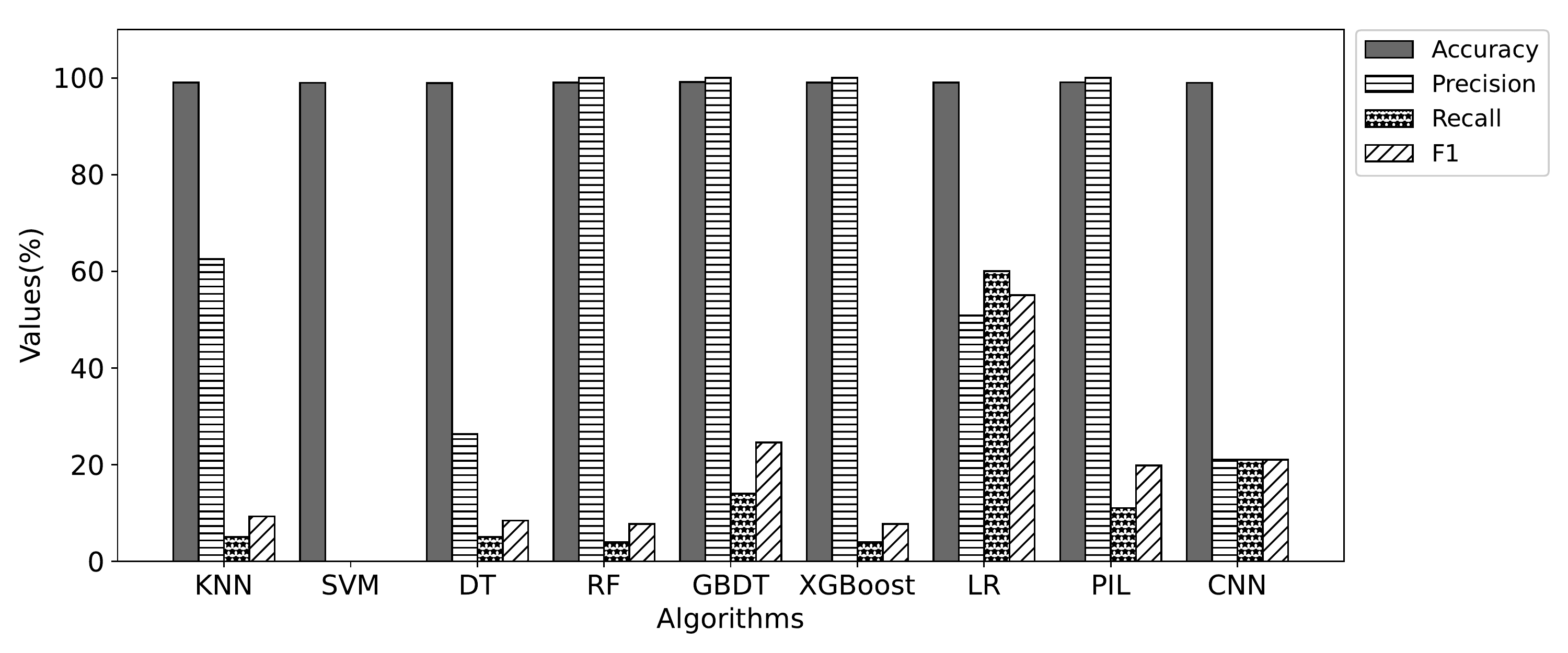}
\caption{Results of algorithms on double stars. Four bars are four evaluation criteria of classification results.}

\label{fig:double_accuracy_bar_bar}   
\end{figure*}


\begin{figure*}

\centering
\includegraphics[height=5cm]{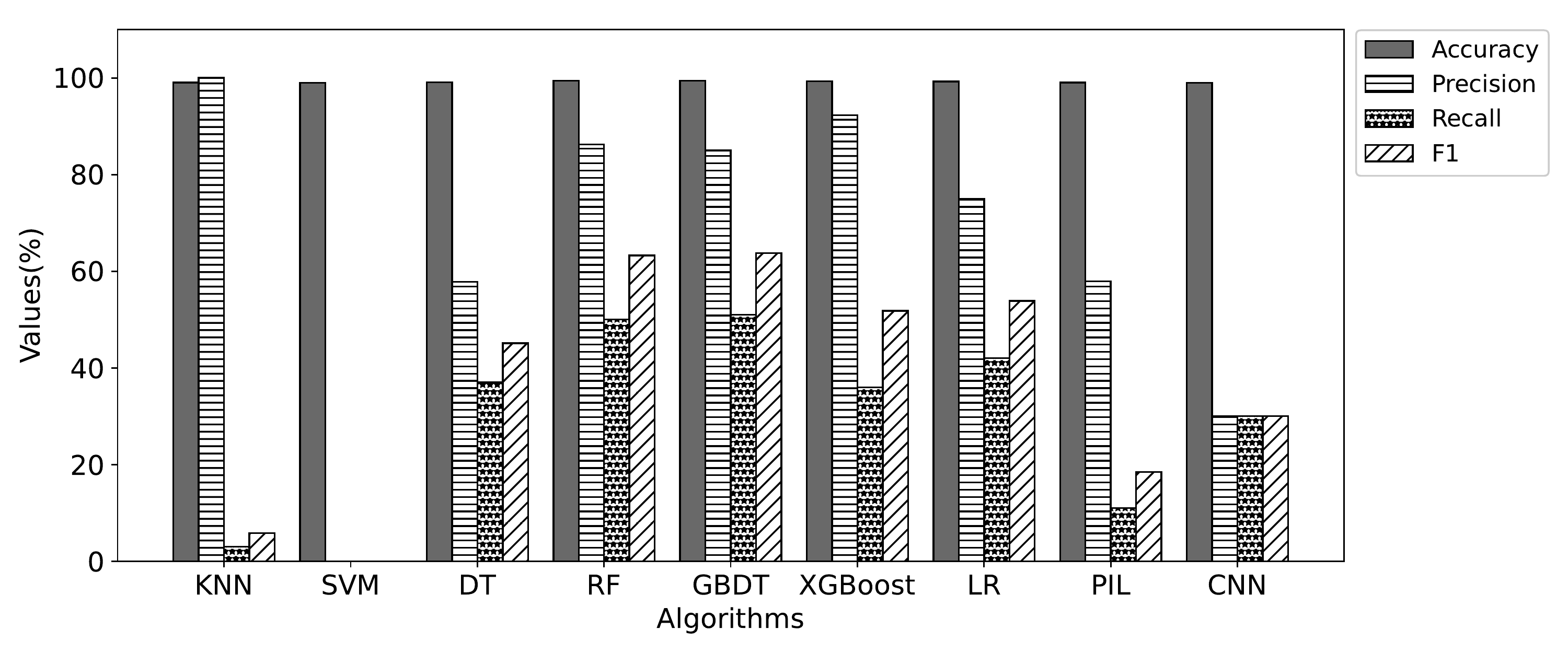}
\caption{Results of algorithms on identifying artefacts. Four bars are four evaluation criteria of classification results.}

\label{fig:red_blue_accuracy_bar_bar}   
\end{figure*}

Compared with the classifications performance of A/F/G/K stars classification on 1D spectra, the classification algorithms perform bad when searching for carbon stars, double stars and identifying artefacts (Figs. \ref{fig:carbon_accuracy_bar_bar}-\ref{fig:red_blue_accuracy_bar_bar}). Because the imbalanced data sets have a bad impact on the classification performance.

Due to the obvious characteristics of carbon stars, classification performance of carbon stars  is better than that of double stars and artefacts. As can be seen from Fig. \ref{fig:carbon_accuracy_bar_bar}, the classic binary classifier SVM and ensemble learning methods(RF, GBDT, XGBoost) perform better than other algorithms.

The classification algorithms have the worst performance in identifying double stars due to the mutual interference of the overlapping parts in double stars. This problem will affect the classification performance on carbon stars. Precision in Fig. \ref{fig:double_accuracy_bar_bar} shows that the ensemble learning (RF, GBDT, XGBoost) can identify some double stars accurately. But the recall in Fig. \ref{fig:double_accuracy_bar_bar} shows that a large number of double stars will be missed.

It can be seen from Fig. \ref{fig:red_blue_accuracy_bar_bar} that classification performance of identifying artefacts is between the carbon stars and double stars. Several ensemble algorithms can also find these rare stars accurately. Compared with the double stars, the recall rate of artefacts is relatively improved. It means that several integration algorithms and KNN can identify more artefacts. But it is inevitable that many artefacts will be missed.

\section{Source Code and Manual}
Source codes used in this paper are provided on \url{https://github.com/shichenhui/SpectraClassification}. Algorithms in the code category are shown in Table \ref{table_sourcemanual}. Because the parameters of algorithms have a significant impact on the classification results, we also provide the parameters of algorithms on each data set and the parameters are optimized by grid-search method provided by sklearn package.

\begin{table*}
\centering
\caption{Source codes notes of classification algorithms}
\label{table_sourcemanual}
\resizebox{\linewidth}{!}{
\begin{tabular}{llllllllll} 
\hline
Algorithms         & KNN    & SVM    & LR    & DT    & RF    & GBDT    & XGBoost    & CNN    & PIL     \\ 
\hline
Source files       & KNN.py & SVM.py & LR.py & DT.py & RF.py & GBDT.py & XGBoost.py & CNN.py & PIL.py  \\ 
\hline
Python version     & \multicolumn{9}{c}{python3.8}                                                     \\ 
\hline
Dependent packages & \multicolumn{9}{c}{NUMPY; PANDAS; SKLEARN; SCIPY; PYTORCH}                        \\
\hline
\end{tabular}
}
\end{table*}

The codes are written in python which is widely used for machine learning and data analysis. Dependent packages of our codes include numpy \citep{harris2020array}, sklearn, matplotlib \citep{hunter2007matplotlib}, pandas, scipy. Each algorithm is organized by the following steps: 1) load training data sets and testing data sets; 2) configure the parameters of classification models; 3) train models on the training data sets; 4) classify the testing data sets by training models; 5) evaluate the performance of training models. To avoid the influence of sample selection on the training data sets, we use 5-fold cross validation to split data sets and evaluate models. But this is not necessary for practical applications.

These codes load data from *.csv files which store tabular data in the form of text. And a row of data is a spectrum. You need to convert your spectra data into this format or modify the data loading mode. Some basic algorithms are directly implemented from sklearn packages.

The parameter K of KNN is not a fixed value (default value in sklearn is 5). Generally, a smaller value is often selected according to the sample distributions. And an appropriate K value can be selected by cross-validation. Besides, it adopts Euclidean distance as distance metrics to get good results in low dimensional space. Other distance metrics can also be applied in KNN to avoid the disadvantage of Euclidean distance.

SVM needs to select kernel functions. There are many kernel functions: linear kernel function, polygon kernel function, RBF kernel function, sigmoid kernel function, etc. The current improvement of SVM is  combined with other methods to classify the large-scale data sets.

Feature selection criteria and feature splitting criteria are two important parameters of decision tree. Different feature selection methods (information entropy, information gain, Gini index) correspond to different decision trees. Features splitting parameters can be "best" or "random". The former is to find the optimal division point from all division points of the features, the latter is to find the local optimal division point from the randomly selected division points. Generally, "best" is often used for the small number of samples and "random" for the large number of samples. Other parameters like tree depth and the number of trees are also needed to be determined. 

Ensemble learning algorithms (i.e. random forest, GBDT and XGBoost) are integrated by decision trees. We need to choose the number of integrated trees. Methods in sklearn use 100 decision trees by default. But GBDT can not be parallel, we need to reduce the number of decision tree appropriately. Other parameters in decision tree can be set up according to the introduction in the previous paragraph. 

Logistics regression is a binary classifier. It integrates multiple LR classifiers for multi-classification tasks. The integration strategy is always "OVR". And it uses "L1" and "L2" regularization to reduce overfitting. "L2" is more commonly used. But for high dimensional data, "L1" penalty can help you reduce the impact of unimportant features.

The good design of neural network structures is important for ANN based methods. We find that one-dimensional convolutional structure can extract spectral features well. So for spectral classification, the performance of CNN is better than that of fully connected neural network. There are many layers in computer vision. But for data in the format of vector, we do not need to stack too many layers in the neural network structures. Likewise, "L1" and "L2" regularization can be used to reduce overfitting. 

\section{DISCUSSION}
In this paper, we investigate the classification methods used for astronomical spectra data. We introduce the main ideas, advantages, caveats and applications of classification methods. And data sets are designed by data characteristics, data qualities and data volumes. Besides, we experiment with nine basic algorithms (KNN, SVM, LR, PIL, CNN, DT, RF, GBDT, XGBoost) on A/F/G/K stars classification, star/galaxy/quasar classification and rare object identification. Experiments on data characteristics also include the comparative experiments on the matching sources from the LAMOST survey and SDSS survey.

For A/F/G/K stars classification, the accuracy on 1D spectra and PCA shows little difference while PCA spends less time in the training stage. Because it reduces the spectra dimensionality. So PCA is often used to classify large-scale and high dimensional data sets. Among nine basic methods, CNN performs best on 1D spectra and PCA, due to its powerful ability for feature selection. For the classification on line indices, KNN shows superiority among other methods. The performance of classification on SDSS is better than that on LAMOST. Because the calibration quality of LAMOST is undesirable, which is affected by many factors (i.e. fiber-to-fiber sensitivity variations). In addition, high-quality spectra and a large number of samples help us to train models. But with the growth of data volumes, the training time of some models will also increase greatly. So it is necessary to improve the classification speed on large-scale data sets.
 
As for star/galaxy/quasar classification, most performance of classification on rest wavelength frame spectra is better than that on original spectra. Because redshift causes feature movement on original spectra. But for some algorithms (PIL, LR, CNN), the performance of classification on the original spectra is better than that on the rest wavelength frame spectra. Because original spectra have much information. These methods can extract feature well and are less influenced by redshift. For this task, SVM which is good at binary classification and CNN with powerful ability for feature selection perform better than other methods.  

It is difficult to identify carbon stars, double stars and artefacts due to the unbalanced data distributions. Among these three rare objects, the performance of identifying carbon stars is better than others due to their obvious characteristics. The performance of searching for double stars is the worst. In short, researchers need to find other methods for rare object identification.

In this paper, we only evaluate the classification performance of nine basic algorithms on astronomical spectra. Other effective methods still need to be analysed in the future. And experimental results in this paper can only provide a reference to researchers. In practical application scenarios, researchers need to choose appropriate methods according to their data characteristics.

\section*{Acknowledgements}

The authors wish to thank the reviewer, Igor V Chilingarian, for his very helpful comments and suggestions.

The Guo Shou Jing Telescope (the Large Sky Area Multi-Object Fiber Spectroscopic Telescope, LAMOST) is a National Major Scientific Project built by the Chinese Academy of Sciences. Funding for the project has been provided by the National Development and Reform Commission. LAMOST is operated and managed by National Astronomical Observatories, Chinese Academy of Sciences.

The work is supported by the National Natural Science Foundation of China (Grant No. U1931209), Key Research and Development Projects of Shanxi Province (Grant No. 201903D121116), and the central government guides local science and technology development funds (Grant No. 20201070). Fundamental Research Program of Shanxi Province(Grant Nos. 20210302123223, 202103021224275).

\section*{Data Availability}


Experimental data used for this work is obtained from The Guo Shou Jing Telescope (the Large Sky Area Multi-Object Fiber Spectroscopic Telescope, LAMOST) Data Release 8 (\url{http://www.lamost.org/lmusers/}) and Sloan Digital Sky Survey (SDSS) Data Release 16 (\url{https://www.sdss.org/}). Codes used in this paper is also available online at \url{https://github.com/shichenhui/SpectraClassification}.



\bibliographystyle{mnras}
\bibliography{example} 




%
%


\bsp	
\label{lastpage}
\end{document}